\documentstyle[12pt,epsfig]{ioplppt}

\begin{document}
\jl{2}

\title{Photon emission by an ultra-relativistic particle channeling
in a periodically bent crystal
\footnote{published in
Int.\ J.\ Mod.\ Phys.\ E, Vol. 8, No. 1 (February 1999) 49--100
}
}

\author{Andrei V Korol\dag\P\ftnote{4}
{E-mail: korol@rpro.ioffe.rssi.ru, \  korol@th.physik.uni-frankfurt.de},
Andrey V Solov'yov\ddag\ftnote{3}{E-mail:
solovyov@rpro.ioffe.rssi.ru},
and  Walter Greiner \P}

\address{\dag Department of Physics,
St.Petersburg State Maritime Technical University,
Leninskii prospect 101, St. Petersburg 198262, Russia}
\address{\ddag A.F.Ioffe Physical-Technical Institute of the Academy
of Sciences of Russia, Polytechnicheskaya 26, St. Petersburg 194021,
 Russia}
\address{\P Institut f\"{u}r Theoretische Physik der Johann Wolfgang
Goethe-Universit\"{a}t,
60054 Frankfurt am Main, Germany}

% Abstract
\begin{abstract}
This paper is devoted to a detailed analysis of the new type of the
undulator radiation generated by an ultra-relativistic charged
particle channeling along a crystal plane, which is periodically bent
by a transverse acoustic wave, as well as to the conditions limiting
the observation of this phenomenon.  This mechanism makes feasible the
generation of electromagnetic radiation, both spontaneous and
stimulated, emitted in a wide range of the photon energies, from $X$-
up to $\gamma$-rays.
\end{abstract}


%%%%%%%%%%%%%%%%%%%%%%%%%%%%%%%%%%%%%%%%%%%%%%%%%%%%%%%%%%%%%%%%%
\section{Introduction}
%%%%%%%%%%%%%%%%%%%%%%%%%%%%%%%%%%%%%%%%%%%%%%%%%%%%%%%%%%%%%%%%%

This paper is devoted to a detailed analysis of the new
type of the undulator radiation (acoustically induced radiation
-- AIR) generated by an ultra-relativistic charged particle
channeling along a crystal plane, which is periodically bent by a
transverse acoustic wave (AW), as well as to the conditions
limiting the observation of this phenomenon.
This mechanism, suggested recently in \cite{air,laser}, makes feasible
the generation of electromagnetic radiation, both spontaneous and
stimulated, emitted in a wide range of the photon energies,
from $X$- up to $\gamma$-rays.

\begin{figure} %%%%%%%%%%%%% Figure 1.1: Acoustically bent channel
\hspace{3cm}\epsfig{file=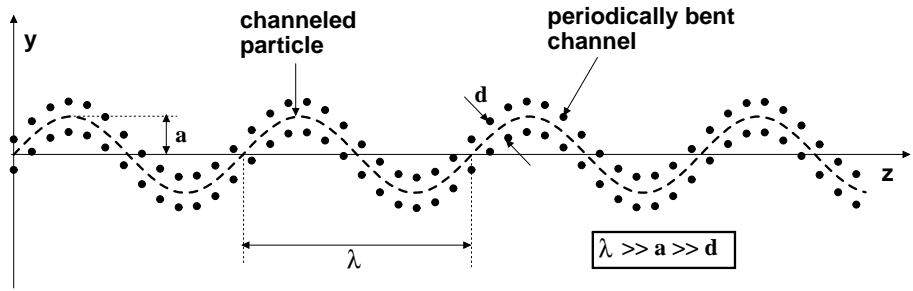,height=12cm,angle=270}
\caption{Schematic representation of the initially linear planar
channel bent by the transverse acoustic wave.
The notations are: $d$ is the channel width, $a, \lambda$ are the AW
amplitude and wavelength, respectively.}
\label{Fig1}
\end{figure}

The mechanism of the AIR generation is illustrated in \fref{Fig1}.
Under the action of a transverse acoustic wave propagating along the
$z$-direction, which defines the center line of an initially straight
channel (not plotted in the figure) the channel becomes periodically
bent.  Provided certain conditions are fulfilled (see below in this
section and in \sref{Undulator}), the beam of positrons, which enters
the crystal at a small incident angle with respect to the curved
crystallographic plane, will penetrate through the crystal following
the bendings of its channel.  It results in the transverse
oscillations of the beam particles while travelling along the $z$
axis.  These oscillations become an effective source of spontaneous
radiation of undulator type due to the constructive interference of
the photons emitted from similar parts of the trajectory.  As we
demonstrate below, the number of oscillations can vary in a wide range
from a few up to a few thousands per $cm$ depending on the the beam
energy, the AW amplitude, $a$, and wavelength, $\lambda$, the type of
the crystal and the crystallographic plane.  In addition to the
spontaneous photon emission by the undulator, the scheme presented in
\fref{Fig1} leads to a possibility to generate stimulated
emission. This is due to the fact, that photons, emitted at the points
of the maximum curvature of the trajectory, travel almost parallel to
the beam and, thus, stimulate the photon generation in the vicinity of
all successive maxima and minima of the trajectory.

%%%%%%%%%%%%%%%%%%%%%%%%%%%%%%%%%%%%%%%%%%%%%%%%%%%%%%%%%%%%%%%%%%%%%%%%
Some aspects of the electromagnetic radiation by a beam of charged
particles channeling in a crystal undergoing the action of an
ultrasonic wave have been discussed in the literature
 \cite{Barysh80}--\cite{Dedkov94}.
We shall mention the main results obtained in the cited papers
and outline the features, both of formal and of principal nature,
which distinguish the AIR radiation described above from the effects
considered in  \cite{Barysh80}--\cite{Dedkov94} at the end of
this section.

Prior to that let us briefly discuss the phenomena closely related to
the mechanism of generating the acoustically induced radiation.
These are the channeling effect, the channeling radiation and the
undulator radiation.

The basic effect of the channeling process in a straight crystal
is that a charged particle can penetrate an anomalously large distance
inside a crystal if travelling nearly parallel to the crystallographic
plane or axis and experiencing the collective action of the electric
field of the lattice ions.
Positively charged particles are steered into the interatomic region
(the planar channeling) while negatively charged projectiles move in
vicinity of the ionic chains (the axial channeling).

Channeling  was discovered in the early 1960s by  computer
simulations of ion motion in crystals \cite{Robinson1,Robinson2}.
Large penetration lengths were obtained for ions incident along
crystallographic directions of low indexes.
It is interesting to note that such a guided motion had already been
predicted \cite{Stark} by Stark in 1912 \cite{Stark}.
A comprehensive theoretical study \cite{Lindhard,Lehmann}
introduced  the important continuum approximation for the interaction
potentials between energetic charged projectiles and lattice atoms
arranged in atomic strings and planes.
These concepts were subsequently widely used in the interpretation of
channeling experiments (see e.g. the review article \cite{Gemmell}).
During recent years particular attention has been paid to studying of
the channeling phenomenon at high energies of the particles
\cite{RelCha, Sorensen89, Instrum}).

The idea of bending the high-energy beams of charged particles,
proposed in \cite{Tsyganov}, has become of great current interest
\cite{RelCha, Instrum},  \cite{Elishev}-\cite{Uggerhoj}, \cite{Biryukov}
because of its practical application for the manipulation of
the beams.
Indeed, the high-energy beams can be bent by crystals
as efficiently as by means of external electric or magnetic fields.
The size of such crystals is typically hundreds times smaller than the size
of the equivalent magnet.  In recent experiments with 450 GeV protons
\cite{Biino} the efficiency of the particle beam deflection was reported on
the level of 60\%.  

Another possible application of bent crystal was suggested in
\cite{Schafer}, where the question whether the channeling effect can
also be used to focus beams, especially heavy-ion beams, was discussed.
To this end a crystal is needed in which the crystal axes are no
longer parallel, but are slanted more and more the farther away they
are from the axis of the beam.
Then the bending angle of the particles far away from the beam axis would
be largest and a general focusing effect will result.
Such a crystal can in principle be produced by varying the germanium
to silicon ratio in a mixed crystal \cite{Schafer}.

The criterion for a stable channeling of an ultra-relativistic
particle in a bent crystal was formulated in \cite{Tsyganov} and has
clear physical meaning: the maximum centrifugal force due to the
channel bending must be less than that of the interplanar
field.
Provided this condition is fulfilled, the beam of (positively charged)
channeling particles at each instant moves inside the channel,
especially parallel to the bent crystal midplane as it does in the
case of linear channeling.

Two realistic ways of ``preparation'' of a periodically bent channel
may be discussed.
It is feasible, by means of mordern technology \cite{Biryukov}, to
grow the crystal with its channels been statically bent according to
a particular pattern.
The channeling phenomenon has been also considered in the periodic
structures such as superlattices  \cite{Ikezi84, Ketterson86}.
THe possibility \cite{air,laser}, \cite{Barysh80} -
\cite{Dedkov94}, of a dynamically bent crystal, arises if one
considers propagation of an acoustic wave along some particular
direction in a crystal as shown in \fref{Fig1}.
Under the action of the AW the crystal channels, being linear initially,
will be periodically bent.
In both cases the passage of an ultra-relativistic particle along the
bent channel gives rise to the AIR phenomenon due to the
curvature of the trajectory, provided that the projectile is trapped
into the channel.
The latter condition is subject to the general criterion for the
channeling process in a bent crystal \cite{Tsyganov} (see also
\cite{Barysh80, Biryukov, Solov}), and can be fulfilled by a proper
choice of projectile energy and maximal curvature of the channel,
as described below in the paper.

The advantage of the static channel is that its parameters are fixed
and, thus, the projectile moves along the fixed trajectory as well.
To calculate the characteristics of the emitted radiation one needs
to know only the number of the periods and the local cuvature radius.
The disadvantage is that when fixing the number of periods then
the parameters of the system can be varied only by changing the energy
of the particle.
This makes the photon generation less tunable.

The motion of a particle in the field of the channel bent by
transverse AW becomes periodic and, therefore, the spectrum of
radiation emitted by a projectile acquires essential
features of the undulator radiation.

It is natural that the AIR emission is accompanied by the ordinary
channeling radiation \cite{Kumakhov1,Barysh77}.
This specific type of electromagnetic radiation arises due to the
transverse motion of the channeling particle inside the channel under
the action of the interplanar field.
The phenomena of the channeling radiation of a charged projectile in
a linear crystal (see eg. \cite{Kniga,Baier,Kumakhov2})
as well as in a ``simple'' (i.e. non-periodically) bent channel
\cite{Solov,Arutyunov}, are known, although in the latter case
the theoretical and experimental data are scarce, at least up to now.

We demonstrate in our paper that the AIR and the ordinary channeling
radiation can be separated provided the condition $a\gg d$ is
fulfilled (see \fref{Fig1}).
Then the frequency of the particle oscillations inside the channel
is much higher than the frequency of the transverse oscillations
caused by the AW.
Therefore, these two motions are well  separated and the AIR mechanism
can be treated independently from the ordinary channeling radiation.
We shall discuss this important issue in more detail in \sref{Undulator}.
Here let us only note that a similar situation occurs in a one-arc
bent crystal, where the channeling charged particle generates
additional synchrotron type radiation due to  the curvature
of the channel \cite{Solov,Arutyunov}.
This component of the total radiation intensity leads to the undulator
effect in the channel periodically bent by AW.
We will show that the intensity of AIR can be made larger than that
of the channeling radiation.

As was pointed out in \cite{air,laser},  the system ``ultra-relativistic
charged particle + periodically bent crystal channel'' represents by
itself a new type of undulator, and, consequently, serves as a
new source of undulator radiation of high intensity,
monochromaticity and with a particular pattern of the
angular-frequency distribution.
The electromagnetic radiation in this undulator arises mainly due to
the bending of the particle trajectory which follows the shape of the
channel.
The parameters of this undulator, as well as the characteristics of
the electromagnetic radiation, depend on the type of the crystal and
on the crystallographic plane, on the type of projectile and its
energy; they also depend on the shape of the bent channel,  and, thus,
can  be varied significantly by varying the enlisted characteristics.

The theory and also the various practical implementations of the
undulator radiation, e.g. the radiation emitted by a charge moving in
spatially periodic static magnetic fields (a magnetic undulator), in
a laser field (a laser-based undulator), etc. have a long history
\cite{Ginz,Motz} and are well elaborated
\cite{Kniga,Baier,Alferov,Fedorov}.
It is also known \cite{Kumakhov3} that a positively charged
ultra-relativistic particle, undergoing a planar channeling in a
linear crystal, radiates electromagnetic waves whose spectral and
angular distributions are those of the undulator (a natural undulator)
due to  the transverse oscillations caused by the action of the
repulsive interplanar potential.
Also the characteristics of the radiation emitted in a natural
undulator are very sensitive  to the distribution of the particles
in the beam over the transverse energy and the incident angle.

The important feature of the dynamically bent crystal by means of an
AW is that it allows  to consider an undulator with the parameters
$N_{\rm u}$ and $p$ (here $N_{\rm u}$ is the number of periods in the
undulator and $p=2\pi\gamma\, a/\lambda$ is its parameter
(see \fref{Fig1}), and $\gamma$ is the relativistic factor)
 varying over a wide range, which is determined not
only by the projectile's energy but also by the AW frequency and
amplitude.
The latter two quantities can easily be tuned resulting in the
possibility of significantly varying the intensity and shape of the
angular distribution of the radiation.
It is important to note that the parameters of the acoustically based
undulator are inaccessible in the undulators based on the motion of
charged particles in periodic magnetic fields and also in the
field of laser radiation \cite{Kniga,Alferov,NIM97}.

In the suggested scheme AIR is generated by the relativistic charged
particles, with relativistic factors $\gamma=\varepsilon/mc^2\gg 1$
($c$ is the velocity of light, $m$ is the mass of the particle and
$\varepsilon$ is its energy).
The large range of $\gamma$ available in modern colliders for various
charged particles, both light and heavy, together with the wide range
of frequencies and the amplitudes possible for AW in crystals
allow to generate AIR photons with the energies up into the GeV region
\cite{air}.

As demonstrated in \cite{laser},  the specific pattern of the
undulator radiation combined with the AIR mechanism allows
to discuss the possibility to create a powerful source of stimulated
monochromatic radiation of high-energy photons.
We analyse this important possibility in more detail in the present article.
Particular attention is paid to the investigation of the influence of
various physical processes, such as the photon attenuation and the
dechanneling effect of positrons in crystals, both leading to a
reduction in the amplification of the stimulated photon emission.

Let us briefly outline here the assumptions, which are adopted
in this paper and formulate the main conditions which allow to consider
the AW bent crystal as an undulator.

Although the general treatment of both the undulator radiation
(\sref{Formalism}) and the channeling process in the AW bent crystal
(\sref{Undulator}) can be applied to axial and a planar channeling,
we consider the latter case because it is known (see
e.g. \cite{Biryukov}) that for bent crystals negatively charged
projectiles are steered along crystallographic axes much less
efficiently than positively charged ones along the crystal planes.
In fact, experimentally the axial channeling of negative particles has not been
observed so far.

Two important conditions which we assume to be fulfilled are the
following:

\begin{enumerate}

\item[a)]\
The bending of the channel becomes significant if the AW amplitude
is noticeably larger than the interplanar spacing, $d$
(see \fref{Fig1}). Thus, the strong inequality
\begin{equation}
a \gg d
\label{AW_0}
\end{equation}
is implied throughout the paper.
Typically, the spacing between the planes, which are characterized by
the low values of the Miller indices (such as the (100), (110) and (111)
planes), lies within the range $0.6\dots2.5$ \AA (see
e.g. \cite{Biryukov} and  \tref{Table1} in \sref{Stimulated}) below.
Therefore, \eref{AW_0} is fulfilled for the AW amplitudes
$a \geq 10$ \AA.

\item[b)]\
The channeling process in a bent crystal takes place if the maximal
centrifugal force in the channel, $m \gamma v^2/R_{\rm min}$
(where $R_{\rm min}$  is  a minimum curvature radius of the bent channel)
is less than the maximal force due to the interplanar field
\cite{Tsyganov,Barysh80,Ell_Pic81,Solov}:
\begin{equation}
m \gamma v^2/R_{\rm min} < q\, U_{\rm max}^{\prime}.
\label{1}
\end{equation}
Here $q$ is the charge of the projectile and the quantity
$U_{\rm max}^{\prime}$ stands for the maximum gradient of the
interplanar field.

\end{enumerate}

Provided \eref{1} is fulfilled, the projectile, incident at small
angle with respect to the mid-plane, will be trapped in the
channel (see \fref{Fig1}).
The condition \eref{AW_0} allows to disregard the oscillations
of the particle inside the channel due to the interplanar force
$q\,U^{\prime}$.
Therefore, the particle trajectory will be primarily harmonic
defined by the equation
\begin{equation}
y(z)=a\sin\left(2\pi{z \over \lambda }\right),
\qquad z=[0\dots L]
\label{AW_1}
\end{equation}
Here $L$ stands for the thickness of the crystal.
The minimum curvature radius of this trajectory is equal to
$R_{\rm min}=(\lambda/2\pi)^2/a$.
Thus, the decrease in  $R_{\rm min}$ and, consequently, the increase
in the maximum  acceleration of the particle in the channel is achieved
by decreasing $\lambda$ and increasing $a$.
As a result, photon emission due to the projectile's acceleration in
the bent channel may  be significantly enhanced.
This radiation is emitted coherently from similar parts of the
trajectory, and may dominate \cite{air} over the channeling radiation.

An adequate approach to the problem of the radiation emission
by an ultra-relativistic particle moving in an external field
was developed by Baier and Katkov in the late 1960s \cite{Baier67}
and was called by the authors the ``operator quasi-classical
method''.
The details of that formalism can be found in
\cite{Baier, Land4} and will not be reproduced here.

The advantage of this method is that it allows to use the
classical trajectory for the particle in an external field and,
simultaneously, it takes into account the effect of the radiative
recoil.

The classical description of the particle motion is valid provided
the characteristic energy of the projectile in an external field,
$\hbar \tilde{\omega}_o$, is much less than its total energy,
$\varepsilon = m \gamma c^2$.
The relation
$\hbar \tilde{\omega}_o/\varepsilon \propto \gamma^{-1} \ll 1$
is fully applicable in the case of an ultra-relativistic projectile.
Typical values of $\hbar \omega_o$ are in the order of magnitude
equal to the continuous potential-well depth, vary from several
eV for the crystals of light materials (e.q. the $LiH$ crystal
\cite{LiH}) up to $10^2$ eV for the crystals of heavy materials
(e.g. the $W$ crystal \cite{Biryukov}).

The role of the radiative recoil, i.e. the change of the projectile
energy due to photon emission, is governed by the ratio
$\hbar \omega /\varepsilon$.
In the limit  $\hbar \omega /\varepsilon \ll 1$ a purely classical
description \cite{Land2} of the radiative process can be used.
For  $\hbar \omega /\varepsilon \le 1$ the quantum corrections
due to the emission of the photon must be taken into account.

The quasi-classical approach neglects the $\hbar \omega_o/\varepsilon$
terms, but it explicitly takes into account the quantum
corrections due to the radiative recoil in the whole range of the
emitted photon energies, except for the
extreme high energy tail of the spectrum.

Using this method the spectra of photons and electron-positron pairs
in linear crystals were successfully described \cite{Baier}.
It was also applied to the problem of synchrotron-type radiation
emitted by an ultra-relativistic projectile channeling in
a non-periodically bent crystal \cite{Solov,Arutyunov}.

We use this general formalism of Baier and Katkov in our paper to
treat the AIR phenomenon.

%%%%%%%%%%%%%%%%%%%%%%%%%%%%%%%%%%%%%%%%%%%%%%%%%%%%%%%%%%%%%%%%%%%%%%%%
To conclude the introductory section let us discuss the results
of the papers \cite{Barysh80}--\cite{Dedkov94} where the
electromagnetic radiation emitted by a charge passing through
a crystal undergoing the action of an ultrasonic wave was
considered.

As far as we are aware of, the first study of the influence of an
external ultrasonic field on the radiation of channeling particles
in a crystal was given in \cite{Barysh80}.
In this article (see also \cite{Barysh82}) spontaneous emission
by channeling particles in a silicon crystal bent by an acoustic wave
of a frequency $\approx 10$ MHz was briefly considered.
The estimates carried out in \cite{Barysh80, Barysh82} were based
on the so-called dipole approximation for the undulator radiation
(see e.g.\cite{Kniga}).
The latter implies that the strong inequality,
$p\ll 2\pi \gamma\, a/\lambda \ll 1$ is valid, so that all the
undulator emission occurs in the fundamental harmonic.
Such an assumption leads to a considerable narrowing of the parameters
of the AIR radiation and, also, disregards the possibility to generate
the emission in higher harmonics.

In papers \cite{Armyane86, Dedkov94} the spontaneous
channeling radiation in the presence of a {\it low-amplitude}
ultrasonic wave was considered by means of classical electrodynamics.
In this limit the AW amplitude is much less than the interplanar
spacing, $a \ll d$, resulting, thus, in low values of the curvatures
of the acoustically bent channels.
This inequality allowed the authors of \cite{Armyane86, Dedkov94}
(see also \cite{Ikezi84}) to consider the  phenomenon
of resonant enhancement of the photon yield due to the coupling
of two mechnisms of the photon emission, the channeling
radiation \cite{Kumakhov1,Barysh77} and that induced by the AW.

In \cite{Ketterson86} both the spontaneous undulator
emission and that of a free electron laser type was investigated in
the case of a relativistic positron beam channeling through a
periodically strained lattice.
In this paper the main attention was paid to the radiation emitted
in an undulator based on a solid state superlattice (see also
\cite{Ikezi84}).
By using the formalism of classical electrodynamics the authors of
\cite{Ketterson86} performed a self-consistent treatment involving
the wave equation for the electromagnetic radiation field and the
kinetic equation for the positron distribution function.
The important result, which is the gain coefficient for the forward
radiating field was obtained analytically in a closed form.
Numerical estimates of the gain in the case of a resonant
coupling of the channeling radiation and that arising from the
superlattice based undulator were presented.

In \cite{Ketterson86} there is also a short comment on the
possibility of using an acoustically bent crystall as an undulator.
In our opinion, the analogy between a superlattice and an acoustic
wave can not be considered as adequate, because
the acoustic wave amplitude, $a$, can greatly exceed the interplanar
distance $d$, whereas in the superlattice based undulator the inverse
condition is assumed, $a \ll d$.
This latter case was analyzed in \cite{Ikezi84, Ketterson86,
Armyane86, Dedkov94}.

It is stressed in \fref{Fig1} and in \eref{AW_0} that we consider here
the opposite case  $a \gg d$.
In this limit there is no resonant coupling between the channeling and
the undulator radiation because the characteristic frequencies of the
photons emitted via these mechanisms are totally incomparable
(see \sref{Undulator}).
As a result it is possible to distinguish two mechanisms of the
radiation formation in an acoustically bent crystal and to
investigate the properties of the AIR, both spontaneous and
stimulated, separately from the channeling radiation.

As we demonstrate below, staying within the regime defined by
\eref{AW_0}, it is possible, by tuning the AW amplitude or frequency,
to vary significantly the intensity of the AIR radiation, to
change the patterns of its spectral and angular distributions.
These characteristics of the radiation, as it is known from the
general theory of undulators (e.g. \cite{Baier}), are crucially
dependent on the magnitude of the undulator parameter $p$.
In our paper we provide the analysis of the AIR radiation
for both limiting cases, $p^2\ll 1$ and $p>1$.

In the case $a \gg d$ it is important to establish the conditions
under which the acoustically bent crystal may serve as an undulator
for the channeled ultra-relativistic particle.
A comprehensive analysis of these conditions, and consequently, the
establishment of the ranges of all parameters in the problem
(these are: the AW amplitude, frequency and velocity, the type of
projectile and its energy, the width of the channel and the magnitude
of the interplanar field) inside which the AIR process is feasible,
are studied in our paper.
In contrast such a discussion is fully omitted in the papers
cite above.

Another point we wish to emphasis is that the quasi-classical
approach used in our paper is more appropriate than the purely
classical treatment utilized in the cited papers.
In our paper we consider the spectral distribution of the radiation
in the whole range of the photon energies: $\hbar\omega=0\dots\varepsilon$.
It is clear that the classical approach is valid only in the
soft-photon limit, i.e. $\hbar\omega \ll \varepsilon$, but is
totally inappropriate in the case $\hbar\omega
\stackrel{<}{\sim} \varepsilon$
where the quantum corrections (the radiative recoil) must be
taken into account.
The emission of high-energy photons becomes important for
energies of projectile positrons in the tens GeV range.

In our paper it is demonstrated, for the first time, that it is
feasible to consider, by means of the AIR mechanism, stimulated
emission within the $10^{-2}\dots 10^{0}$ MeV range.
To establish this interval we carried out a realistic analysis of the
influence of the dechanneling and the photon attenuation effects
on the spectrum of the AIR radiation.

All the features mentioned above distinguish the results presented
below from the earlier works \cite{Barysh80}-\cite{Armyane86},
\cite{Dedkov94}.

Finally, let us note that we do not consider
the electromagnetic radiation \cite{BaryshJPC90}
arising during the projectile channeling in an ultrasonically
excited crystal due to specific diffraction of the photons
by sets of periodically bent planes.
%%%%%%%%%%%%%%%%%%%%%%%%%%%%%%%%%%%%%%%%%%%%%%%%%%%%%%%%%%%%%%%%%

The paper is organized as follows:

In \sref{Formalism} the general formalism of the quasi-classical
description of the undulator radiation is presented.
\Sref{Undulator} is devoted to a detailed analysis of the conditions
which allow to consider the acoustically bent crystal as an
undulator.
Numerical results for the angular and spectral distributions of the
undulator radiation as well as the energy loss due to the AIR
are presented in \sref{Results} for various crystals and
projectile energies.
In \sref{Stimulated} the estimates, both analytical and numerical,
for the possibility to generate the stimulated emission by means
of the acoustically based undulator are presented.
In \sref{Conclusions} we outline the relevant problems which,
to our mind, deserve to be thoroughly investigated.

%%%%%%%%%%%%%%%%%%%%%%%%%%%%%%%%%%%%%%%%%%%%%%%%%%%%%%%%%%%%%%%%%
\section{Quasi-classical description of the undulator
radiation}\label{Formalism}
%%%%%%%%%%%%%%%%%%%%%%%%%%%%%%%%%%%%%%%%%%%%%%%%%%%%%%%%%%%%%%%%%

Within the framework of the quasi-classical approach the
distribution of the energy radiated in given direction by an
ultra-relativistic particle (of a spin $s=1/2$) and summed over the
polarizations of the photon and the projectile,
is given by the following expression
\begin{equation}
\d E_{\omega}({\bf n}) =
\biggl({q c \over 2 \pi}\biggr)^2 \d {\bf k}\
\int \d t_1\  \d t_2 \
\e^{\i \omega^{\prime} \varphi(t_1,t_2)}
\ f(t_1,t_2)
\label{wkb1}
\end{equation}
Here $\d {\bf k}=c^{-3} \omega^2 \d\omega \d \Omega_{\bf n}$,
${\bf n} = c\, {\bf k}/\omega$ is the unit vector in the direction of
the photon emission, $c$ is the light velocity and $q$ is the
projectile charge.
The functions $\varphi(t_1,t_2)$ and $f(t_1,t_2)$ are defined as follows
\numparts
\begin{eqnarray}
 \varphi(t_1,t_2) = t_1 - t_2 - {1 \over c}\,
{\bf n}\cdot ({\bf r}_1 - {\bf r}_2)
\label{wkb2a} \\
 f(t_1, t_2)= {1 \over 2}
\biggl\{
\left( 1+(1+u)^2 \right)\,
\left(
{ {\bf v}_1\cdot {\bf v}_2 \over c^2} -1
\right)
+{u^2 \over \gamma^2}
\biggr\}
\label{wkb2b}
\end{eqnarray}
\endnumparts
The notations used are ${\bf r}_{1,2}= {\bf r}(t_{1,2})$,
${\bf v}_{1,2}= {\bf v}(t_{1,2})$, with ${\bf r}$ and ${\bf v}$
standing for projectile's  radius vector and  velocity, respectively.

Expression (\ref{wkb1}) looks almost like the classical formula
\cite{Kniga, Baier}, although with quantum corrections:
\begin{equation}
\omega\longrightarrow
\omega^{\prime} =
{ \varepsilon \over
\varepsilon - \hbar \omega}\, \omega,
\qquad
u=
{ \hbar \omega \over
\varepsilon - \hbar \omega},
\qquad
\label{wkb3}
\end{equation}
which take into account the radiative recoil.

Let us now obtain a general quasi-classical expression
for the radiated energy, $\d E_{\omega}({\bf n})$,
in the case of a particle moving in an undulator which contains
$N_{\rm u}$ periods.
The  radius vector and the velocity of the particle in an undulator
satisfy the conditions
\begin{equation}
{\bf r}(t+T) ={\bf r}(t) + {\bf v}_0\, T,
\qquad
{\bf v}(t+T) ={\bf v}(t),
\label{quasi_1}
\end{equation}
where $T$ is the time interval during which a projectile passes
one period of the undulator, and the quantity ${\bf v}_0$
\begin{equation}
{\bf v}_0 = {1 \over T}\, \int_0^T {\bf v}(t) \d t
\label{quasi_3}
\end{equation}
is the mean velocity along the undulator axis.

Following \cite{Baier} and introducing the quantities
\begin{equation}
\fl
{\bf v}_T = \int_0^T \d t \, {{\bf v}(t) \over c} \,
\exp\left(\i\Phi(t)\right),
\quad
{\rm v}_T^0 = \int_0^T \d t \,
\exp\left(\i\Phi(t)\right),
\quad
\Phi(t) =\omega^{\prime}
\left(t - { {\bf n}\cdot{\bf r}(t)
\over c} \right)
\label{quasi_6}
\end{equation}
expression (\ref{wkb1}) can be presented in the form:
\begin{equation}
{\d E_{\omega}({\bf n}) \over \d\omega\,
\d\Omega_{\bf n}} =
{q^2 \omega^2 \over 8 \pi^2 c }
\, N_{\rm u}^2\, D(\eta) \,
\biggl\{
\left( 1+(1+u)^2 \right)\,
\left(
|{\bf v}_T|^2 - | {\rm v}_T^0|^2
\right)
+ {u^2 \over \gamma^2}\, |{\rm v}_T^0|^2
\biggr\}
\label{quasi_5}
\end{equation}
where the function $D(\eta)$ and its argument are
\begin{equation}
D(\eta) =
\left(
{\sin N_{\rm u}\pi \eta \over N_{\rm u}\, \sin \pi \eta}
\right)^2,
\quad
\eta = {\omega^{\prime} \over \omega_0}
\left(
1 - {{\bf n}\cdot{\bf v}_0 \over c}
\right),
\label{quasi_8}
\end{equation}
respectively.
The parameter $\omega_0 = 2\pi / T$ is for the undulator frequency.

Expression (\ref{quasi_5}) together with the definitons (\ref{quasi_8})
clearly exhibit the features typical for the undulator radiation
\cite{Alferov}.
The radiation intensity is proportional to the square of the total
number of the undulator periods, $N_{\rm u}^2$,
reflecting the coherence effect of radiation.

The spectral and angular dependence of the radiation are
determined mainly by the function $D(\eta)$, which is well-known
in a classical theory of diffraction \cite{Land4}.
This function has sharp maxima at the points $\eta=K=1,2,3\dots$.
It is most clearly seen in the case of infinite $N_{\rm u}$:
\begin{equation}
\lim_{N_{\rm u}\to\infty}
N_{\rm u}\, D(\eta) =
\sum_{K} \delta(\eta-K)
\label{D1}
\end{equation}
For $N_{\rm u}<\infty$ the function $D(\eta)$ has main maxima
in the points $\eta=K=1,2\dots$ where its magnitude
$D(k)=1$. In the interval $\eta = [K, K+1]$
$D(\eta)$ has minima in the points
$\eta_m^{\rm min} = K + m/N_{\rm u}$, $m = 1,\dots, N_{\rm u}-1$ where
$D(\eta_m^{\rm min})=0$.
The local maxima are located at
$\eta_m^{\rm max}$ ($m = 1,\dots, N_{\rm u}-1$) which can be
found from the equation
$N_{\rm u}\ {\rm tg}\,\pi\eta N_{\rm u} = {\rm ctg}\,\pi\eta$.
The values $D(\eta_m^{\rm max})$ rapidly decrease
with $m$ and are much smaller than 1. For $m=1, 2$ the corresponding
values are
$D(\eta_1^{\rm max})\approx 1/22$,
$D(\eta_2^{\rm max})\approx 1/62$.
The width $\Delta \eta$ of the main maxima of $D(\eta)$, estimated as
$\Delta K = 2(\eta_1^{\rm min} -K)$, is equal to
$\Delta K = 2/N_{\rm u}$ and does not depend on $K$.

Since $\d E_{\omega}({\bf n}) \propto D(\eta)$ the
frequency-angular distribution of the radiation is represented by
the sets of harmonics $\omega^{\prime}_K$ (each of
the width $\Delta \omega^{\prime}_K$ independent on $K$) which
are effectively emitted at the angle
$\theta$ (measured with respect to the undulator axis)
\begin{equation}
\omega^{\prime}_K = {\omega_0\, K
\over \left(1 - {{\bf n}\cdot{\bf v}_0 \over c}\right)},
\qquad
\Delta \omega^{\prime} = {2 \over N_{\rm u}} \, {\omega^{\prime}_K \over K}
\qquad K=1,2,3\dots
\label{quasi_18}
\end{equation}

\hspace*{0.2cm}

Let indices $\parallel$ and $\perp$ indicate,
respectively, the parallel and the perpendicular (with
respect to the undulator axis) components of the projectile
radius-vector and velocity:
\begin{equation}
{\bf r}(\tau) = {\bf r}_{\parallel}(\tau) + {\bf r}_{\perp}(\tau),
\quad
{\bf v}(\tau) = {\bf v}_{\parallel}(\tau) + {\bf v}_{\perp}(\tau),
\quad
\tau=[0,T]
\label{quasi_10}
\end{equation}
For $\gamma\gg 1$ the following relations are valid
with the accuracy up to the $\gamma^{-2}$ terms:
\begin{equation}
{{\rm v}_{\parallel} \over c} \approx
1 -
{1 \over 2 \gamma^2}
\left(1+ {{\rm v}_{\perp}^2 \over c^2}\gamma^2\right),
\qquad
{\overline{{\rm v}_{\parallel}} \over c} \equiv
{{\rm v}_{0} \over c} \approx
1 -
{1 \over 2 \gamma^2}
\left(1+ {\overline{{\rm v}_{\perp}^2} \over c^2}\gamma^2 \right)
\label{quasi_11}
\end{equation}
where
$\overline{{\rm v}_{\perp}^2} =
T^{-1}\, \int_0^T {\rm v}_{\perp}^2(t) \d t$.

By inserting (\ref{quasi_11}) into (\ref{quasi_6})
and introducing the dimensionless variable $\psi = \omega_0\, \tau$,
one obtains
\numparts
\begin{eqnarray}
\Phi(\tau) \equiv
f(\psi) = \eta\psi
+ {\omega^{\prime} \over \omega_0}\cos\theta \,
{\Delta(\psi) \over 2}
- {\omega^{\prime} \over c}\,  {\bf n}\cdot{\bf r}_{\perp}(\tau)
\label{quasi_12a} \\
\Delta(\psi)
=
\int_0^{\psi} \left[
{{\rm v}_{\perp}^2 \over c^2} -
{\overline{{\rm v}_{\perp}^2} \over c^2}
\right]\, \d\psi
\label{quasi_12b}
\end{eqnarray}
\endnumparts

\noindent
and in the following expression for $\d E_{\omega}({\bf n})$:

\begin{equation}
\fl
{\d E_{\omega}({\bf n}) \over \d\omega\,
\d\Omega_{\bf n}} =
{q^2\, N_{\rm u}^2 \over 4 \pi^2 c }
\, D(\eta) \,
{\omega^2 (1+u) \over \omega_0^2 \gamma^2}\,
\biggl\{
\gamma^2 \left( 1+ {u^2 \over 2(1+u)} \right)\,
\left[
|{\bf I}_{\perp}|^2 - {\rm Re} {\rm I}_0 {\rm I}_{\parallel}
\right]
-|{\rm I}_0|^2
\biggr\},
\label{quasi_13}
\end{equation}

\noindent where the quantities ${\rm I}_{\perp},\, {\rm I}_0$ and
${\bf I}_{\parallel} $ stand for the integrals

\begin{equation}
{\rm I}_0  = \int_0^{2\pi}
\e^{\i f(\psi)}
\, \d\psi,
\quad
{\rm I}_{\parallel}  = \int_0^{2\pi}
{ {\rm v}_{\perp}^2 \over c^2}\,
\e^{\i f(\psi)}
\, \d\psi,
\quad
{\bf I}_{\perp}  = \int_0^{2\pi}
{ {\bf v}_{\perp} \over c }\,
\e^{\i f(\psi)}
\, \d\psi\ .
\label{equiv_14}
\end{equation}

\hspace*{0.2cm}

Within the same level of accuracy as in (\ref{quasi_11}) the
parameter $\eta$ from (\ref{quasi_8}) reads
\begin{equation}
\eta=
{\omega^{\prime} \over \omega_0}\,
\left(
{1 \over 2 \gamma^2}
+
{p^2 \over 4 \gamma^2}
+
{\theta^2 \over 2}
\right)
=
{p^2 \over 4 \gamma^2}\,
{\omega^{\prime} \over \omega_0}\,
\left(
{2 \over p^2} + 1
+
{\theta^2 \over \theta_0^2}
\right)
\label{quasi_15}
\end{equation}

The quantity $p$, introduced here, is called the
undulator parameter and is defined by the following relation
\begin{equation}
p^2 = 2 \gamma^2\,{\overline{{\rm v}_{\perp}^2} \over c^2}
\label{quasi_16}
\end{equation}
The parameter $\theta_0$ is related to $p$ and $\gamma$ through
\begin{equation}
\theta_0^2 = {p^2 \over 2 \gamma^2} =
{\overline{{\rm v}_{\perp}^2} \over c^2}
\label{quasi_17}
\end{equation}
and has the following physical meaning:
in the case $p^2 \gg 1$ (so-called, non-dipole limit),
$\theta_0$ defines the cone,
along the undulator axis, $0 \le \theta \le \theta_0$,
into which the radiation is emitted.
For $p^2 < 1$ (the dipole case) the emission effectively occurs
into the cone $0 \le \theta \le \gamma^{-1}$.

The formalism sketched above
will be now applied to a planar harmonic undulator, in which the
quasi-periodic trajectory (\ref{quasi_1}) lies in a plane.
Let the undulator axis coincide with the $z$-direction and
let its plane be the $(yz)$-plane.
In this type of an undulator the particle trajectory is purely harmonic
and is defined as
\begin{equation}
y(z) = a\sin kz
\label{yz}
\end{equation}
We assume that the parameters $a$ and $k$ satisfy the condition
$\xi\equiv ka \ll 1$. In the next section, where the undulator
based on the acoustic wave transmission in a crystal is described
in detail, it is demonstrated that this condition is fulfilled.

For an ultra-relativistic projectile, moving along the trajectory
(\ref{yz}) the following formulas are valid for the quantities
introduced above (the expressions below are written in the lowest
orders in $\xi$ and $\gamma^{-1}$):
\numparts
\begin{eqnarray}
{\rm v}_{\parallel} \simeq {\rm v}_0  \simeq c,
\quad
{\rm v}_{\perp} \simeq  c\, \xi \cos (kz) \ll {\rm v}_{\parallel},
\quad
\overline{{\rm v}_{\perp}^2} \simeq
\xi^2\, {c^2 \over 2}\ ,
\label{AW_9a} \\
\Delta(\psi) \simeq {\xi^2 \over 4}\, \sin 2\psi,
\qquad
p^2 = \gamma^2 \xi^2,
\qquad
\omega_0 = c k\ ,
\label{AW_9b} \\
f(\psi) = \eta\psi
+ {\omega^{\prime} \over \omega_0} {\theta_0^2 \over 4} \,
\sin2\psi
- {\omega^{\prime} \over \omega_0}\, \sqrt{2} \,
\theta_0\, \theta\, \cos\varphi \sin\psi \ .
\label{AW_9c}
\end{eqnarray}
\endnumparts
Here $\theta$ and $\varphi$ are the polar and the azimuthal
angles of the photon emission, respectively.

By substituting (\ref{AW_9a}-\ref{AW_9c}) into
(\ref{quasi_13}) and (\ref{equiv_14}) one obtains general quasi-classical
expressions for the spectral-angular distribution of the radiation
in a planar undulator \cite{Baier}.
These expressions can be simplified in two limiting cases:
$p^2 \ll 1$ and $p^2 \gg 1$.
For the sake of further reference below the corresponding
formulas are exhibited.

%%%%%%%%%%%%%%%%%%%%%%%%%%%%%%%%%%%%%
\subsection{Non-dipole case: $p^2\gg1$}

In the limit  $N_{\rm u}, p^2 \gg 1$ the spectral-angular distribution of
the undulator radiation is given by the following asymptotic
expression valid for $\theta < \theta_0$:
\begin{eqnarray}
\fl {\d E({\bf n}) \over \d\omega\, \d \Omega_{\bf n}}
=
{16 q^2 \over c} \, {N_{\rm u}^2 \gamma^2 \over p^2}\,
{D(\eta) \over 1+u }\,
\left({4x^2 \over \mu }\right)^{2/3}\,
\Biggl\{
\biggl[
{\theta^2 \sin^2\varphi \over 2\, \theta_0^2} +
\Delta\cdot \zeta\, \left({\mu \over 4x^2}\right)^{1/3}
\, \biggr]\, \cos^2\beta\, {\rm Ai}^2(\zeta)
\nonumber \\
\lo{+}
(1+\Delta)\, \left({\mu \over 4x^2}\right)^{1/3}\,
\sin^2\beta\, {\rm Ai}^{\prime\,2 }(\zeta)
\Biggr\}
\label{Ang1}
\end{eqnarray}
where ${\rm Ai}(\zeta),\, {\rm Ai}^{\prime}(\zeta)$
 are the Airy function and its derivative, respectively, and
the abbreviations were introduced
\begin{eqnarray}
x = {\omega^{\prime} \over \omega_0}\, {p^2 \over 4\gamma^2},
\qquad
\zeta = \left({4x^2 \over \mu}\right)^{1/3}\left[{1\over p^2} +
{\theta^2 \sin^2\varphi \over 2\theta_0^2} \right],
\qquad
\mu = 1 - \delta^2
\nonumber \\
\delta =  {\theta \cos\varphi \over \sqrt{2}\, \theta_0},
\quad
\beta = \eta\left( {\pi \over 2} - {\rm arcsin}\,\delta\right)
-3x\,\delta\,\mu^{1/2},
\quad
\Delta = {u^2 \over 2(1+u)}
\label{Ang2}
\end{eqnarray}

Other notations used in (\ref{Ang1}) have been defined earlier.

The Airy function and its derivative, both satisfying the conditions
 ${\rm Ai}(\zeta),|{\rm Ai}^{\prime}(\zeta)|\sim 1,\
\ \zeta \le 1$ and
${\rm Ai}(\zeta),|{\rm Ai}^{\prime}(\zeta)|
\longrightarrow 0,\ \zeta \gg 1$,
define the frequency of the radiated intensity maximum,
$\omega_{\rm max}^{\prime}$.
For $\omega^{\prime}\gg \omega_{\rm max}^{\prime}$ the intensity of
radiation exponentially decreases. It can be shown that
$\omega_{\rm max}^{\prime} \sim p\gamma^2\,\omega_0$. It corresponds
to $K_{\rm max} \sim p^3$ which is the largest number
of the radiated harmonics.

From (\ref{Ang1})-(\ref{Ang2}) follows (in accordance with general
theory of a planar undulator \cite{Kniga, Baier, Alferov})
that only odd harmonics ($K=1,3,\dots$) are radiated along the
undulator axis in the soft-photon (classical) limit,
$\hbar\omega/\varepsilon = 0$.
Indeed, by putting $\theta=0$ and $u=0$
in (\ref{Ang1})-(\ref{Ang2}) one gets
$\left[\d E({\bf n})/ \d\omega\, \d \Omega_{\bf n}\right]_{\theta=0}
\propto D(\eta) \sin^2(\pi\eta/2)$.
Keeping in mind that $\eta = K \pm1/N_{\rm u}$, one gets $K=1,3,\dots$.

Equations \eref{Ang1} and \eref{Ang2} are simplified considerably
if $\hbar\omega/\varepsilon = 0$. It can be demonstrated that in
this case  \eref{Ang1} represents by itself the classical result
for $\d E({\bf n})/ \d\omega\, \d \Omega_{\bf n}$
(see e.g. \cite{Kniga}) written in the limit $p^2\gg 1$.
It is worth noting that for $\hbar\omega/\varepsilon = 0$
the asymptotic expression \eref{Ang1} and the exact classical
formula written in terms of Bessel functions produce close
results starting with the value $p^2=2$.

The spectral intensity of the radiation emitted by a projectile
moving along the path (\ref{yz}) in an undulator of the total length
$L$  can be obtain by integrating (\ref{Ang1}) over $\varphi =[0,
2\pi]$ and $\theta=[0,\theta_0]$,
or by deducing it from the general expression given in \cite{Baier}.
The result is \cite{air}:
\begin{equation}
{1 \over L}\, {\d E \over \d \omega}
=
{q^2 \over c}\,
N_{\rm u} p\, {\omega^{\prime} \over \omega}\, \left(
G_1(s) + \left[1+{u^2 \over 2(1+u)}\right] G_2(s)\right)
\label{Sp1}
\end{equation}
where the parameter $s$ is defined as
$s=\left( \omega^{\prime} /  \omega_0 \gamma^2 p\right)^{2/3}$.
The functions $G_{1,2}(s)$ are equal to
\begin{eqnarray}
G_1(s) =  - 2 s^{5/2} \int_1^{\infty}
\d x \left[\pi - \arccos\left(1-{2 \over x^3}\right)
\right] {\rm Ai}(s x)
\nonumber \\
G_2(s) = - 8 s^{1/2} \int_0^{\infty}
{ \d x \over ({\rm ch} x)^{5/3}}\,
{\rm Ai}^{\prime}(s ({\rm ch} x)^{2/3})
\label{Sp2}
\end{eqnarray}

In the extreme limit $p^2 \longrightarrow \infty$ expressions
(\ref{Ang1}) and (\ref{Sp2}) reduce to the analogue ones for
synchrotron radiation \cite{Baier}.

%%%%%%%%%%%%%%%%%%%%%%%%%%%%%%%%%%%%%
\subsection{Dipole case: $p^2\ll 1$}

The general quasi-classical expression one finds in \cite{Baier}.
We will need here its classical limit
$\hbar\omega/\varepsilon = 0$.
The angular distribution of the radiation is given by the
following expression \cite{Kniga}
\begin{equation}
{\d E({\bf n}) \over \d\omega\, \d \Omega_{\bf n} }
=
{q^2 \over c} \,
{N_{\rm u}^2 \gamma^2 p^2
\over
\left(1+(\gamma\theta)^2\right)^2}\,
\left\{
\sin^2\varphi + \cos^2\varphi \,
\left(
{1-(\gamma\theta)^2 \over 1+(\gamma\theta)^2 }
\right)^2
\right\}\,
D(\eta)
\label{Ang3}
\end{equation}
Note, that in the limit $p^2 \ll 1$ the parameter $\eta$
reduces to (see (\ref{quasi_15})):
\begin{equation}
\eta \simeq
{\omega \over \omega_0}\,
\left(
{1 \over 2 \gamma^2}
+
{\theta^2 \over 2}
\right),
\qquad
\theta = [0, \gamma^{-1}]
\label{quasi_15a}
\end{equation}

Expression (\ref{Ang3}) is written in the lowest order in $p^2$ and
describes the angular distribution of the first harmonic, $K=1$.
Thus $\eta = 1 \pm 1/N_{\rm u}$ is implied in (\ref{Ang3}). The
higher-harmonics emission is strongly supressed in the limit $p^2 \ll 1$
\cite{Kniga, Baier}.
The spectral distribution of the radiation in a dipole case
can be obtained by a direct integration of
(\ref{Ang3}) over  $\varphi =[0, 2\pi]$ and $\theta=[0,\gamma^{-1}]$.

%%%%%%%%%%%%%%%%%%%%%%%%%%%%%%%%%%%%%%%%%%%%%%%%%%%%%%%%%
\section{Acoustically bent channel as an undulator}
\label{Undulator}
%%%%%%%%%%%%%%%%%%%%%%%%%%%%%%%%%%%%%%%%%%%%%%%%%%%%%%%%%

In this section we formulate the conditions which, in addition to
\eref{AW_0}and \eref{1}, must be fulfilled to consider the system
``AW + channeled particle'' as an undulator.

Let us consider the case of planar channeling of an
ultra-relativistic positively charged particle in a channel
bent by a monochromatic transverse acoustic wave (AW) which is
transmitted in the $z$-direction (see \fref{Fig1}).
The shape of the channel centerline is described by
the dependence \eref{AW_1}.

Another condition which we impose on the AW is
\begin{equation}
\xi \equiv k a =2\pi \, {a \over \lambda} \ll 1
\label{AW_4}
\end{equation}
where the quantity $k$ denotes the AW wave number, $k=2\pi/\lambda$.
It will be demonstrated below that the condition (\ref{AW_4})
holds in all cases which are of interest.

%%%%%%%%%%%%%%%%%%%%%%%%%%%%%%%%%%%%%%
\subsection{Motion of the particle in a channel bent by the AW}

Let a particle enter the crystal at the point $z=0$ having a small
incident angle with respect to the (curved) crystallographic plane.
The corresponding quantitative criterion will be given below.

In the case of small-angle injection one may use the continuous
potential approximation to describe the interaction of the particle
with the crystallographic plane \cite{Gemmell, Biryukov}.
This potential, $U(\rho)$ depends only on the transverse
displacement, $\rho$,
of the particle from the mid-plane of the channel ( $|\rho| \leq d/2$).

In the case of an ultra-relativistic projectile the classical
approach adequately  describes the projectile motion,
so that one may totally disregard the quantum corrections
to  the transverse oscillations of the projectile in the channel due to
the action of the interplanar potential $U$.
The following estimate proves this statement.
Let $U_o$ denote the depth of the interplanar potential well.
The separation of the neibouring energy levels,
$\hbar\tilde{\omega}_o$, of the transverse oscillations is equal,
by the order of magnitude, to
$\hbar\tilde{\omega}_o\approx \hbar \sqrt{ q\, U^{\prime\prime}/m\gamma} \sim
\hbar \sqrt{ q\,  U_o/d^2 m\gamma}$.
Hence, the number of these levels, estimated as
\begin{equation}
{U_o \over \hbar \omega_o } \sim
{d \over \lambda_c}\, \sqrt{q\, U_o \, \gamma \over m c^2}
\propto \sqrt{\gamma},
\qquad \lambda_c = {\hbar \over mc}
\label{Nlevels}
\end{equation}
is large when $\gamma \gg 1$, and trajectories can  be applied to
describe the  projectile dynamics.

The particle's trajectory lies in the $(yz)$-plane and is governed
by the equation of motion
\begin{equation}
{\d {\bf p} \over \d t} = - q\, {\rm grad}\,U
\label{AW1}
\end{equation}
where ${\bf p}=m {\bf v} \gamma$ is the kinematic momentum,
$\gamma^{-2}=(1 - v_z^2/c^2 -  v_y^2/c^2)$.

We assume that the dependence $U(\rho)$ on $\rho$ in the case of a
bent channel is the same as for a linear one. Below we demonstrate that
this assumption is adequate.

The Hamiltonian of the particle,
$H= \sqrt{p^2c^2 + m^2 c^4} + q U(\rho)$ does not explicitly depend
on $t$, hence, the relation
$\gamma = H/m c^2=\gamma_0\, (1+q U/m c^2 \gamma_0)$
is valid, with $\gamma_0=(1 - v^2/c^2)^{-1/2}$ and $v$ standing for
the initial particle velocity.
The condition $q U \ll m c^2 \gamma_0$
allows to consider $\gamma$ as an integral of motion.

To proceed further and to obtain all the
relations analytically we consider the harmonic approximation for the
interplanar potential,  $U(\rho) = \kappa \rho^2/2$.

Taking into account (\ref{AW_4}) the transverse coordinate  $\rho$
satisfies the equality $\rho = (y - a\sin kz)$ which is valid
up to terms $\sim \xi$. With the same accuracy the equation of
motion  (\ref{AW1}) reads
\begin{equation}
{\d \over \d t}(\dot{\rho} + \dot{z} \xi \cos kz) =
-{q \kappa \over m \gamma}\, \rho,
\qquad
{\d^2 z \over \d t^2} =
-{q \kappa \over m \gamma}\, \xi\  \cos kz
\label{AW1a}
\end{equation}

In lowest order in $\xi$ the system (\ref{AW1a}) yields $z=v t$
(with $\quad v\approx c$) and \eref{AW_1} for the particle
trajectory.
Within the linear approximation in $\xi$ the second equation
(\ref{AW1a}) defines the deviation from this trajectory through
\cite{Ell_Pic81}
\begin{equation}
m \gamma {\d^2 \rho  \over \d^2 t} = - {p v \over R(z)}
- q\, \kappa \rho
\label{AW_2}
\end{equation}
The term $- p v / R(z)$ represents the centrifugal force due to the
curvature of the bent channel. $R(z)$ is the curvature radius of the
curve (\ref{AW_1}) and is equal to
\begin{equation}
R(z) =
- {1 \over k^2 a \sin kz}
\label{AW_3}
\end{equation}
When writing (\ref{AW_3}) we neglected the terms $\sim \xi^2$ and
higher.

Equation (\ref{AW_3}) combined with the inequalities (\ref{AW_0}) and
(\ref{AW_4}) justifies the approximation made above (see the paragraph
after \fref{AW1}) concerning the equivalence of the interplanar
potentials in the bent channel and in the linear one.
Indeed, the deviation of the function $U(\rho)$ due to the channel
curvature is proportional to the ratio $d/R$, which can be neglected:
$d/R \propto (d/\lambda) (a/\lambda) \ll \xi^2 \ll 1$.

The general solution of (\ref{AW_2}) reads
\begin{equation}
\rho(z) = \rho_0 \sin(\chi z + \phi)
+ {a k^2 \over \chi^2 - k^2}
\sin kz, \qquad z=v t
\label{AW_5}
\end{equation}
where we abbreviated $\chi^2 = q\, \kappa / \left(m \gamma v^2\right)$.
The first term on the right-hand side of (\ref{AW_5}) describes the
oscillations due to the action of the interplanar force $-q U^{\prime}$
and is responsible for the traditional channeling radiation
\cite{Kumakhov1}.
The constants $\rho_0$ and $\phi$ are subject to the initial conditions
of the projectile injection into the channel.
In the case when the vector of the initial velocity of the particle
is aligned with the tangent of the channel centerline
one may put $\rho_0=0$, and, consequently,
$\rho_0 \sin(\chi z + \phi) \equiv 0$.

The second term in (\ref{AW_5}) corresponds to the oscillations of
the projectile in the vicinity of the mid-plane due to the action of
the (driving) centrifugal force.
The channeling in a bent channel is stable
against the driving oscillations provided their amplitude,
$a k^2/\left(\chi^2 - k^2\right)$,
is less than  half the width of the channel:
\begin{equation}
{a k^2 \over \chi^2 - k^2} \ll {d \over 2}
\label{AW_6}
\end{equation}
Let us introduce
a minimum curvature radius of the channel bent by the
transverse AW
\begin{equation}
R_{\rm min} = { 1 \over a k^2}
=
\left({\lambda \over 2\pi}\right)^2\, {1 \over a}
=
\left({V \over 2\pi}\right)^2\, {1 \over \nu^2 a}
\label{AW_7}
\end{equation}
Here $V$ and $\nu$ are the velocity and the frequency
of the AW, respectively.

Recalling (\ref{AW_0}),
the condition  (\ref{AW_6}) is presented in the form equivalent
to \eref{1}:
\begin{equation}
{\varepsilon \over R_{\rm min}\,
q U_{\rm max}^{\prime}}
< 1
\label{AW_8}
\end{equation}
Here the notation $U_{\rm max}^{\prime}$ is used to substitute
$2 \kappa (d/2)$, and we have put $pv \approx p c = \varepsilon$.

The relation (\ref{AW_8}) is written in the form which can be used
for an arbitrary potential $U(\rho)$.
It establishes the condition for a channeling process in a bent
channel \cite{Tsyganov}:
the maximal centrifugal force, $\varepsilon / R_{\rm min}$,
must be less than the maximal interplanar force
$q U_{\rm max}^{\prime}$.

The parameter on the left-hand side in (\ref{AW_8}) defines also
a critical incident angle $\Theta_c$ \cite{Biryukov} of the particle
injection into acoustically bent crystal:
$\Theta_{\rm c}= \Theta_{\rm L}\, (1-
\varepsilon/ R_{\rm min}\, q U_{\rm max}^{\prime})$, where
$\Theta_{\rm L}= \sqrt{2 q\, U_{max}/m \gamma\, c^2}$
is the critical angle for the linear channel
\cite{Lindhard}.

Provided \eref{AW_0} and (\ref{AW_8}) are fulfilled, the projectile,
entering the crystal under an angle $\Theta<\Theta_{\rm c}$ (with
respect to the centerline) will be trapped in the channel and move
along the trajectory defined by (\ref{AW_1}).  The deviation,
(\ref{AW_5}), of the real trajectory from (\ref{AW_1}) can be
neglected.

By taking into account the last equality from (\ref{AW_7}) the condition
(\ref{AW_8}) may also be denoted as
\begin{equation}
\nu^2\, a < C \equiv \gamma^{-1}\cdot
\left({ V \over 2\pi} \right)^2 \cdot
\left({ q\, U_{\rm max}^{\prime} \over mc^2}\right)
\label{AW_8a}
\end{equation}
\noindent
and determines the ranges of $\nu=V/\lambda$ ($\nu$ is the AW
frequency), $a$ and $\gamma$ for which the channeling process as well
as the undulator radiation, can occur for a given crystal and
crystallographic plane (the parameters $U_{\rm max}^{\prime}$ and $V$
depend on the choice of the particular crystal and the
crystallographic plane) and for given projectile type, characterized
by its rest mass $m$ and charge $q$.

Figures \ref{Fig3.1} illustrate the ranges of $\nu$ and $a$ in which
the channeling process is possible for a positron and a proton both of
the same energy (as indicated) in $C$ (diamond), $Si$, $Ge$ and $W$
crystals near the $(110)$ crystallographic plane.  The solid thick
lines represent the boundary $\nu^2\, a = C$, so that the range of
validity of (\ref{AW_8a}) lies below these lines.  To obtain $V$ we
used its relation \cite{Mason} to the crystal elastic constants
$c_{11}$ and $c_{12}$ and the density $\rho$: $V =
\sqrt{(c_{11}-c_{12})/ 2 \rho}$.  Using the data from \cite{Mason}
ones gets the following values of $V$ for $C$, $Si$, $Ge$ and $W$
crystals respectively: $V = 11.6\cdot10^5,\ 4.67\cdot10^5,\
2.76\cdot10^5,\ 2.81\cdot10^5,\ $ cm/s.

%%%%%%%%%%%%%%%%%%%%%%%% AW  $\nu$ and $a$ ranges
\begin{figure}
\hspace{3cm}\epsfig{file=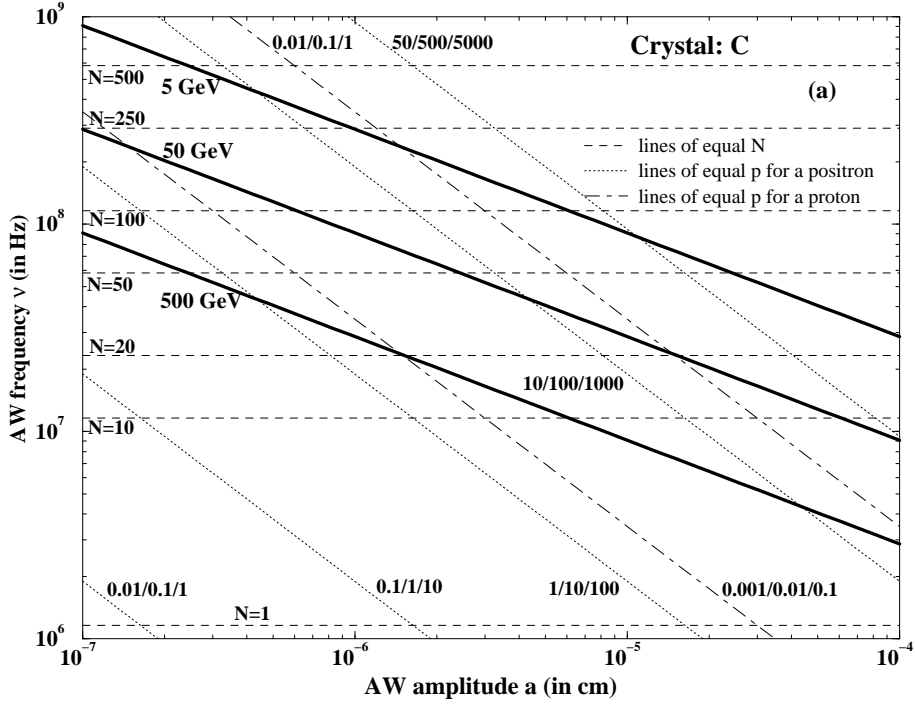,height=12cm,angle=270}
\caption{The ranges of $\nu$ (in Hz) and $a$ (in cm) in which the
channelling process is possible for a positron and proton along the
(110) plane in
{\bf (a)} a diamond,
{\bf (b)} a silicon crystal,
{\bf (c)} a germanium crystal,
{\bf (d)} a tungsten crystal.
The heavy full lines represent the boundaries, $\nu^2 a = C$ (see
(\ref{AW_8a})), for a 5/50/500 GeV projectile as indicated.  The
dashed lines correspond to the constant values of the parameter $N\,
({\rm cm^{-1}}) = \lambda^{-1}$ (see also the explanations in the
text). The dotted and the chain lines indicate the constant values of
the undulator parameter $p=2\pi \gamma a\nu/V$ for a positron and a
proton, respectively. The values of $p$, $p_1, p_2$ and $p_3$
correspond to the projectile energies 5, 50 and 500 GeV,
respectively.}
\label{Fig3.1}
\end{figure}

\setcounter{figure}{1}
\begin{figure}
\hspace{3cm}\epsfig{file=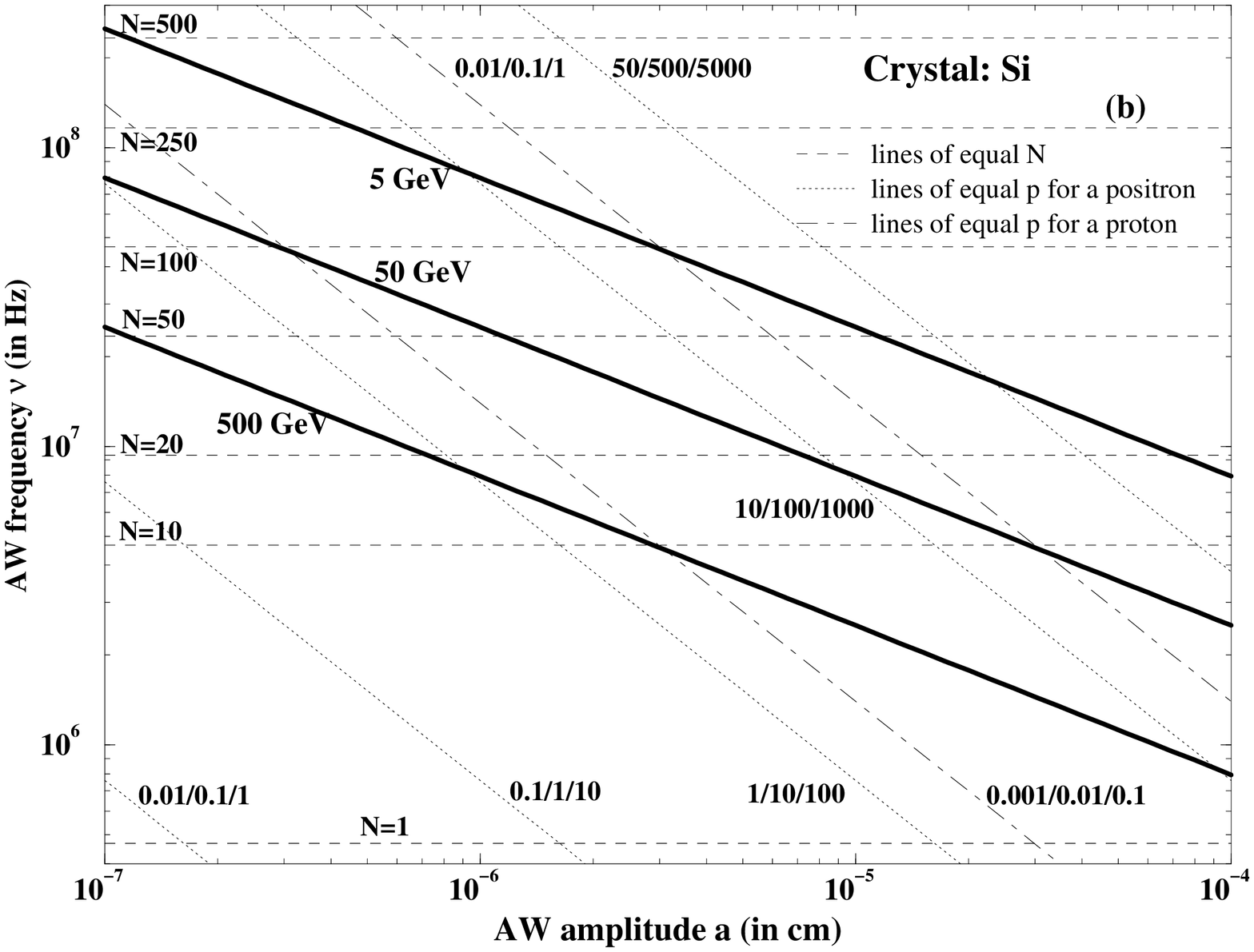,height=11cm,angle=270}
\caption{(\emph{continued})}
\end{figure}

\setcounter{figure}{1}
\begin{figure}
\hspace{3cm}\epsfig{file=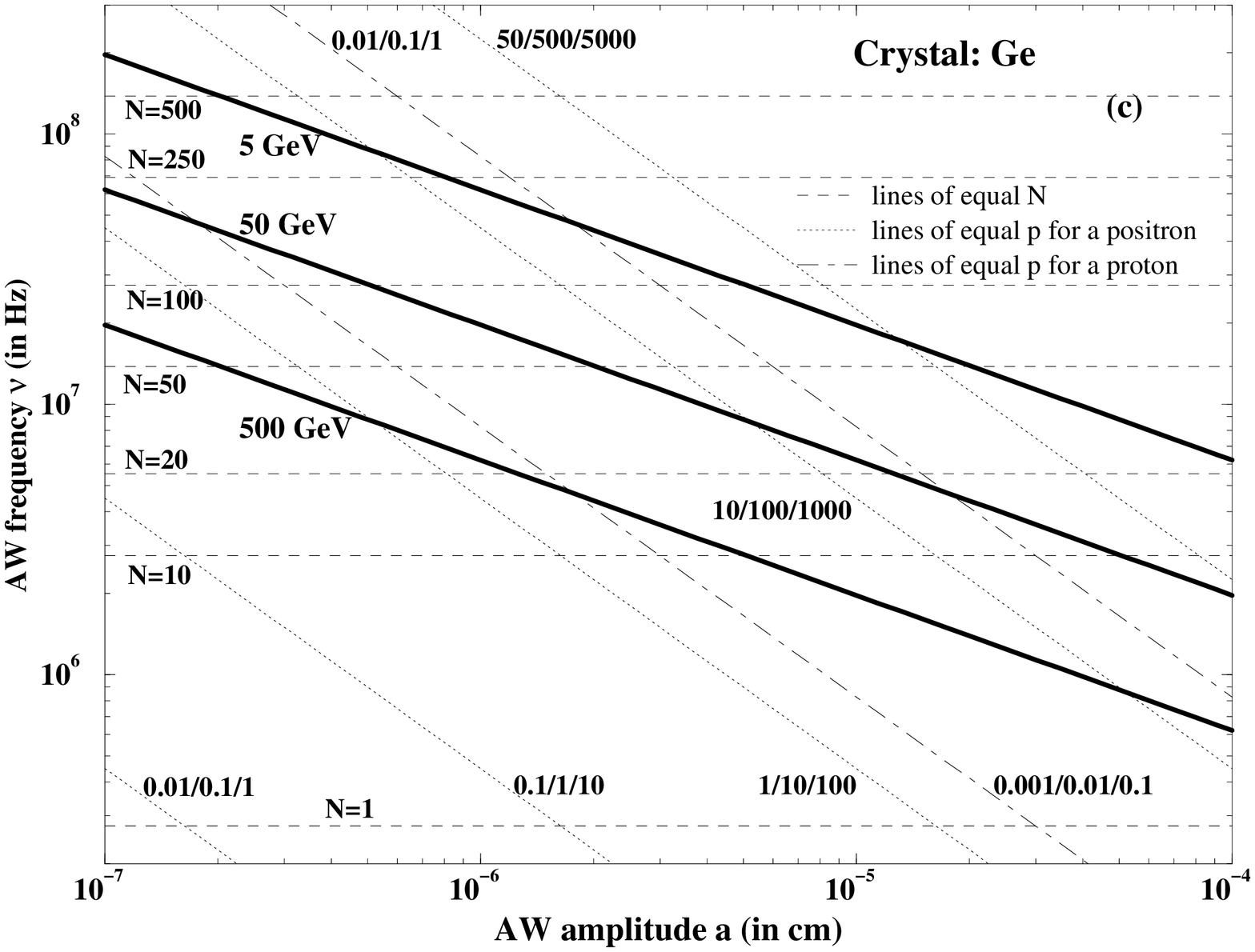,height=11cm,angle=270}
\caption{(\emph{continued})}
\end{figure}

\setcounter{figure}{1}
\begin{figure}
\hspace{3cm}\epsfig{file=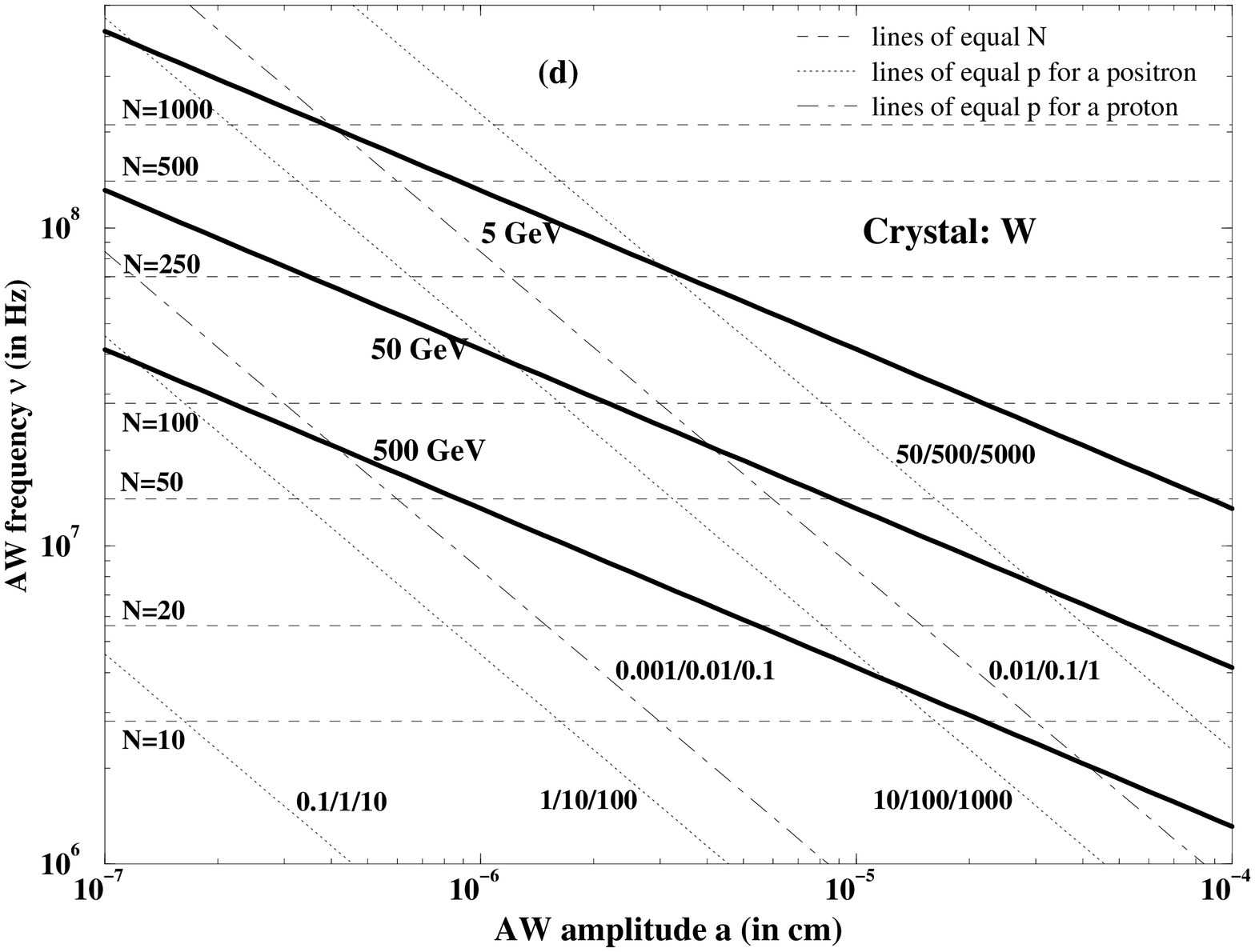,height=11cm,angle=270}
\caption{(\emph{continued})}
\end{figure}

The data for $q\, U_{\rm max}^{\prime}$ (for a positron) are
$q\, U_{\rm max}^{\prime} = 12.0, \ 5.7,\ 10.4,\ 43.0$ Gev/cm,
and for the interplanar distance $d = 1.26, 1.92, 2.00, 2.24$ 10$^{-8}$cm
for $C$, $Si$, $Ge$ and $W$ crystals, respectively.
They were taken from \cite{Biryukov, Baier}.

Taking into account the double-log scale  in these figures
it may be concluded, that the channeling condition (\ref{AW_8a})
is fulfilled within wide ranges of the AW amplitudes and frequencies,
for different types of projectiles and wide ranges of their
energies, and for a variety of crystals.
One can also deduce from figures \ref{Fig3.1} that
within the range of validity of (\ref{AW_8a}) the condition (\ref{AW_4})
is well fulfilled.

%%%%%%%%%%%%%%%%%%%%%%%%%%%%%%%%%%%%%%
\subsection{The undulator conditions}

Now we formulate the conditions (in addition to the channeling
condition established above, eqs. (\ref{AW_8})-(\ref{AW_8a}))
which allow to consider the system ``AW + channeled particle''
as an undulator.

Let $L$ denote a crystal thickness. Both the motion of the
projectile in the acoustically bent channel and the spectrum
of the generated radiation are of the undulator type, only
if
\begin{equation}
\lambda \ll L,
\label{Cond1}
\end{equation}
i.e. if the channeling particle oscillates
many times within the length of the crystal.
As with any other undulator, the suggested undulator is characterized
by the frequency, $\omega_0$, and the undulator parameter, $p$ (see
(\ref{quasi_16}, \ref{AW_9b}).
These quantities depend on the AW amplitude and the wave length,
and on the relativistic factor, and are expressed as follows:
\begin{equation}
\omega_0 = c k =2\pi\, {c \over \lambda},
\qquad
p = \gamma \xi \equiv \gamma k a =  2\pi\,\gamma {a \over \lambda} =
 2\pi\,\gamma {a\, \nu \over V}
\label{Cond2}
\end{equation}

For further reference let us note the difference in functional
form of $p$ for the AW undulator and that for
the undulators based on magnetic fields (both helical and planar).
In the latter case the undulator parameter is equal to (see eg.
\cite{Colson85}) $p_{\rm B}= q B \lambda_{\rm B} / 2 \pi mc^2$,
where $B$ is the amplitude value of the magnetic induction and
$\lambda_{\rm B}$ is the period of the magnetic field. The quantity
$p_{\rm B}$ is independent on $\gamma$ whereas $p$ from (\ref{Cond2})
is linearly proportional to it.

In figures \ref{Fig3.1} the dotted and the chain lines
indicate the constant values of the undulator parameter $p$.
Note that each dotted and chain line corresponds to different
values of $p$ depending on the energy of the projectile.
Values of $p$ corresponding to the energy 5/50/500 GeV are shown in
figure 2 in the vicinity of each line.
The broken lines correspond the constant values of the
parameter $N$ both for positron and proton, which is defined as the
number of the AW periods per 1 cm:  $N\, ({\rm cm^{-1}}) = \lambda^{-1}$.
Figures \ref{Fig3.1} demonstrate that, within the range
of validity of the channeling condition (\ref{AW_8a}), the parameters $p$
and  $N$ vary over  wide ranges: $N=1...100$, $p=0.1-500$ for
a positron projectile and  $p= 0.001-0.1$ for a proton.
The upper limiting values of the $p$'s are larger by more than an
order of magnitude than those accessible in the undulators based
on the motion of charged particles in periodic magnetic fields
\cite{NIM97}.

The undulator parameter $p$ is proportional to $a$ (see (\ref{Cond2})).
In this connection the question arises of the validity
of the assumption that $a=const$ during the time of flight,
$\tau = L/c$, of the particle through the crystal.
Hoevwr, for a running or standing AW the amplitude is time dependent.
To formulate a quantitative criterion let us consider the crystal
bending by a standing AW.
In this case the constant value $a$ must be substituted
with $\tilde{a}(t) = a_o \cos (2\pi\, \nu\, t)$.
For time scales $\tau \ll 1/\nu$, one gets the following relation
for the deviation, $\Delta \tilde{a}=a_o -\tilde{a}(t)$,
of $\tilde{a}(t)$ from its amplitude on the timescale $\tau$:
$\Delta \tilde{a} /a_o = 2 \pi^2 (\tau\nu)^2$.
Thus, thie time dependence of the AW is negligible provided
\begin{equation}
{2\pi^2 \over c^2}\, (L \nu)^2 =
2\pi^2  \left({ V \over c}\right)^2
\left({L \over \lambda}\right)^2
\ll 1
\label{Cond3}
\end{equation}
This condition allows to disregard the variations
in the shape (\ref{AW_1}) of the channel on the timescale $\tau$.

To estimate the corresponding values of $L$ let us consider
$\nu \le 100$ MHz (see figures\ref{Fig3.1}).
Then one gets $L \ll 70$ cm which is more than well-fulfilled for
any realistic $L$-value.

Finally, let us demonstrate that the characteristic frequencies
of the undulator radiation due to the particle motion along
the trajectory (\ref{AW_1}) and those of the ordinary channeling radiation
are well separated, provided conditions
(\ref{AW_0}) and  (\ref{AW_8a}) are fulfilled,

The ordinary channeling radiation is due to the transverse oscillations
(the first term in (\ref{AW_5})) of the particle inside the
channel under the action of the interplanar force. The characteristic
frequencies of this radiation may be estimated as
\cite{Kumakhov2}: $\omega_{\rm c}\sim \gamma^2 \chi\, c$, where
$\chi = (q U^{\prime}_{\max}/ \varepsilon\, d)^{1/2}$
(see (\ref{AW_5}) and (\ref{AW_8})).

The characteristic frequencies of the radiation formed
in the acoustically based undulator are given by
$\omega_{\rm u} \sim \gamma^2 k\, c = \gamma^2 \omega_0$
(see (\ref{Cond2})).
Then, the ratio $\omega_{\rm u}/\omega_{\rm c}$ is equal to
\begin{equation}
{\omega_{\rm u} \over \omega_{\rm c}}
\sim
\left(
{\varepsilon \over R_{\rm min}\,
q U_{\rm max}^{\prime}}
\right)^{1/2}
\,
\left(
{d \over a}
\right)^{1/2}
\ll 1,
\label{Cond4}
\end{equation}
provided the conditions \eref{AW_0} and (\ref{AW_8}) are fulfilled.
Indeed, for $a \sim 10 d$ and
$R_{\rm min}\sim 10\, \varepsilon/q U_{\rm max}^{\prime}$
one gets $\omega_{\rm u}/\omega_{\rm c}\sim 10^{-1}$.

This condition allows to distinguish two mechanisms for the
radiation formed in the channeling process in an acoustically bent
crystal, and to consider all the radiation in the spectral
region $\omega \ll \omega_{\rm c}$ as formed via the undulator
mechanism.

We note here, that the inequality (\ref{Cond4}) is fulfilled in the
whole range of the AW parameters consistent with the channeling
condition (\ref{AW_8a}), only if $d < a$ (eq. (\ref{AW_0}))
is valid too. In the opposite case, $a < d$\,
\cite{Ikezi84, Ketterson86, Armyane86}, which can be realized in the
undulators based on a superlattice \cite{Ketterson86}
or/and on a low-amplitude AW \cite{Ikezi84, Ketterson86, Armyane86}
the frequencies $\omega_{\rm u}$ and $\omega_{\rm c}$
can be of the same order of magnitude and, thus, two mechanisms
of the radiation formation must be treated simultaneously.
As mentioned several times already, in the present paper only
the case $d < a$ is considered.

When discussing the motion of the particle in a bent channel and
the undulator conditions, we ignored so far the effects
of dechanneling and the photon attenuation.
The influence of these effects on the undulator radiation will be
 discussed in the next sections.

Let us summurize our results so far:
Provided conditions \eref{AW_0}, (\ref{AW_8a}, \ref{Cond1}, \ref{Cond3})
are fulfilled, the system ``ultrarelativistic charged particle
+ acoustically bent crystal'' represents
a new type of planar undulator, and, consequently, serves as a new
source for high-frequency undulator radiation of high intensity,
monochromaticity and a particular pattern of the
angular-frequency distribution.
The latter as it will be illustrated  in the next sections.
The validity of the classical treatment of the particle motion in
a bent channel allows to apply the quasi-classical approach,
described in the previous section, for calculating the characteristics
of the undulator radiation.

%%%%%%%%%%%%%%%%%%%%%%%%%%%%%%%%%%%%%%%%%%%%%%%%%%%%%%%%%
\section{Numerical results for $\d E({\bf n})/ \d\omega\, \d
\Omega_{\bf n}$ and $\d E / \d\omega$} \label{Results}
%%%%%%%%%%%%%%%%%%%%%%%%%%%%%%%%%%%%%%%%%%%%%%%%%%%%%%%%%

In this section we present numerical results for the spectral
and the angular distrbutions of the spontaneous undulator
radiation formed during a passage of an ultra-relativistic
particle through an acoustically bent crystal.
The calculations are performed by using the formalism
outlined in  \sref{Formalism}.
We resctrict ourselves to the case of {\it positron} planar
channeling.

\subsection{The angular distribution of the radiation}

We start with the illustration of the general features of the
angular distribution of the radiation formed in an acoustically
based undulator.

\begin{figure}
\hspace{3cm}\epsfig{file=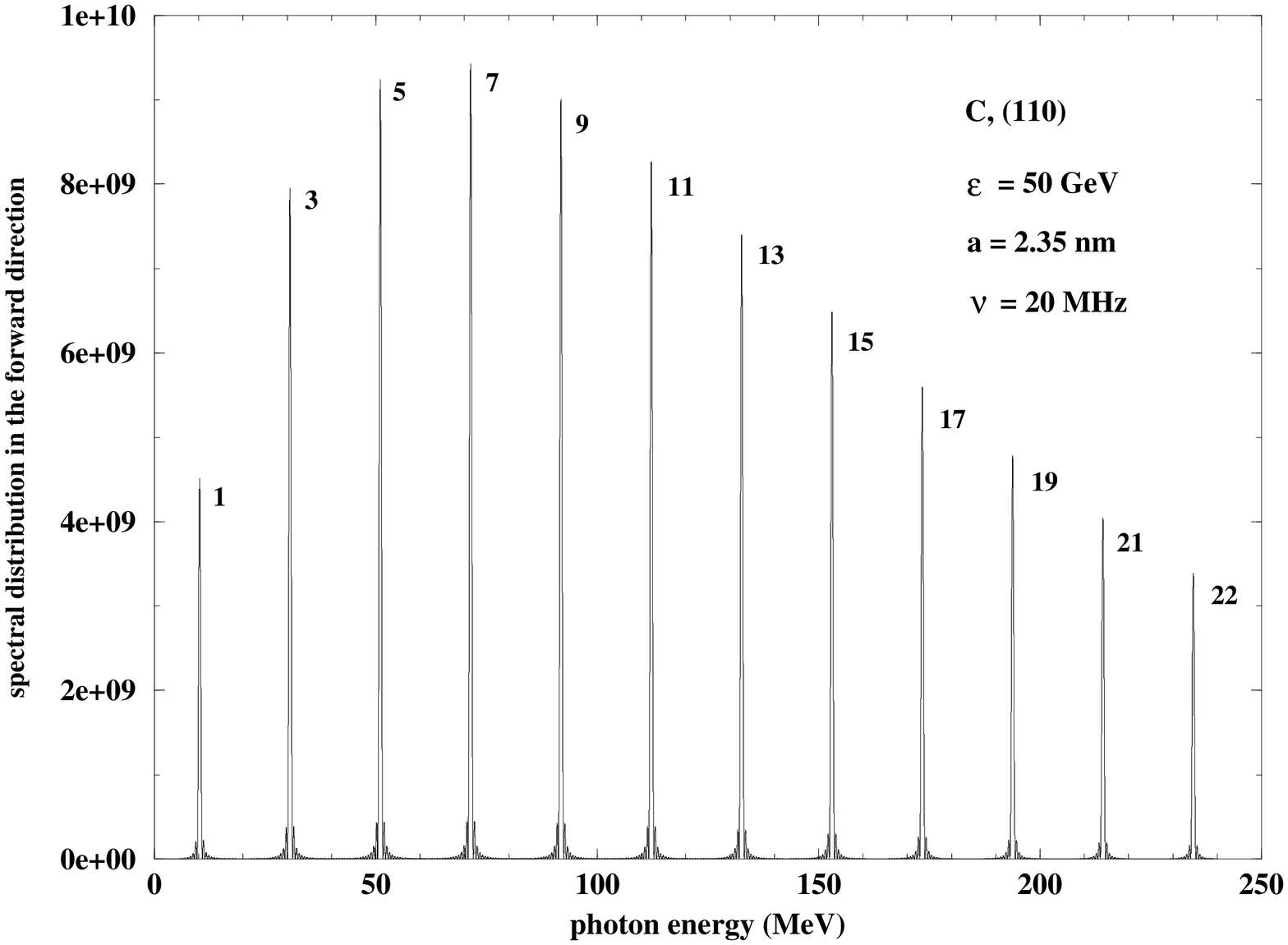,height=11cm,angle=270}
\caption{The spectral dependence of the radiated energy for a 50 GeV
positron channeling in a diamond.
The polar angle of the emission $\theta=0^{\circ}$.
Other parameters are as indicated.
See also the explanations in the text.}
\label{Fig.4_1}
\end{figure}

\begin{figure}
\hspace{3cm}(a)\epsfig{file=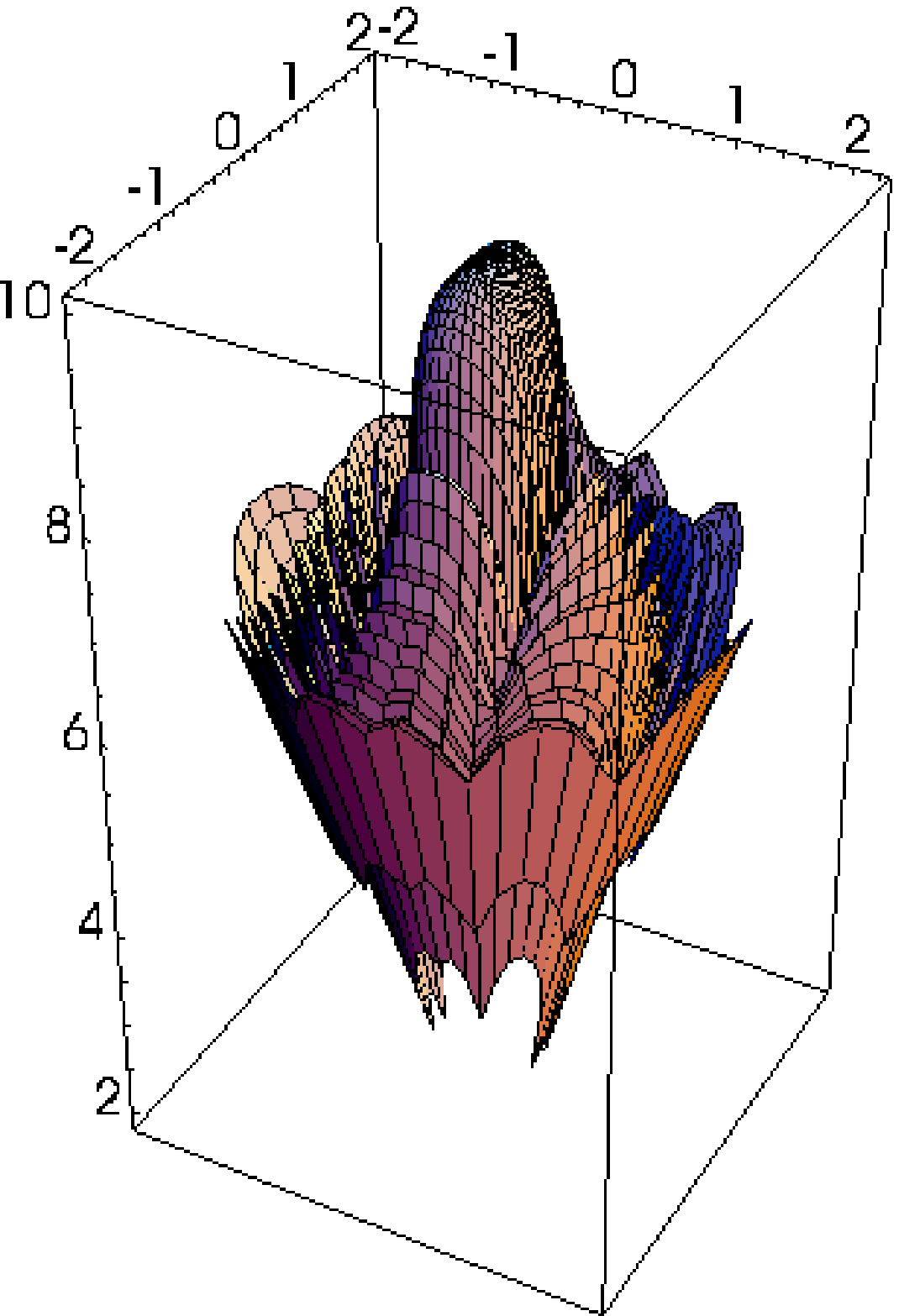,width=5cm}

\vspace{0.5cm}

\hspace{3cm}(b)\epsfig{file=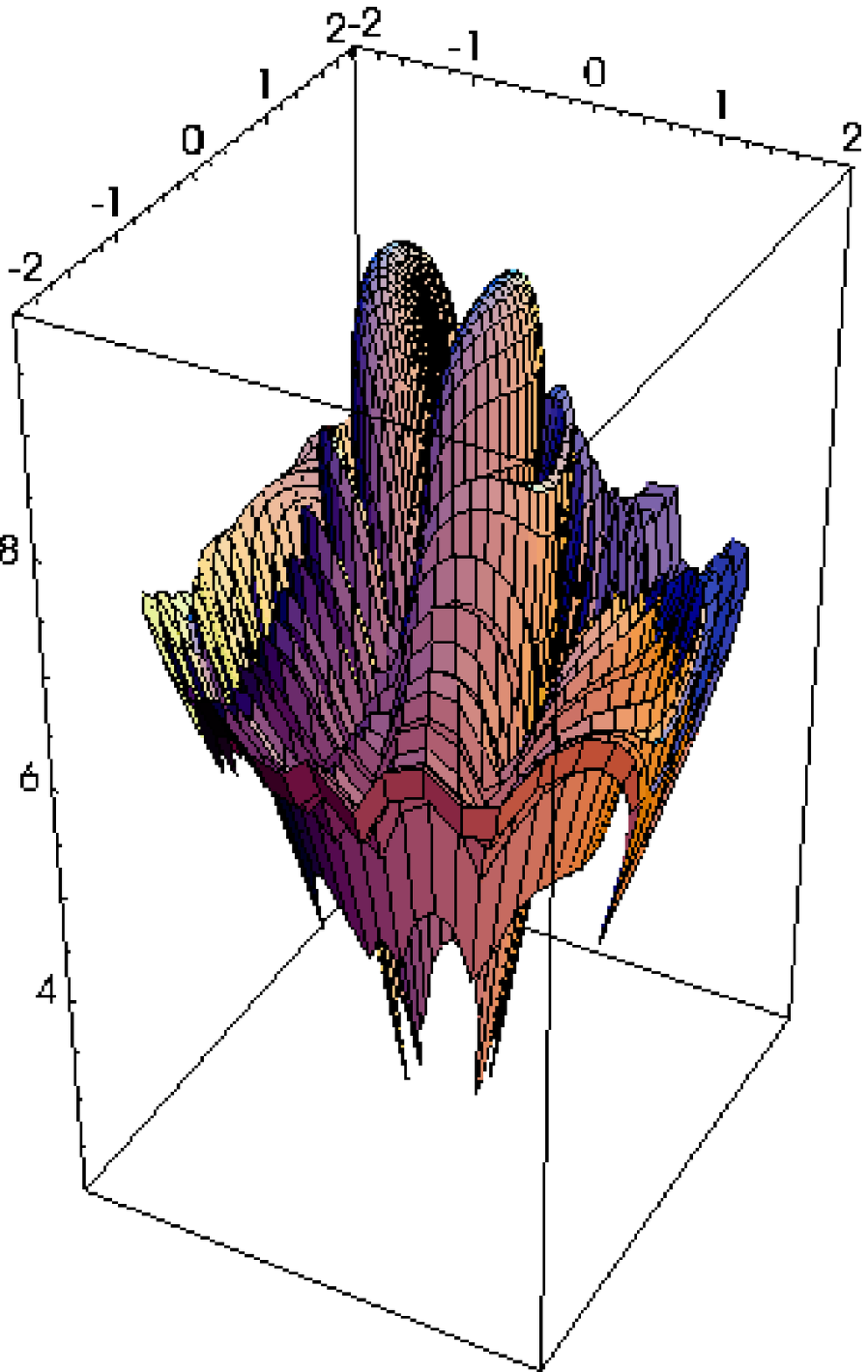,width=5cm}
\caption{The spatial distribution of the radiation emitted in:
(a) the odd harmonic with $K=7$, (b) the even harmonic with $K=6$.
Other parameters are as in \fref{Fig.4_1}.  The undulator axis
lies along the $z$ direction. The undulator plane is $yz$.  The $x$
and $y$ axis are scaled with respect to the dimensionles parameter
$\theta/\theta_0$.  See also the explanations in the text.}
\label{Fig.4_2}
\end{figure}

Figures \ref{Fig.4_1} and \ref{Fig.4_2} correspond
to the case of a 50 GeV positron channeling along (110) plane in
a carbon crystal of thickness $L=1$ cm.
For chosen values of the AW amplitude $a=2.35\cdot10^{-7}$ cm
and frequency $\nu = 20$ MHz, the corresponding
magnitudes of the total number of the undulator periods, the
undulator parameter and the fundamental harmonic energy are
as follows: $N_{\rm u}=17,\ p^2=10.1,\ \hbar\omega_1=9.9$ MeV.
In a diamond crystal the spacing for the (110) planes is
$d=1.26\cdot10^{-8}$; therefore the condition \eref{AW_0}
is well fulfilled.

%%%%%%%%%%%%%%%%%%%%%%%% Angular distributions

\Fref{Fig.4_1} represent the spectral distribution of the radiation
emitted along the undulator axis,
$\hbar^{-1}\left(\d E({\bf n})/\d\omega\,\d\Omega_{\bf n}\right)_
{\theta=0^{\circ}}$ (see \eref{Ang1}).
Each peak corresponds to the emission into the odd harmonics
\cite{Kniga, Baier, Alferov}, the energies of which are  deduced
from the relation
\begin{equation}
\omega^{\prime}_K =
{4\gamma^2 \omega_0\, K \over p^2 +2}
\label{eq4.1}
\end{equation}
which is obtained from \eref{quasi_18} and \eref{quasi_11}
by putting $\theta=0^{\circ}$.

It is seen that all harmonics are well separated: the distance
between two neighbouring peaks is $2 \hbar\omega_1 \approx 20$ MeV
whilst the width of each peak is
$\hbar\omega_1/N_{\rm u} \approx 1.2$ MeV.
In the vicinity of each main maximum the shape of the
spectral distribution is governed by the behaviour of the function
$D(\eta)$ defined in \eref{quasi_8}.

Two 3D plots in figures \ref{Fig.4_2}(a) and \ref{Fig.4_2}(b)
allow to visualize the spatial behaviour of the angular distribution
of the radiation (at fixed frequency)
 as a function of the azimuthal, $\varphi$, and
the polar, $\theta$, angles of the photon emission.
In these figures the quantity
${\rm lg}\left(\d E({\bf n})/\d(\hbar\omega)\,\d\Omega_{\bf n}\right)$
(measured along the $z$-axis) as a function of the azimuthal,
$\varphi$, and the polar, $\theta$, angles of the photon emission is
plotted for two central harmonics.
\Fref{Fig.4_2}(a) represents the distribution
of the radiation in the odd harmonic $\hbar\omega_K \approx 71.4$ MeV
with $K=7$, \fref{Fig.4_2}(b) corresponds to the emission in the
even harmonic $\hbar\omega_K \approx 61.2$ MeV, $K=6$.

The undulator axis lies along the $z$ direction, and $yz$ is the
undulator plane. The positive $y$ direction corresponds to
$\varphi=0^{\circ}$.
In these figures the dimensionless variable $\theta/\theta_0$ is
used to characterize the distribution in of the radiation with respect
to the polar angle.
In the considered case the quantity $\theta_0$, defined by
\eref{quasi_17},  is equal to $\theta_0 =23\ \mu$rad.

Figures \ref{Fig.4_2}(a,b) illustrate general features intrinsic
to the planar undulator radiation in the case $p^2\gg 1$.
We first note that the intensity of the radiation in the odd harmonic
is governed by a powerful maximum in the forward direction, whereas
there is no radiation in even harmonics for $\theta= 0^{\circ}$
The latter reaches its maximal values in nearly forward, but not in
forward, directions.
Apart from the intensity peak in the forward (for $K$ odd)
or nearly forward ($K$ even) direction, the radiation in a particular
harmonic is emitted in a wide range of the polar angles. In both
of the figures there are several clearly distinct non-forward peaks
in which the radiation intensity reaches the maxima, although the
magnitudes of $\d E({\bf n})/\d(\hbar\omega)\,\d\Omega_{\bf n}$
in the maxima at $\theta\neq 0^{\circ}$ rapidly decrease with the
polar angle (recall the log scale along the $z$-axis).

Another feature to be mentioned is the abscence of the axial symmetry
in the angular distribution shape.
More specifically, the radiation emitted within the undulator plane
is concentrated in the cone $\theta_{\parallel}\sim \theta_0 =
p/(\sqrt{2}\gamma)$, whereas the cone angle for the photon
emission in the direction perpendicular to the $yz$ plane equals
$\theta_{\perp}\sim 1/\gamma$.
Thus, the ratio $\theta_{\parallel}/\theta_{\perp} \sim p/\sqrt{2}$
characterizes the above mentioned asymmetry in the angular
distribution with respect to the atimuthal angle $\varphi$.
This ratio is equal to 2.2 in the considered case.
This peculiarity most clearly exhibits itself in the shape of the
spatial distribution of the radiation in the odd harmonics in the
forward direction, see \fref{Fig.4_2}(a).

By changing the parameters of the acoustically based undulator (which
are the AW amplitude and frequency, the type of a crystal, and
the relativistic factor $\gamma$ of the projectile positron)
it is possible to vary the intensity of the radiation, the
energy, width and the number of emitted harmonics.
To illustrate this statement in figures \ref{Fig.4_3}(a,b,c) we
present the dependencies of the spectral distribution of the radiation
emitted at $\theta=0^{\circ}$ for a 2, 5 and 10 GeV positron
channelling along (110) planes in $Ge$ and $Si$ crystals.
In addition to the parameters indicated in the figures there are
other ones, i.e. the crystal length, the number of the undulator
periods, the undulator parameter, the fundamental harmonic energy
and the width of the line. They  are the following:

\vspace*{0.2cm}
\noindent
\Fref{Fig.4_3}(a):\
$L=0.1$ cm,\
$N_{\rm u}=36$,\
$p^2=3.25$,\
$\hbar\omega_1 = 53$ keV,\
$\hbar\,\Delta\omega = 2.9$ keV;

\vspace*{0.2cm}
\noindent
\Fref{Fig.4_3}(b):
$L=1$ cm,\
$N_{\rm u}=43$,\
$p^2=6.9$,\
$\hbar\omega_1 = 0.23$ MeV,\
$\hbar\,\Delta\omega = 10.5$ keV;

\vspace*{0.2cm}
\noindent
\Fref{Fig.4_3}(c):
$L=1$ cm,\
$N_{\rm u}=43$,\
$p^2=27.8$,\
$\hbar\omega_1 = 0.27$ MeV,\
$\hbar\,\Delta\omega = 12.7$ keV.

\begin{figure}
\hspace{3cm}\epsfig{file=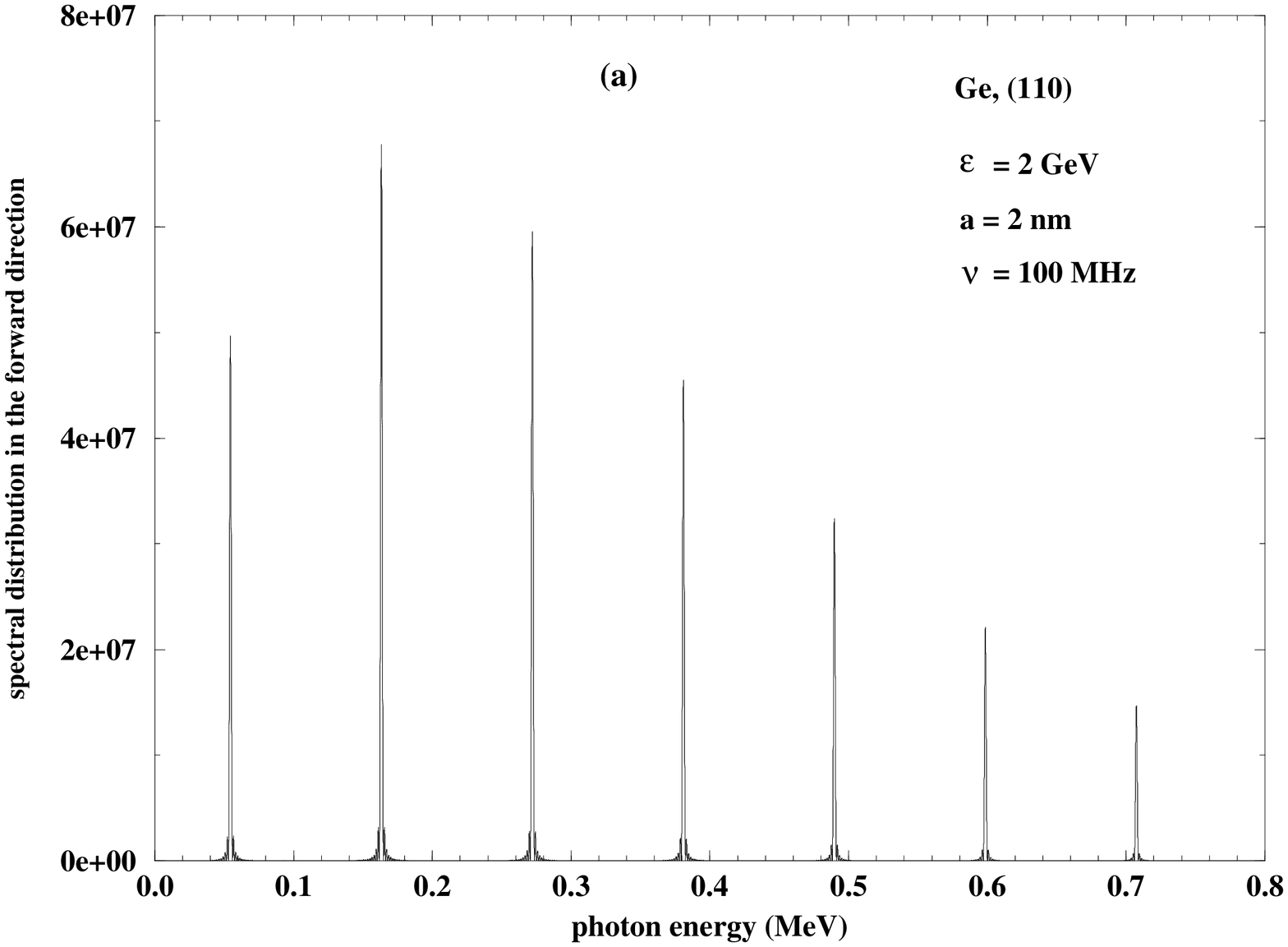,height=11cm,angle=270}
\caption{The dependence of
$\hbar^{-1}\, \d E({\bf n})/\d\omega\,\d\Omega_{\bf n}$
at $\theta=0^{\circ}$ on the photon energy $\hbar\omega$
calculated for a positron channeling in an acoustically bent
crystals. The types of the channels, the positron energies and
the AW parameters are as indicated. Other parameters are
enlisted in the text.}
\label{Fig.4_3}
\end{figure}

\setcounter{figure}{4}
\begin{figure}
\hspace{3cm}\epsfig{file=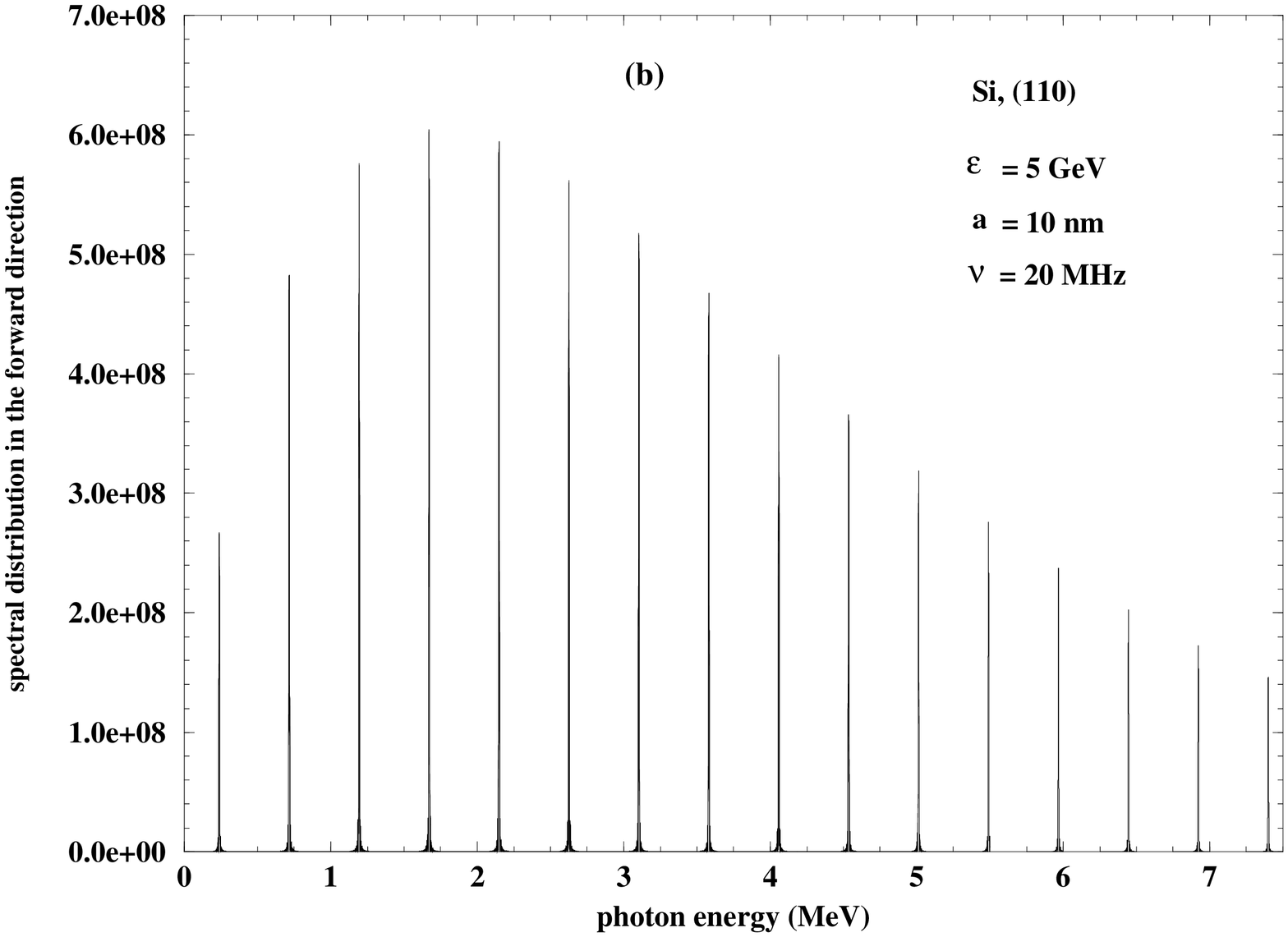,height=11cm,angle=270}
\caption{(\emph{continued})}
\end{figure}

\setcounter{figure}{4}
\begin{figure}
\hspace{3cm}\epsfig{file=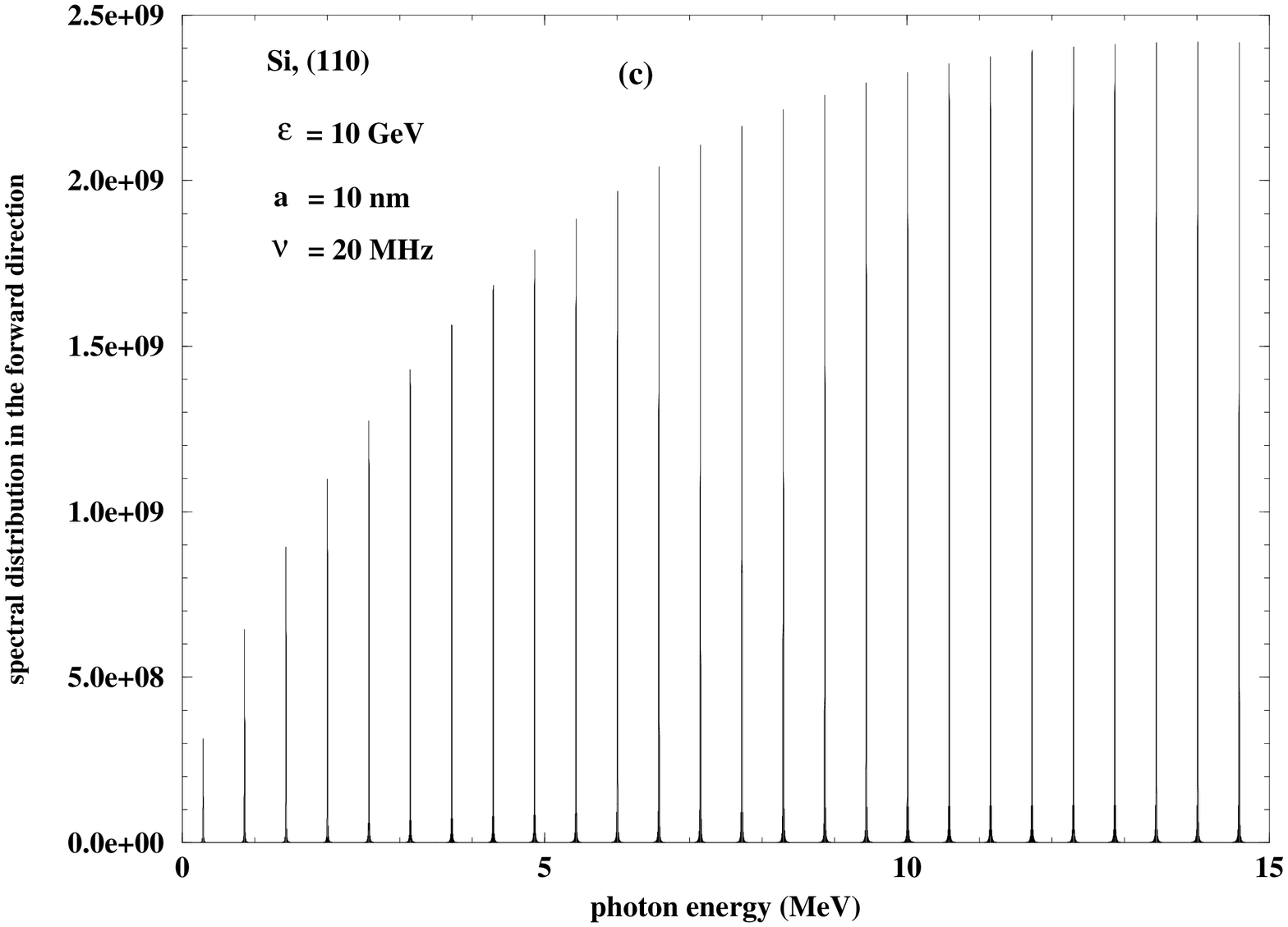,height=11cm,angle=270}
\caption{(\emph{continued})}
\end{figure}

\subsection{The spectral distribution of the radiation}

For large undulator parameters, $p> 1$, the spectral distribution
of the radiation emitted by a charged projectile channeling in a
crystal bent by an acoustic wave is obtained from eqs.
(\ref{Sp1},\ref{Sp2}).
Using these formulae, we calculated the spectral distributions for
different parameters of the AW and for various crystals.
The results of the calculations are presented in figures
\ref{Fig.4_4}--\ref{Fig.4_6}.

\begin{figure}
\hspace{3cm}\epsfig{file=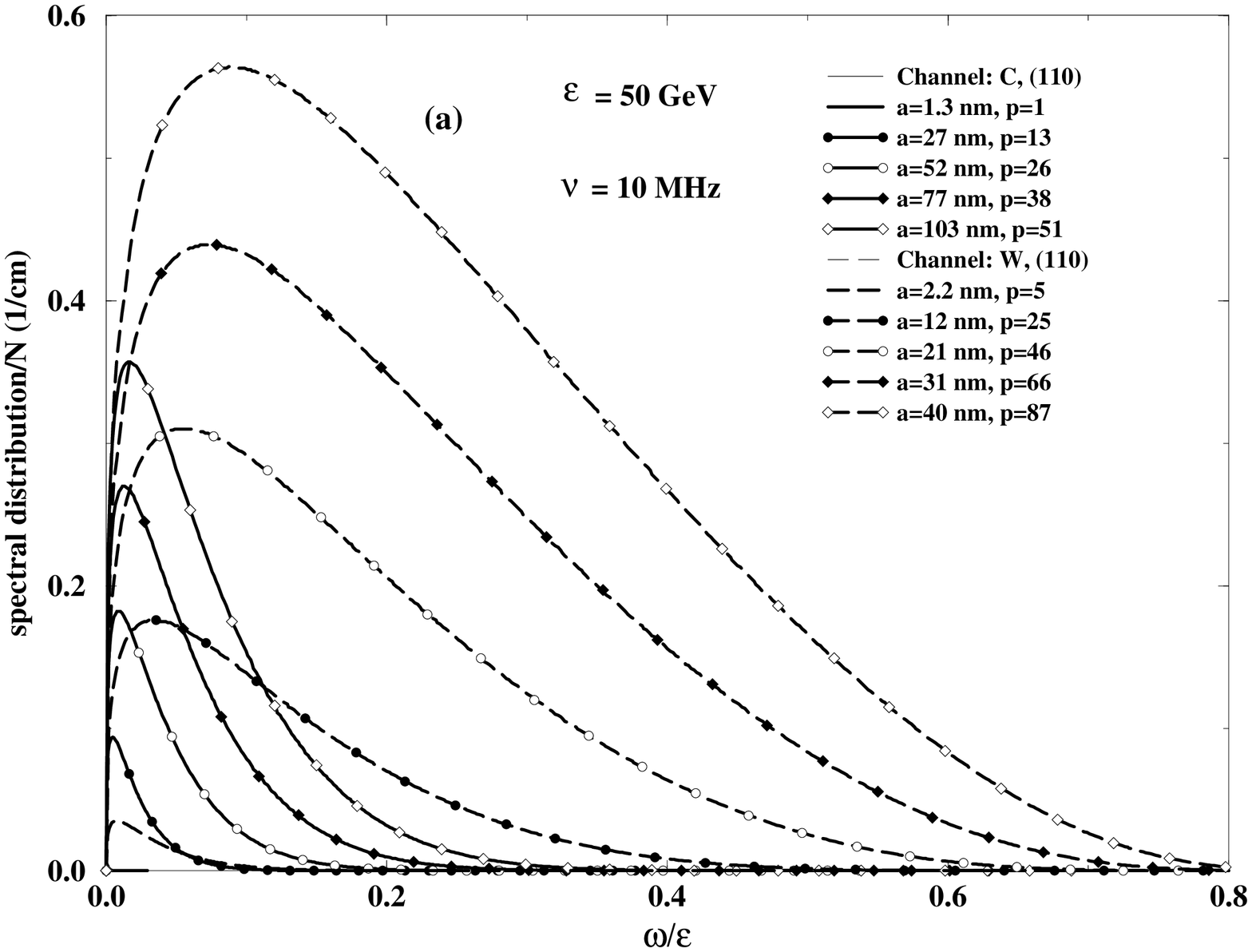,height=11cm,angle=270}
\caption{The spectral intensity (per one period) of the AIR,
$\d E/\d(\hbar \omega)/L/N_u$, emitted by:
{\bf (a)} a $50$ GeV  positron,
{\bf (b)} a $500$ GeV  positron,
channeling along the (110) plane in $C$ and $W$
crystals calculated for the fixed AW frequency $\nu = 10$ MHz
and for various values of the AW amplitude $a$ and the
corresponding undulator parameters $p$ (as indicated).
Further explanations are given in the text.}
\label{Fig.4_4}
\end{figure}

\setcounter{figure}{5}
\begin{figure}
\hspace{3cm}\epsfig{file=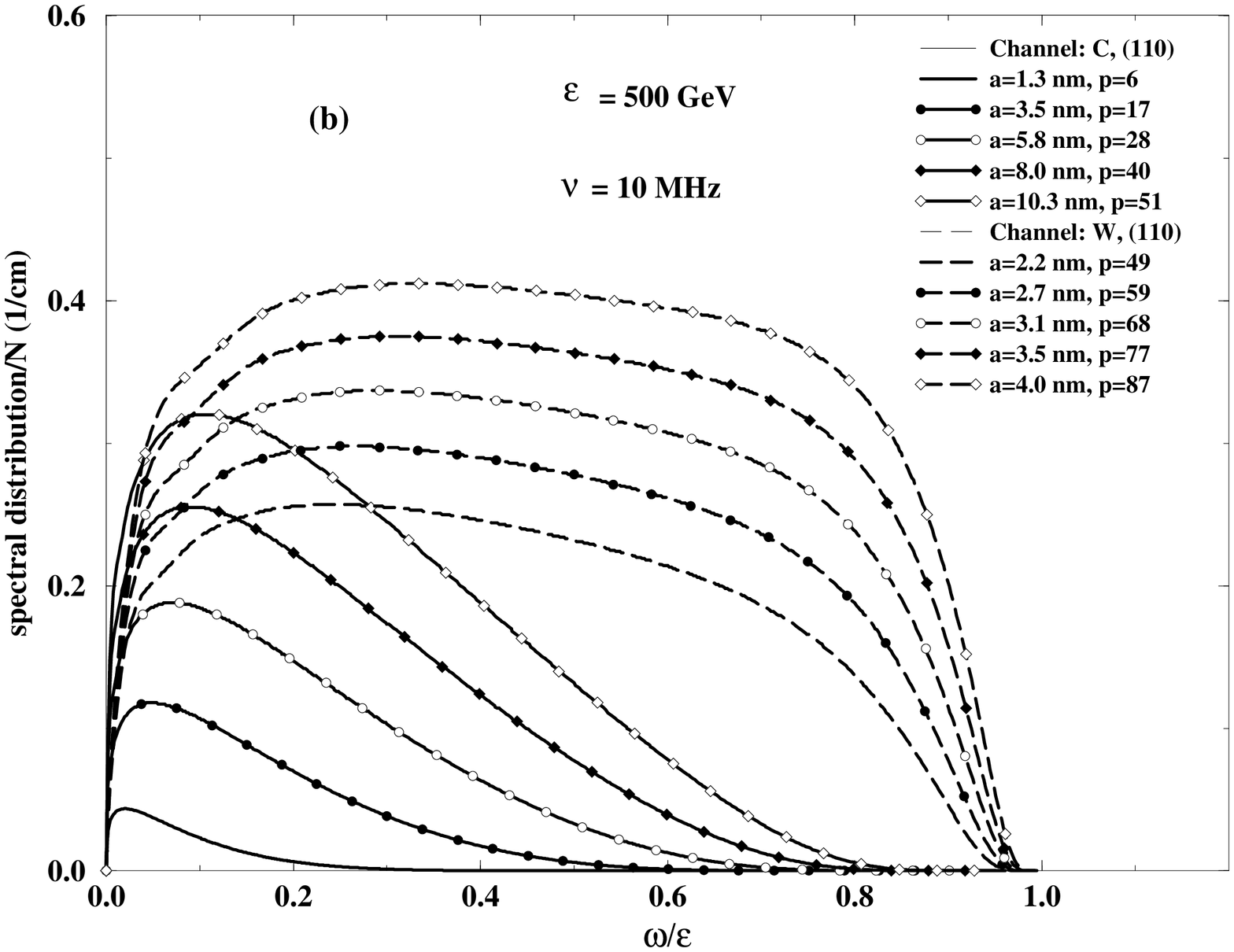,height=11cm,angle=270}
\caption{(\emph{continued})}
\end{figure}

%%%%%%%%%%%%%%%%%%%%%%%% Spectral distributions

\begin{figure}
\hspace{3cm}\epsfig{file=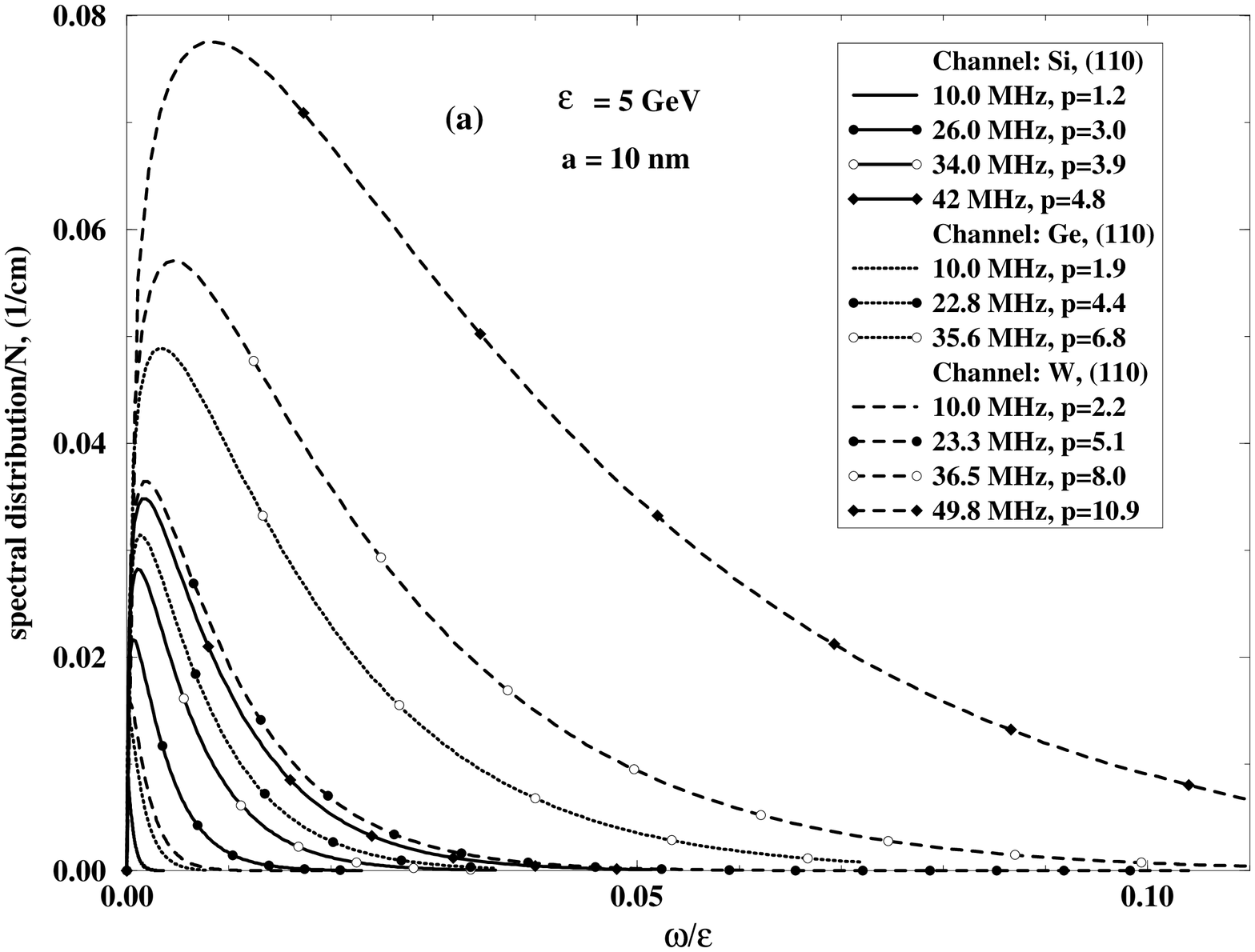,height=11cm,angle=270}
\caption{The spectral intensity (per one period) of the AIR,
$\d E/\d(\hbar \omega)/L/N_u$, emitted by:
{\bf (a)} a $5$ GeV  positron,
{\bf (b)} a $50$ GeV  positron,
channeling along the (110) plane in  $Si$, $Ge$ and $W$
crystals calculated for the fixed AW  amplitude $a = 10$ nm
and for various values of the AW  frequency $\nu$ and the
corresponding undulator parameters $p$ (as indicated).
See also explanations in the text.}
\label{Fig.4_5}
\end{figure}

\setcounter{figure}{6}
\begin{figure}
\hspace{3cm}\epsfig{file=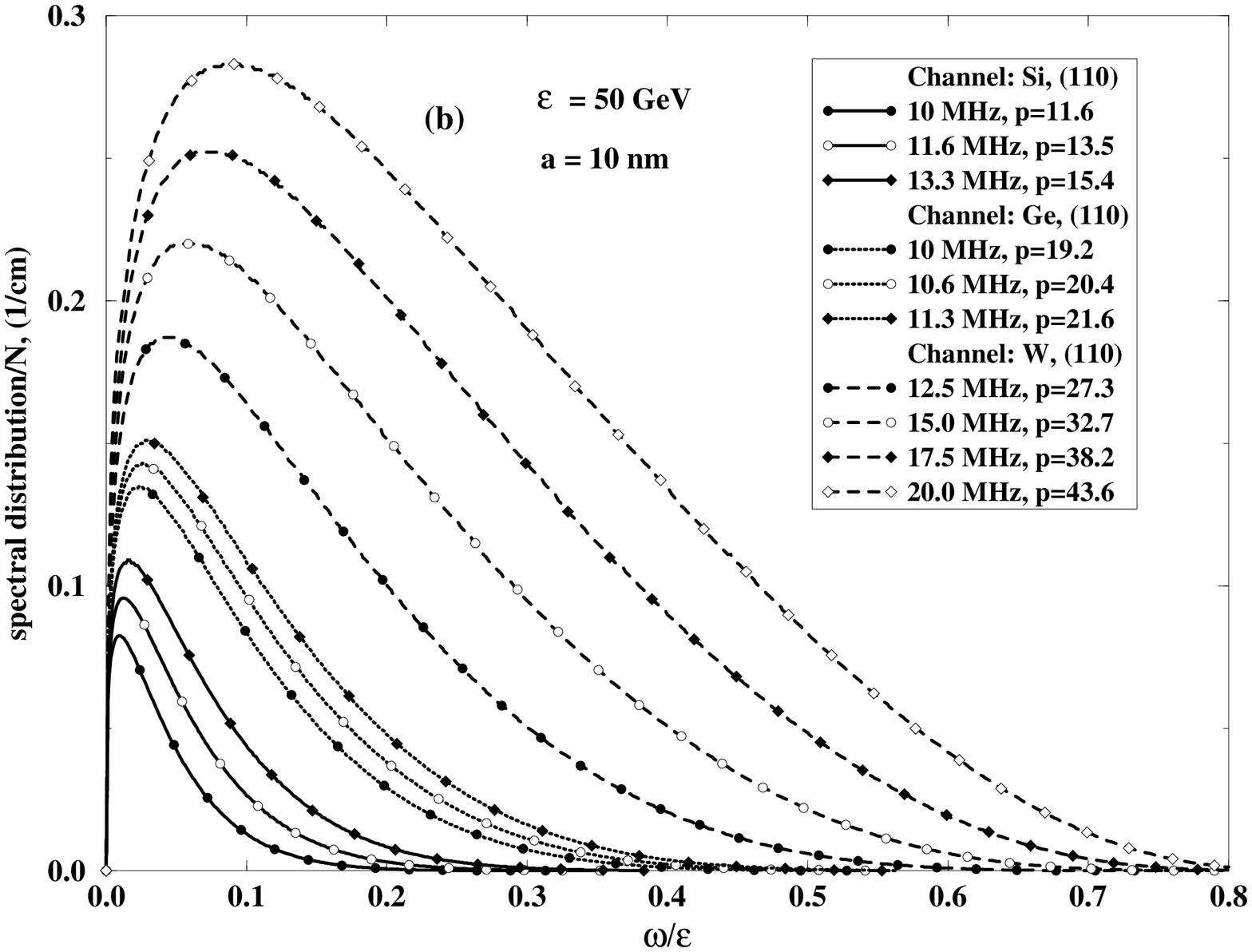,height=11cm,angle=270}
\caption{(\emph{continued})}
\end{figure}

\begin{figure}
\hspace{3cm}\epsfig{file=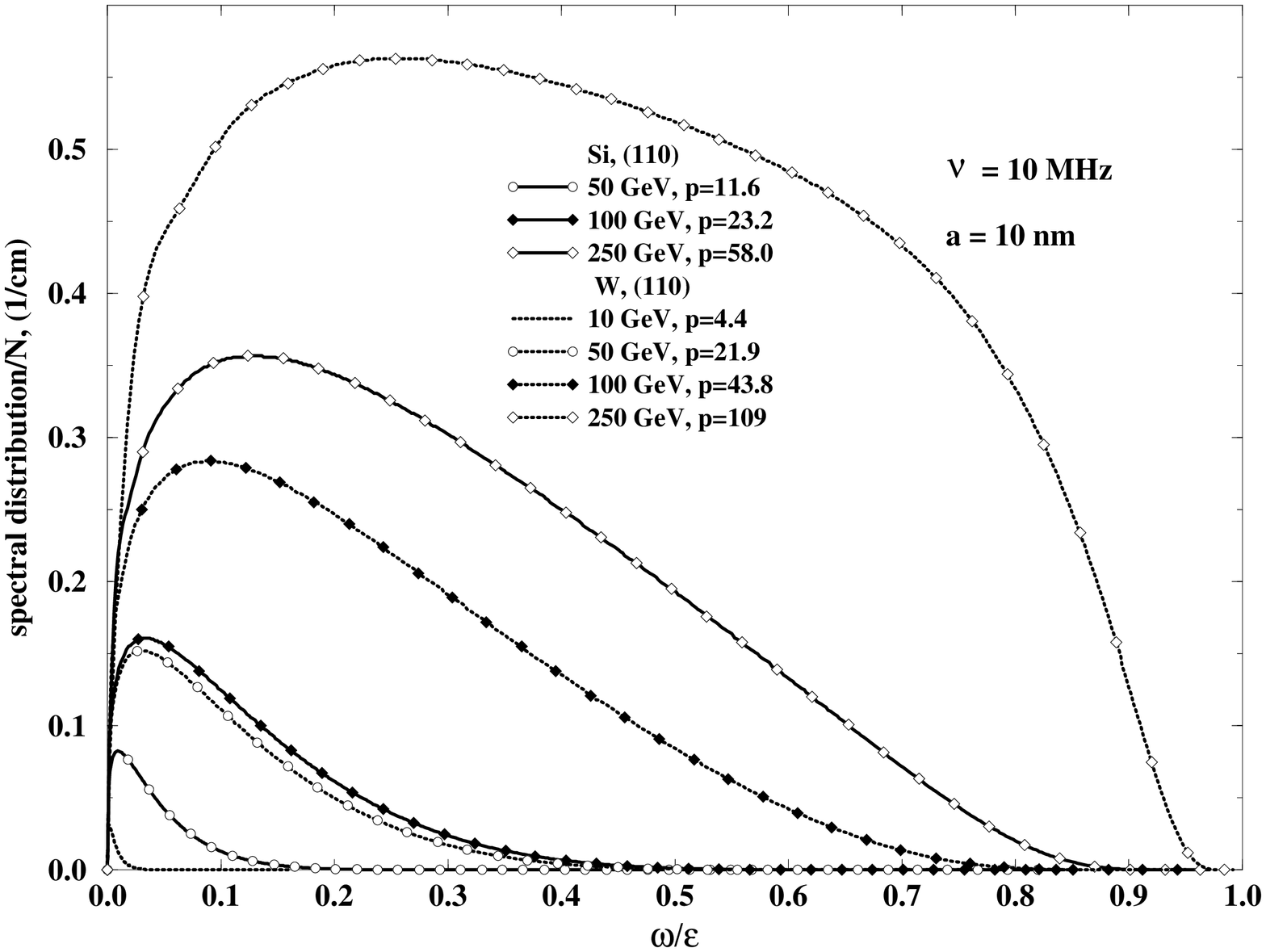,height=11cm,angle=270}
\caption{The spectral intensity (per one period) of the AIR,
$\d E/\d(\hbar \omega)/L/N_u$, emitted by a positron of various
$\varepsilon$ (as indicated) channeling along the (110) plane
in $Si$ and $W$ crystals calculated for the fixed AW amplitude
$a = 10$ nm and frequency $\nu=10$ MHz.
The corresponding undulator parameters $p$ are indicated.}
\label{Fig.4_6}
\end{figure}

Figures \ref{Fig.4_4}(a,b) exhibit dependencies of
$\d E/\d(\hbar \omega)/L/N_u$ (in cm$^{-1}$) for $\varepsilon=50$
and $\varepsilon=500$ GeV {\em positrons} moving along the (110)
channels in a diamond and a tungsten crystal.
The AW frequency is fixed at $\nu=10$ MHz.
In these figures each curve shows the $\omega$-dependence obtained
at a particular value of the AW amplitude and the corresponding  value
of the undulator parameter $p$, as indicated.
The lowest values for the AW amplitude are chosen to be consistent
with the condition (\ref{AW_0}), where the parameter $d$ is equal to
1.26 \AA \ for a diamond and 2.24 \AA \ for a tungsten (110) channels
\cite{Biryukov}.
The largest $a$-values correspond to the relation
$\varepsilon /(R_{\rm min}\, e U_{\rm max}^{\prime}) \approx 1/6$
(see (\ref{AW_8})).

These figures demonstrate that at a certain frequency $\nu$ the
intensity of radiation can be varied in a wide range by altering the
AW amplitude, which is proportional to the value of the undulator
parameter as $a= p/(\gamma k)$, or by choosing the type of a crystal.
The latter option affects the value of the undulator parameter via
its dependence on the sound velocity $V$: $p = \gamma\, (2\pi\nu/V) a$.
Therefore, for equal $a$, $\nu$ and $\gamma$ the AIR intensity,
being proportional to $p$ (see (\ref{Sp1})), is approximately
$(V_{C}/V_{W}) \approx 4$ times higher for a tungsten crystal as
compared with that for a diamond.

The radiation intensity can also be varied by changing the  AW
frequency at some constant value of $a$.
This possibility is illustrated by figures \ref{Fig.4_5}(a,b)
where the dependencies of
$\d E/\d(\hbar \omega)/L/N_u$ on $\omega$ for $\varepsilon=5$
and $\varepsilon=50$ GeV positron channeled in $Si$, $Ge$ and $W$
crystals are presented.

The spectral distributions calculated at different energies
$\varepsilon$ of a positron channeling in $Si$ and $W$ crystals
are compared in \fref{Fig.4_6}.

The features to be discussed in connection with all these figures are
the pattern of the spectral distribution shape and its
transformation with the increase in $\varepsilon$.
It was noted above (see the discussion below \eref{Ang1}) that for
$\omega^{\prime} \gg \omega_{\rm max}^{\prime} \approx
p\gamma^2\,\omega_0$ the intensity of the undulator radiation
(both the angular and the spectral ones) exponentially decreases to zero.
Within the frame of the quasi-classical approximation the photon
frequency $\omega$ is related to the  quantity $\omega^{\prime}$
through  the first equality in (\ref{wkb3}).
Thus, the intensity of radiation effectively falls off for the photon
energies satisfying the relation
\begin{equation}
{\hbar\omega \over \varepsilon} >
{ 1 \over  1 + \varepsilon/ p\gamma^2\,\hbar\omega_0}
=
{ 1 \over  1 + a/\lambda_c\, p^2}
\label{additional1}
\end{equation}
Here $ \lambda_c = \hbar/ mc = 2.42\cdot 10^{-10}$ cm is the
Compton wavelength of the positron. To obtain the equality in
(\ref{additional1}) we made use of $p=\gamma k a$ and $\omega_0=k c$.

In the case of $(a/\lambda_c)\, p^{-2} \gg 1$, which is achieved by
choosing relatively low values of the parameters $\gamma$, $a$ or
$\nu$, the spectrum of radiation is concentrated
in the region of low photon energies, $\hbar\omega/\varepsilon \ll 1$
whereas for $\hbar\omega/\varepsilon \leq 1$ it rapidly drops to zero,
as follows from the estimation (\ref{additional1}).
Such a behaviour most vividly is illustrated by \fref{Fig.4_5}(a)
(see also the curves for the C (110)  channel in \fref{Fig.4_4}(b)
and for the Si (110) and Ge (110) channels in \fref{Fig.4_5}(b)).

By increasing $\gamma$ the parameter
$(a/\lambda_c)\, p^{-2}$ becomes $<1$, leading to a prominent increase
in the intensity of radiation in the high-energy photon region,
$\hbar\omega/\varepsilon \leq 1$. The curves for the W (110) channel
in \fref{Fig.4_4}(b) and the curve for a 250 GeV positron
channelling in  W (110) (see \fref{Fig.4_6}) clearly exhibit
the transformation of the spectral distribution shape: it becomes
flatter, thus reflecting that, except for the regions
$\hbar\omega/\varepsilon \longrightarrow 0$ and
$\hbar\omega/\varepsilon \longrightarrow 1$, the probability of the
photon emission is almost independent on the photon energy.

\subsection{The energy loss due to the AIR}

The stability of spontaneous undulator radiation formed during
the passage of the projectile through an acoustically bent crystal
of total length $L$ to a great extend is subject to the condition
$\Delta \varepsilon/\varepsilon \ll 1$, where $\Delta \varepsilon$
is the absolute decrease in the projectile energy, $\varepsilon$,
due to various inelastic processes.
In particular, the above-written inequality implies, that the
energy loss, $\Delta  \varepsilon$, due to spontaneous AIR only must be
small compared with $\varepsilon$.

To estimate this quantity in the case $p^2\gg 1$ it is necessary to
integrate $\d E /\d\omega$ from (\ref{Sp1}) over the interval
$\hbar\omega =0\dots \varepsilon$.
On the other hand, to obtain $\Delta \varepsilon $ for $p^2<1$
it is sufficient to use
the classical expression (\ref{Ang3}) for the spectral and angular
distribution of the radiation and to carry out the integrations over
the angles $\varphi$ and $\theta$ and over the photon energies.

In figures \ref{Fig.4_7}(a,b) the dependencies of the relative energy
loss, $\Delta \varepsilon/\varepsilon$, on the energy of the
projectile $\varepsilon$ is plotted for a positron channeling in (110)
channels for $C$, $Si$ and $W$ crystals and for various AW frequencies
as indicated.  The curves in \fref{Fig.4_7}(a) correspond to the AW
amplitude $a = 2$ nm, which is equal, by the order of magnitude, to
$10 d$ for all crystals considered.  In \fref{Fig.4_7}(b) $a = 10$ nm.
For all curves in these figures the AW frequency and amplitude are
subject to the strong inequality $\nu^2 a \ll C$ (see
eq. (\ref{AW_8a})).  Both figures correspond to a crystal length $L=1$
cm.  The total number of the undulator periods, $N_{\rm u}$ is related
to $L$, $\nu$ and the AW velocity $V$ according to $N_{\rm u}= L\cdot
V/\nu$.

\begin{figure}  %%%%%%%% Energy loss
\hspace{3cm}\epsfig{file=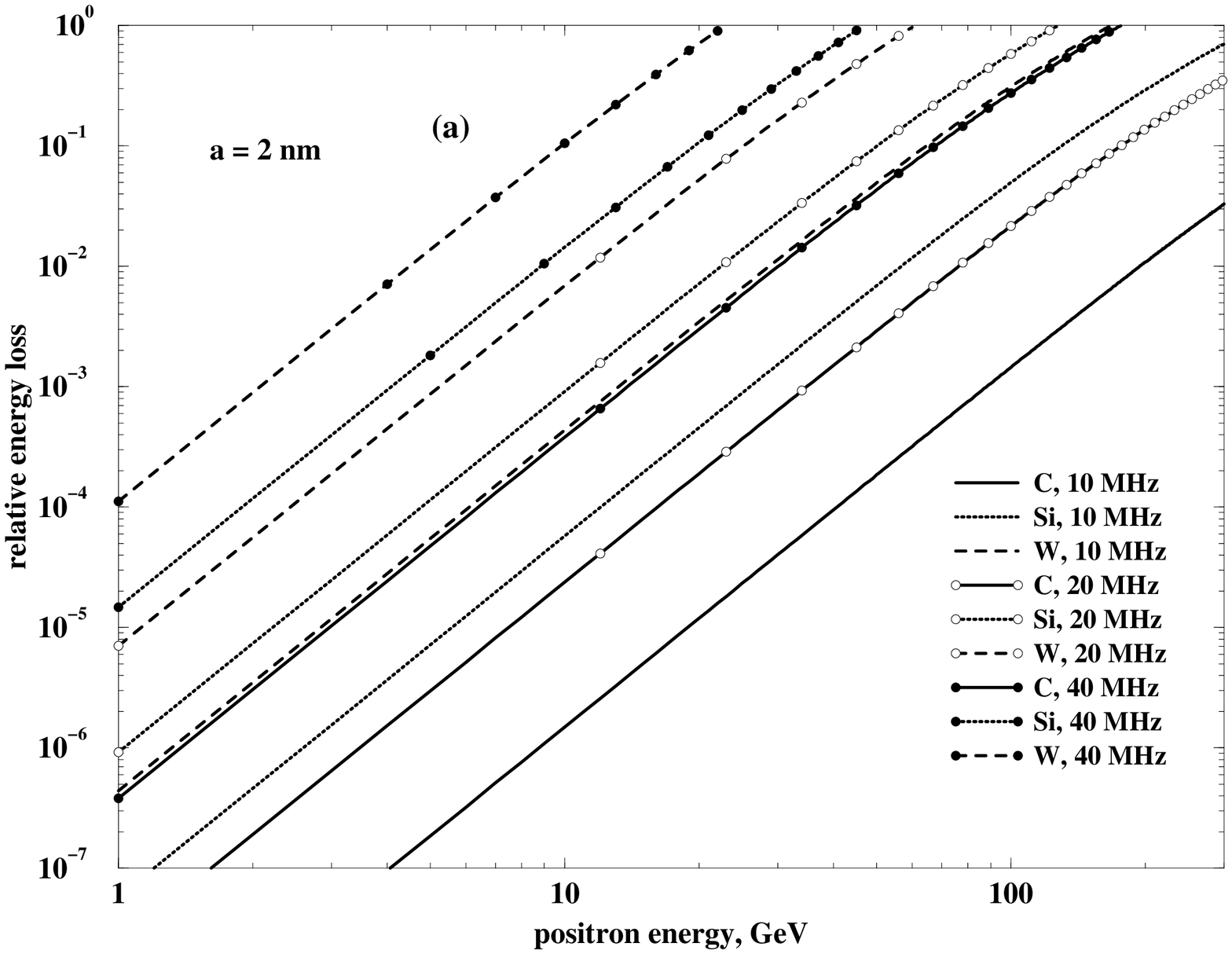,height=11cm,angle=270}
\caption{The relative energy loss, $\Delta \varepsilon/\varepsilon$, due
to the acoustically induced radiation versus the energy of a
positron channeling along the (110) plane in  $C$, $Si$ and $W$
crystals calculated for the fixed AW amplitude:
{\bf (a)} $a = 2$ nm,
{\bf (b)} $a = 10$ nm,
and for various values of the AW  frequency $\nu$  (as indicated).
The curves correspond to the crystal length $L=1$ cm.}
\label{Fig.4_7}
\end{figure}

\setcounter{figure}{8}
\begin{figure}
\hspace{3cm}\epsfig{file=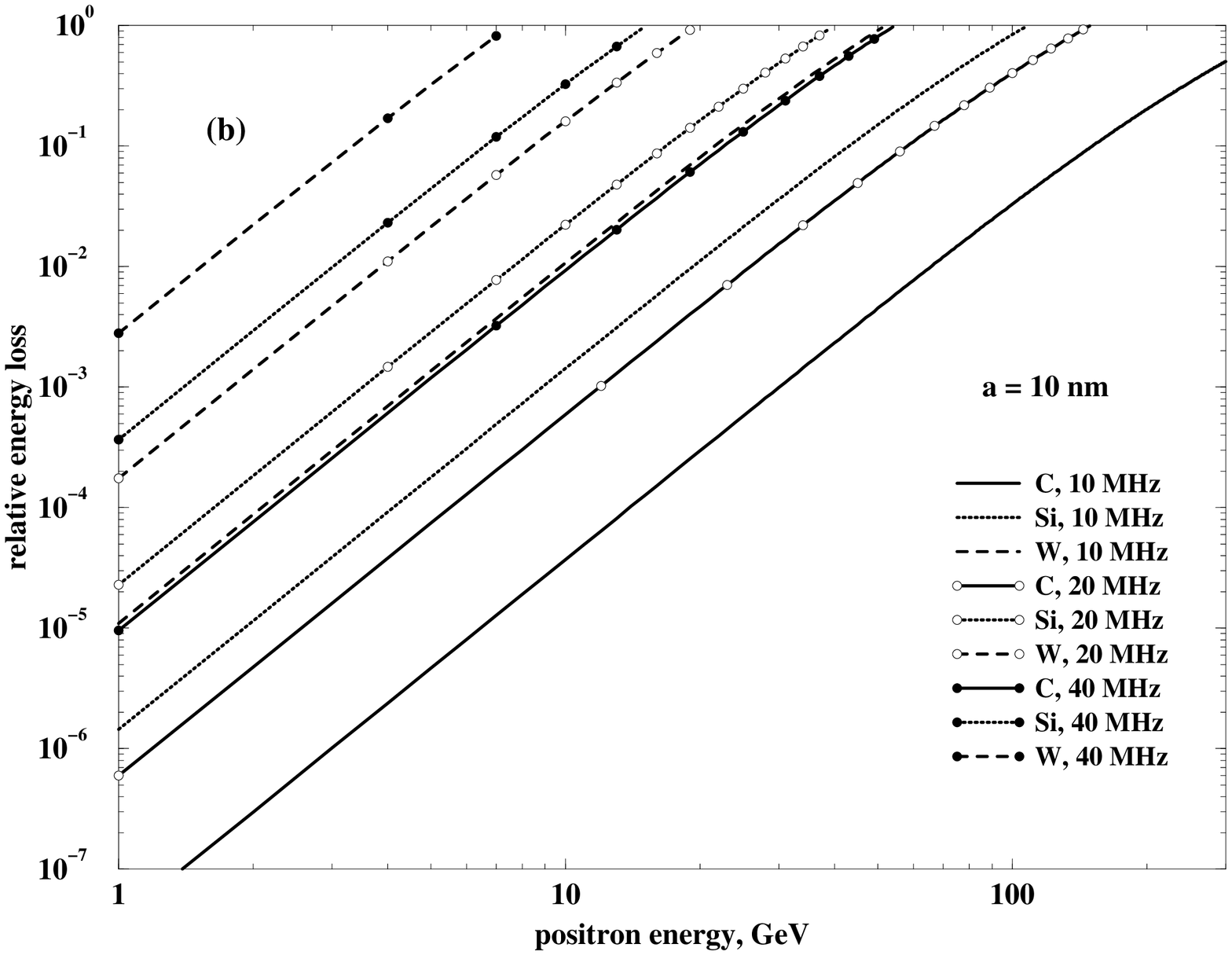,height=11cm,angle=270}
\caption{(\emph{continued})}
\end{figure}

Except for the case $p^2 < 1$, it is hardly feasible to establish
by explicit analytical means the ranges of parameters inside which
the inequality $\Delta \varepsilon /\varepsilon \ll 1$ is valid.
Indeed, expressions (\ref{Sp1}) and (\ref{Sp2}) exhibit quite
complicated dependence of  $\d E /\d\omega$, and, consequently,
that of $\Delta \varepsilon$ on $\varepsilon$, $a$ and $\nu$.
In addition to these the energy loss is dependent (indirectly,
through the conditions (\ref{AW_0}), (\ref{AW_8a}) and (\ref{Cond1}))
on the parameters of the crystal, i.e. on
 $d$, $U_{\rm max}^{\prime}$ and $V$  and on its length $L$.

Thus, for any particular set of the above-mentioned parameters, the
magnitude of the energy loss must be calculated numerically.
Nevertheless, figures \ref{Fig.4_7}  allow
to {\em estimate} the range of validity of
$\Delta \varepsilon /\varepsilon \ll 1$.
It is seen that for $a = 10\dots 100\, d$ and $L \leq 1$ cm
there are wide ranges of positron energy ($\varepsilon$ from
$\leq 1$ GeV up to several tens of GeV) and  AW frequency
($\nu$ from several MHz up to several tens of MHz) for which all
the undulator conditions  (\ref{AW_8a}, \ref{Cond1}, \ref{Cond3})
are fulfilled and, simultaneously, the relative energy loss due to
the acoustically induced radiation is negligibly small.

%%%%%%%%%%%%%%%%%%%%%%%%%%%%%%%%%%%%%%%%%%%%%%%%%%%%%%%%%
\section{Stimulated emission}\label{Stimulated}
%%%%%%%%%%%%%%%%%%%%%%%%%%%%%%%%%%%%%%%%%%%%%%%%%%%%%%%%%

In this section we discuss the possibility to generate
stimulated emission of high energy photons by means
of a bunch of ultra-relativistic particles of
length $L_{\rm b}$ moving in a channel bent by a
transverse acoustic wave.

\begin{figure}
\hspace{3cm}\epsfig{file=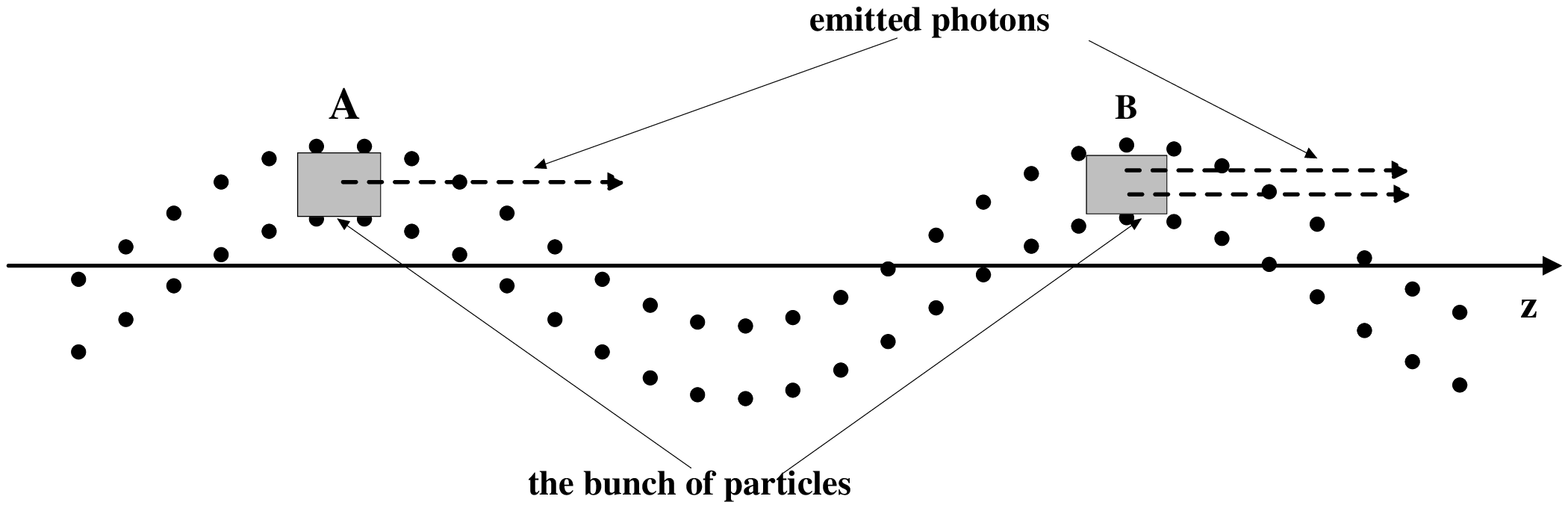,height=11cm,angle=270}
\caption{Mechanism of the radiation stimulation in the AW based
undulator. Photons, which are emitted in the forward direction by the
positron bunch (dark block) as it passes the undulator maximum (marked
with A), stimulate the emission by the same bunch in the vicinity of
the next maximum (marked with B). It is assumed that the length of the
bunch, $L_b$ , its velocity, $v$, and the undulator period $\lambda$,
satisfy the condition $(1-v/c)\lambda \ll L_b$, which means that the
photon slippage against the positron bunch during one undulator period
is much less than the latter.}
\label{Fig5.1}
\end{figure}

The mechanism of the emission stimulation is illustrated in figure
\ref{Fig5.1}. The photons emitted in the nearly forward direction at some
maximum/minimum point of the trajectory by a group of particles of the
bunch stimulate the emission of the same photons by another
(succeding) group of particles of the same bunch when it reaches next
maximum/minimum.

This scheme implies that the stimulation is due to
the motion of the same bunch along the trajectory (\ref{AW_1}).
Therefore, the bunch velocity $v$, its length $L_{\rm b}$ and
the length of the crystal $L$ must satisfy  the condition
\begin{equation}
{L \over c}  \geq {L-L_{\rm b} \over v},
\qquad {\rm or}\qquad
{v \over c}  \geq 1- {L_{\rm b} \over L}
\label{stim1}
\end{equation}
which means that the electromagnetic wave does not slip against the
bunch of particles by more than $L_{\rm b}$ on the lenghtscale 
$L \gg L_{\rm b}$
and, thus, the amplification occurs during each of the undulator
periods of total number $N_{\rm u}=L/\lambda$.

In the theory of free-electron lasers (FEL) this principle is called
``Self-Amplified Spontaneous Emission (SASE)''
\cite{Pellegrini1984, Pellegrini1985} and usually is referred to the
FEL operation in the high gain (or collective instability) regime
\cite{Baier80, Murphy85, Kim93, Saldin95}.

An accurate treatment of the evolution of the photon pulse in the
SASE mode implies that the space charge fields, energy spread of
the particles in the beam and diffraction effects are taken into
account \cite{Saldin95}. In the case when the beam undergoes
channeling the picture must be additionally complicated by
considering the dechanneling dynamics \cite{Biryukov} as well as
the photon attenuation \cite{Hubbel}.

We do not aim to fulfil this program in the present paper.
Instead, for the suggested undulator, we shall present
an estimate of the gain exploiting the general relationship
between gain and spontaneous spectrum of undulator radiation
\cite{Kniga,Lebedev78}. Thus, we disregard
the dynamics of the particles in the undulator and consider the
case of a monoenergetic beam of independent particles.
Additionally we assume that there is no beam divergence.
The effects of dechanneling and photon attenuation are
discussed qualitatively on a phenomenological level.

%%%%%%%%%%%%%%%%%%%%%%%%%%%%%%%%%%%
\subsection{General expression for the gain}

It is our aim to establish the range of parameters
(which are: the AW amplitude and frequency (wavelength), the type of
the particle and its energy, and the type of a crystal) within which
there is a principle possibility to consider the SASE radiation
by means of the acoustically bent crystal.

It had been established long ago \cite{Madey71} that in a
free-electron device the stimulation of emission occurs if only
the photon frequency is equal to that of the spontaneous undulator
radiation, $\omega_K$ (see (\ref{quasi_18})).

Let us estimate the gain factor, $g_{K}$, which defines the increase
(i.e. the case $g_K>0$) or the decrease (if $g_K<0$) per 1 cm in the
total number, $N_K$, of the emitted photons at a frequency
$\omega_K$
\begin{equation}
\d N_K = g_K\, N_K\,\d z,
\label{gain}
\end{equation}
 due to stimulated emission (absorption) by the particles in the
beam.
We shall do this within the frame-work of
low-gain approximation (see eg. \cite{Saldin95}).
Also we restrict our consideration to the case of the stimulated
emission of soft photons,
$\hbar\omega \ll \varepsilon$, and, hence, $\omega^{\prime}=\omega$
(see (\ref{wkb3})).

The general expression for the quantity $g_{K}$ is
\begin{equation}
g_{K} = n\,
\left[
\sigma_{\rm e}( \varepsilon, \varepsilon-\hbar\omega_{K})
-
\sigma_{\rm a}( \varepsilon, \varepsilon+\hbar\omega_{K})
\right]
\label{stim2}
\end{equation}
Here $\sigma_{\rm e}( \varepsilon, \varepsilon-\hbar\omega_{K})$
and $\sigma_{\rm a}( \varepsilon, \varepsilon+\hbar\omega_{K})$
are the cross sections of, correspondingly,
the spontaneous emission and absorption of the
$\omega_{K}$-photon by a particle of the beam, $n$ stands for
the volume density of the beam particles.
By using the known relations between the cross section of the
photon emission/absorption and the spectral-angular intensity of
the emitted radiation \cite{Land4}, and taking into account
the relation $\hbar\omega \ll \varepsilon$, one gets the following
expression for the gain, corresponding to the increase in the
number of photons of the frequency  within the interval
$\omega_K \pm \Delta \omega$ emitted in the cone  $\Delta\Omega_K$
(see below) with the axis along the the $z$-direction:
\begin{equation}
g_K =
-(2\pi)^3\, {c^2 \over \omega_K^2}\, n\,
{\d \over \d\varepsilon }
\left[
{\d E \over \d \omega_K\, \d \Omega }\right]_{\theta=0}
\, \Delta \omega \, \Delta\Omega_K
\label{stim3}
\end{equation}
The width $\Delta \omega$ one gets from (\ref{quasi_18}) by
substituting $\omega^{\prime}$ with $\omega$,
$\Delta \omega = (2/N_{\rm u})\left(4\gamma^2\omega_0/(2+p^2)\right)$.

Making use of (\ref{quasi_18}) (see also (\ref{quasi_15}))
one gets the estimation of the quantity  $\Delta\Omega_K$:
\begin{equation}
\Delta \Omega_K = {2\pi (1 + 2\, p^{-2}) \over K}\, \theta_0^2
\label{stim4}
\end{equation}
It defines the maximal solid angle along the undulator axis
inside which the $K$th harmonic is well resolved.
The quantity $\theta_0^2$ is given in (\ref{quasi_17}).
Expression (\ref{stim4}) is valid for both $p^2\gg 1$ and  $p^2\ll 1$.
In the latter case one may set
$(1 + 2\, p^{-2})\theta_0^2 \approx \gamma^{-2}$.

For the energy emitted in the forward direction
($\theta=0$) the formulas (\ref{Ang1}) and (\ref{Ang3})
yield:
\numparts
\begin{eqnarray}
\left[
{\d E \over \d \omega_1\, \d \Omega }
\right]_{\theta=0} & = &
{q^2 \over c} \, N_{\rm u}^2 \gamma^2 p^2  \, D(\eta),
\qquad {\rm for}\ \ p^2 \ll 1
\label{stim5a} \\
\left[
{\d E \over \d \omega_K\, \d \Omega }
\right]_{\theta=0} & = &
{q^2 \over c} \, N_{\rm u}^2\, {\gamma^2 \over p^2}\,
(2K)^{2/3} \, D(\eta)
\qquad {\rm for}\ \ p^2 \gg 1
\label{stim5b}
\end{eqnarray}
\endnumparts
Expression (\ref{stim5a}) is denoted for the fundamental harmonic,
$\omega_1=2\omega_0\gamma^2$ in accordance with (\ref{Ang3}).
In (\ref{stim5b}) we assumed that $K \ll K_{\rm max} \sim p^3$
and took into account that in this case the argument $\zeta$
(see (\ref{Ang2})) of the derivative of the Airy function
from (\ref{Ang1}) is small, so that
$16 {\rm Ai}^{\prime\, 2}(0) \approx 1$ \cite{Abramowitz}.

When carrying out the derivative of
$[\d E / \d \omega_1\, \d \Omega]_{\theta=0}$
with  respect to $\varepsilon$ the main contribution comes
from the term
$\d D(\eta)/ \d \varepsilon = (\d \eta /\d \varepsilon)\cdot
(\d D(\eta)/ \d \eta)$.
To obtain maximal positive gain it is necessary to choose that slope of
$D(\eta)$ where it is negative.
Then the extremal value of this derivative is equal to
\numparts
\begin{eqnarray}
{\d D(\eta)\over \d \varepsilon} =
 {\d \eta /\d \varepsilon} {\d D(\eta)\over \d \eta}
& \approx &
{-2 \over  \varepsilon} \cdot {N _{\rm u}\over 2}
\qquad {\rm for}\  p^2 \ll 1
\label{dDNa} \\
{\d D(\eta)\over \d \varepsilon} =
 {\d \eta /\d \varepsilon} {\d D(\eta)\over \d \eta}
& \approx &
{-4 K \over  \varepsilon p^2} \cdot {N_{\rm u} \over 2}
\qquad {\rm for}\  p^2 \gg 1
\label{dDNb}
\end{eqnarray}
\endnumparts
The first factors on the right-hand sides of (\ref{dDNa}) and (\ref{dDNb})
correspond to $(\d \eta /\d \varepsilon)$ and, as it is seen,
their magnitude depends on $p^2$.
The last factor $(N_{\rm u}/2)$ equals to the maximum value of
the derivative
$(\d D(\eta)/ \d \eta)$ with $\eta$ lying within the interval
$K\pm 1/N_{\rm u}$.
We concentrate our attention on these details in order to point out the
nature of the difference in the $p$-dependence of $g_K$ for the AW
undulator and that for the undulator based on magnetic field. As it
was noted above (see the paragraph after(\ref{Cond2})) in the latter
case the undulator parameter is independent of $\gamma$ resulting in
$\d \eta /\d \varepsilon = - 2K/\varepsilon$ for both
$p_{\rm B}^2 \ll 1$  and $p_{\rm B}^2 \gg 1$ cases
($K=1$ if $p_{\rm B}^2 \ll 1$).
In turn it leads to the proportionality of the gain to $p_{\rm B}^2$
regardless to the magnitude of the latter  \cite{Colson85, Coisson81}.
In our case the gain $g_K$, which is equal to
\numparts
\begin{eqnarray}
g_K\ (\, {\rm cm}^{-1}\, ) \approx
(2\pi)^3\ {q^2 \over m}\   r_{\rm cl}\,
{ N_{\rm u}^2\, \lambda \over  \gamma^3}\, n\ p^2
\qquad {\rm for}\ \ p^2 \ll 1
\label{stim6a} \\
g_K\ (\, {\rm cm}^{-1}\, ) \approx
{2 (2\pi)^3 \over (2K)^{4/3} }\
{q^2 \over m}\
 r_{\rm cl}\,
{ N_{\rm u}^2\, \lambda \over  \gamma^3}\, n
\qquad {\rm for}\ \ p^2 \gg 1,
\label{stim6b}
\end{eqnarray}
\endnumparts
is independent on $p$ if $p^2 \gg 1$.
In the above equations
$\lambda_{\rm c}=\hbar/m_{\rm e}c=3.9\cdot10^{-11}$ cm
is the Compton wavelength and
$r_{\rm cl}= {e^2 /m_{\rm e} c^2}=2.8\cdot10^{-13}$ cm
is the classical radius of the electron,
$q$ and $m$ are the charge and the mass of the projectile
measured in units of the elementary charge $e>0$ and electron mass
$m_{\rm e}$, respectively.
The quantity $\lambda$ is measured in cm and  $n$ in cm$^{-3}$.

Note the strong inverse dependence of $g_K$ on $\gamma$ which is due to
the radiative recoil.
The gain is proportional to the factor $N_{\rm u}^2$,
which reflects the coherence of radiation.
The proportionality of the gain to $\lambda$ means that the
increase in $\lambda$ leads to an enchancement of the radiation
intensity in the forward direction.

Expressions (\ref{stim6a}) and  (\ref{stim6b}) were obtained
by using the asymptotic formulae (\ref{Ang1}) and (\ref{Ang3})
valid for $p^2\gg 1$ and $p^2\ll 1$, respectively.
Extrapolating (\ref{stim6a}) and  (\ref{stim6b}) to the region
$p^2 \sim 1$ and considering in (\ref{stim6b}) the fundamental
harmonic, $\omega_1 \approx 4\gamma^2\omega_0/p^2$, one recognizes
that in the point $p=1$ (\ref{stim6b}) is within a factor
$2^{-1/3} = 0.79$ of (\ref{stim6a}). Such a discrepancy may be
ignored having in mind the approximate character of the initial
formulae (\ref{Ang1}) and (\ref{Ang3}).

Apart from the absence of the $p^2$ dependence in case $p^2 \ll 1$
(due to the reasons mentioned above) expressions (\ref{stim6a}) and
(\ref{stim6b}) have the same  functional form as those obtained for
the undulator based on the action of periodic magnetic field
\cite{Madey71, Hopf76, Coisson81, Colson85}.

%%%%%%%%%%%%%%%%%%%%%%%%%%%%%%%%%%%%%%%%
\subsection{Estimations of $g_K$ for the AW based undulator}

Expressions (\ref{stim6a}) and  (\ref{stim6b}) allow to make
quantitative estimates concerning the possibility of
stimulated emission of the high-energy photons.

Before carrying out this analysis let us note two main qualitative
features which distinguish the AW based undulator and that based on
the  action of the periodic magnetic field. In the latter case
both the particle and the photon beams propagate in vacuum.
Therefore, there are no strong mechanisms leading to the
decrease in the beam density.

In the scheme considered here both beams propagate in a crystal.
Thus, an adequate description of the stimulated emission with
necessity must account for perticle dechanneling and for photon
attenuation.

Random scattering of the channeling particle by the electrons and nuclei
of the crystal leads to the gradual increase of the particle energy
associated with the transverse oscillations in the channel. As a
result, the transverse energy at some distance, $L_d$, from the
entrance point exceeds the depth of the interplanar potential well,
and the particle leaves the channel. The quantity $L_d$ is called
the dechanneling length \cite{Biryukov}.

The dynamics of the photon flux propagation, apart from the
stimulation effect, is strongly influenced by a variety of processes
occuring in a crystal.
These are the atomic and the nuclear photoeffects, the coherent and the
incoherent scattering with electrons and nuclei, the
electron-positron pair production (in the case of high energy
photons).
All these processes lead to the decrease in the density of the photon
flux as it propagates through the crystal.

The rigorous treatment of the stimulated emission in the system
``channelng beam + photon flux'', with both the dechanneling effect and
the photon attenuation accounted for,
must include solving the coupled equations of the Fokker-Planck
form for the particle beam  \cite{Biryukov,Kumakhov2}
and for the photon beam \cite{Ternov2}.
This will be done at a later time.
In this paper we present the quantitative analysis of the stimulated
emission by a beam channeling in the AW bent crystal by using
phenomenological arguments when treating the dechanneling and
the photon attenuation.

Let us first estimate the role of the {\it dechanneling effect}.

\noindent
To simplify the expressions, we will only consider the stimulated
emission of the fundamental harmonic
\begin{equation}
\omega = {4\, \omega_0 \gamma^2 \over 2+p^2},
\label{omega}
\end{equation}
and omit the index ``1'' in the notations $\omega_1$ and $g_1$.
Also, we restrict the treatment to the case of a positron
channeling, $M=Z=1$.

By expressing the total number of the undulator periods $N_{\rm u}$
through the undulator length, $N_{\rm u}= L/\lambda$, and using
(\ref{stim6a}) and  (\ref{stim6b}), one gets the following estimate
for  $g$:
\begin{equation}
g  \equiv g(L)  \approx
 \Theta(p)\, (2\pi)^2\  r_{\rm cl}\, n\,
{k\, L^2 \over  \gamma^3}
\label{stim7}
\end{equation}
where the quantity $\Theta(p)$ is defined as
\begin{equation}
 \Theta(p) = \cases{1 & for $p^2 \geq 1$ \\
p^2 & for $p^2 < 1$ \\}
\label{stim7a}
\end{equation}
In (\ref{stim7a}) we have omitted the factor $2^{-1/3} = 0.79 \approx 1$
in the case $p^2 \geq 1$.
In (\ref{stim7}) the argument $L$ in the notation $g(L)$ stresses the
dependence of the gain factor on the crystal length.

Provided the dechanneling is neglected, one may unrestrictedly increase
$g$ by considering larger $L$-values. In reality,
the volume density of the channeling particles decreases with the
penetration distance, $z$, and, roughly, satisfies the
exponential decay law \cite{Biryukov}
\begin{equation}
n(z) = n_0\, \exp\left(-  z/ L_d(\gamma, R)\right)
\label{stim8}
\end{equation}
where $n_0$ is the volume density at the entrance,
$L_d(\gamma, R)$ is the dechanneling length. For given crystal
and channel $L_d(\gamma, R)$ depends on a positron energy
(relativistic factor) and on the curvature radius $R$.

As argued in \cite{Biryukov}, the decrease of $n$ with $z$ acquires
the exponential form (\ref{stim8}) at a considerable penetration
depth, $z\geq L_d(\gamma, R)$. If the crystal length is less than
$L_d(\gamma, R)$ then the dependence $n(z)$ is strongly influenced
by the initial conditions (the incident angle, the beam divergency)
of the beam impacting on the crystal.
Therefore, for estimation purposes, we set the length of the crystal
equal to the dechanneling length, $L=L_d(\gamma, R)$,
and assume that the beam density for $z<L$ is equal to its initial
value $n_0$.

For a bent crystal with a constant curvature radius $R$ the dechanneling
length $L_d(\gamma, R)$ satisfies the relation \cite{Biryukov}
\begin{equation}
L_d(\gamma, R)
= (1 - x^2)^2 L_d(\gamma, \infty)
\equiv (1 - {R_c \over R})^2 L_d(\gamma, \infty)
\label{stim9}
\end{equation}
where $L_d(\gamma, \infty)$ is the dechanneling length of a positron
of the same energy in a straight channel ($R=\infty$) and
\begin{equation}
R_c = {\varepsilon \over e\, U_{\rm max}^{\prime} }
\label{stim10}
\end{equation}
is the critical (minimal) radius consistent with the channeling
condition in a bent crystal,
``the centrifugal force $<$ the interplanar force'' (\ref{1}).

In an acoustically bent crystal the curvature $1/R(z)$ is not constant
(see (\ref{AW_3})). Therefore it is natural to consider the mean
curvature $1/\bar{R}$ which is obtained by averaging
$1/|R(z)|$  over the undulator period
\begin{equation}
{1 \over \bar{R}} = {1 \over \lambda} \int_0^{\lambda}
k^2 a |\sin kz| \d z = {2  \over \pi}\, {1 \over R_{\rm min}}
\label{stim10a}
\end{equation}
with $R_{\rm min}$ from (\ref{AW_7}).

\begin{figure}
\hspace{3cm}\epsfig{file=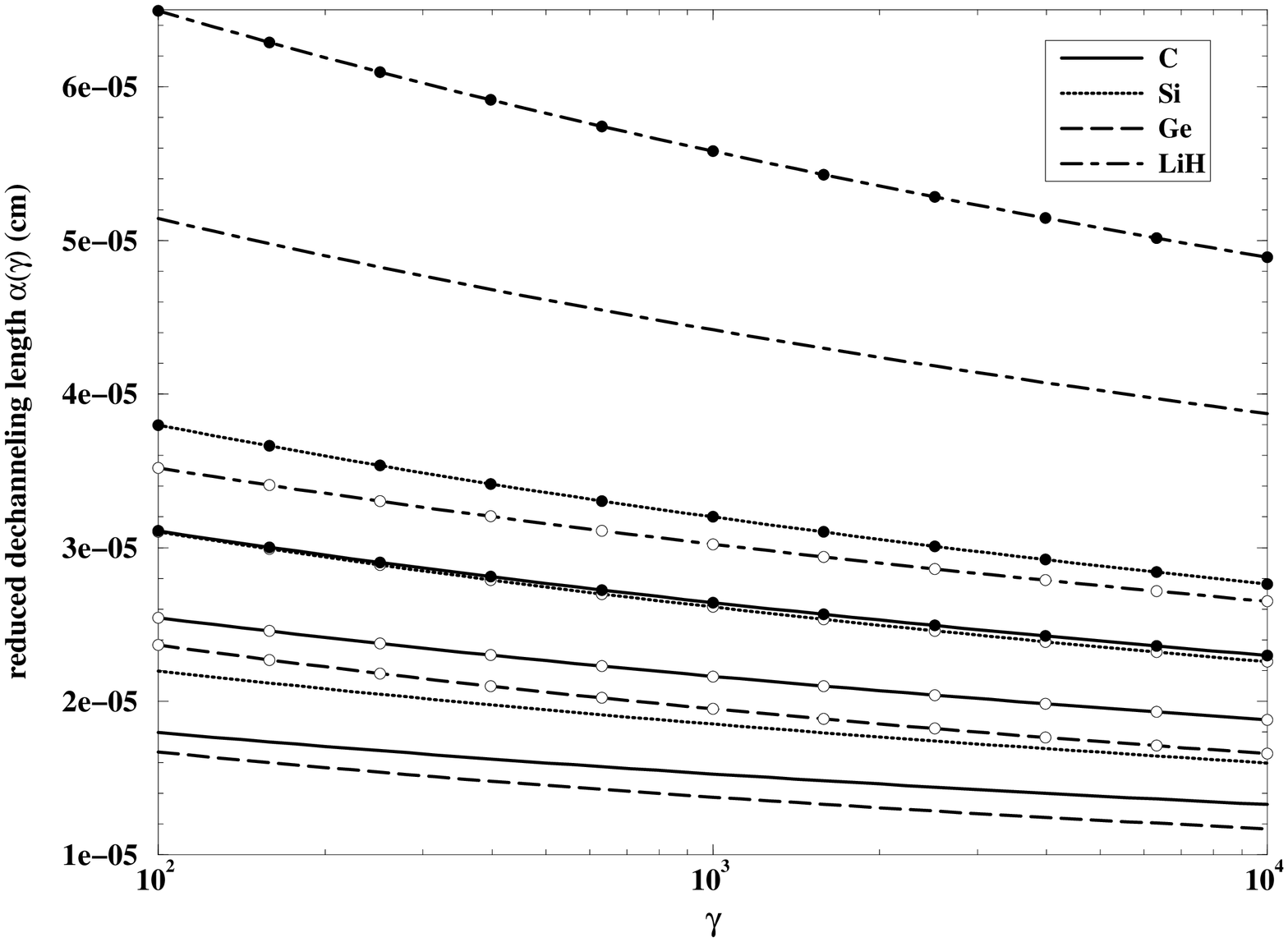,height=11cm,angle=270}
\caption{Dependence of a straight channel reduced dechanneling
length, defined as $\alpha \equiv L_d(\gamma,\infty)/\gamma$ (see also
(\ref{stim11})), on the relativistic factor $\gamma$ of a
positron channelled in (100), (110) and (111) planar channels of $C$,
$Si$, $W$ and $LiH$ crystals as indicated.  The full-circle curves
correspond to (111) channels, the open-circle curves to (110), the
curves without circles stand for (100) channels.  The analogue curves
for the $Ge$ crystal (not presented in the figure) almost coincide
with those for the $C$.}
\label{Fig5.2}
\end{figure}

To proceed further we make use of the fact, that for a given crystal
and crystallographic plane the reduced dechanneling length,
$\alpha\equiv L_d(\gamma, \infty)/\gamma$, depends on
$\gamma$ weakly.
Its explicit expression, calculated by using the Lindhard approximation
for the  potential of a planar channel, reads \cite{Biryukov}
\begin{equation}
\alpha\,  (\gamma)  = {256 \over 9\pi^2}
\,{ a_{\rm TF} \over r_{\rm cl} }\,
{ d \over \ln\left(2\varepsilon /I\right) - 1 }
\label{stim11}
\end{equation}
Here $a_{\rm TF}=0.8853 Z_c^{-1/3} a_0$ and $I = 16 Z_c^{0.9}$ eV are
the Thomas-Fermi atomic radius and ionization potential,
respectively. $Z_c$ is the atomic number of the crystal atoms, and
$a_0$ is the Bohr radius.  The dependences $\alpha(\gamma)$ for
various planar channels are illustrated in Figure \ref{Fig5.2}.  For
the $LiH$ crystal we used the average atomic number $Z_c=2$, and the
interplanar spacings $d$ for its (100), (110) and (111) channels were
deduced from \cite{LiH}.

By introducing (\ref{stim9}) and (\ref{stim11}) in (\ref{stim7}),
and expressing $k$ through $x$, $R_c$ and $a$
as $k=x\sqrt{\pi/2 R_c a}$, one gets
\begin{equation}
g(L_d)  =
(2\pi)^2\, \sqrt{{\pi \over 2}}\,  \Theta(p)\,
r_{\rm cl}\, n\, {\alpha^2 \over \gamma}\,
{\sqrt{e  U_{\rm max}^{\prime} \over \varepsilon  a}}
\ x\cdot(1-x^2)^4
\label{stim12}
\end{equation}
This expression conveniently exposes the dependences of the
gain factor on the projectile energy, on the parameters
of a crystal, $ U_{\rm max}^{\prime}$ and $\alpha$, and on the
AW amplitude $a$. The factor $x\cdot(1-x^2)^4$ contains the
dependence of $g(L_d)$ on the average curvature radius of the channel
since $x^2$ is defined as $x^2 = R_c/\bar{R}$.

By analysing (\ref{stim12}) one obtains the ranges of parameters
inside which the process of the emission stimulation occurs
most efficiently.
The factor $ x\cdot(1-x^2)^4$ reaches its maximum value of $0.208$ at
\numparts
\begin{equation}
x_0^2 = {\varepsilon \over e U_{\rm max}^{\prime} \bar{R}}
= { 1 \over 9},
\label{stim13a}
\end{equation}
defining, thus, for given $\gamma$ and crystal,
{\it the optimal value} of a minimum curvature radius
of the channel bent by the AW (recall (\ref{AW_7})):
\begin{equation}
R^{opt}_{\rm min} =  {18 \over \pi}\,
 {e U_{\rm max}^{\prime} \over  \varepsilon}
\label{stim13aa}
\end{equation}
\endnumparts

Our analysis of an undulator based on an acoustically bent channel
assumes the validity of the condition (\ref{AW_0}).
It follows then, that the range of the $a$-values,
being subject to (\ref{stim13a})),
is restricted from below to some $a_{\rm min} \gg d$.

Another condition to be fulfilled follows from (\ref{Cond1}) and
implies that the total number of the undulator periods on the lengthscale
$\left[L_d(\gamma, \bar{R})\right]_{x=x_0} =
(1-x_0^2)^2 L_d(\gamma, \infty) =
0.79\cdot L_d(\gamma, \infty)$ is large.
From (\ref{stim9}),  (\ref{stim10}) and (\ref{stim13a}) it follows then
\begin{equation}
N_{\rm u} = {0.79 x_0 \over 2 \sqrt{2\pi}}\,
{ \alpha \gamma \over \sqrt{R_c a} } \ge N_{\rm min} \gg 1
\label{stim13b}
\end{equation}
This inequality defines the upper limit of the AW amplitude:
\begin{equation}
a \le a_{\rm max} = {\pi \over 2}\,
\left({0.79 x_0 \over 2\pi}\right)^2
{ \alpha^2 \gamma^2 \over R_c N^2_{\rm min} }
\label{stim13c}
\end{equation}

The inequalities $a_{\rm min} \gg d$ and (\ref{stim13c}) must be
satisfied simultaneously, leading to a natural condition
$a_{\rm min} \le a_{\rm max}$.
The latter results in  the relation
\begin{equation}
\alpha \gamma^2  \ge  {2 \over \pi}\,
 \left({0.79 x_0 \over 2\pi}\right)^2
{ m c^2  \over e U_{\rm max}^{\prime} }\,  N^2_{\rm min} a_{\rm min},
\label{stim13d}
\end{equation}
giving the lowest value $\gamma_{\rm min}$ (which is obtained from
\eref{stim13d}) with the equality sign). It allows to obtain
the non-zero mesh defined by (\ref{stim13a}).

In \tref{Table1} the $\gamma_{\rm min}$-values are presented for
a positron channeling near various crystallographic
planes in $C,\ Si,\ Ge,\ W$ and $LiH$ crystals. The data
correspond to $a_{\rm min}= 10 d$, $N_{\rm min}= 10$.
The $d$ values for $C,\ Si,\ Ge,\ W$ were taken from
\cite{Biryukov,Baier}, the $LiH$ data were adopted from \cite{LiH}.
The values of  $ e U_{\rm max}^{\prime}$ were calculated
by using the Moli\`ere approximation \cite{Gemmell} for an interplanar
potential at the temperature $T=150$ K.

\begin{table}
\caption{The values of $\gamma_{\rm min}$ (see (\ref{stim13d}))
calculated for a positron channelling in (100), (110) and (111) planar
channels in C, Si, Ge, W and LiH crystals.  The minimum AW amplitude,
$a_{\rm min}$, is set to $10 d$ ($d$ is the interplanar distance).  $e
U_{\rm max}$ stands for a maximum gradient of the interplanar
potential.  $L_d(\gamma_{\rm min},\bar{R})$ and $\bar{R}$ are,
respectively, the dechanneling length and the average bending radius,
both corresponding to $\gamma_{\rm min}$.}
\begin{indented}
\item[]\begin{tabular}{@{}rrrrrr}
\br
Channel &  $d$ & $e U_{\rm max}^{\prime}$ & $\gamma_{\rm min}$&
$\bar{R}$ & $L_d(\gamma_{\rm min},\bar{R})$  \\
       & \AA   &  GeV/cm  &  & mm  & mm \\
\br
   C   &      &  &  &  &  \\
(100)  & 0.89 & 4.8 & 1570 & 1.68 & 0.184 \\
(110)  & 1.26 & 7.4 &  630 & 0.43 & 0.111 \\
(111)  & 1.54 & 9.5 &  380 & 0.20 & 0.084 \\
\br
  Si   &      &  &  &   &  \\
(100)  & 1.36 & 4.8 & 1650 & 1.76 & 0.223 \\
(110)  & 1.92 & 7.1 &  700 & 0.51 & 0.148 \\
(111)  & 2.35 & 8.8 &  430 & 0.25 & 0.115 \\
\br
  Ge   &      &  &  &  &  \\
(100)  & 1.41 & 9.6 & 1200 & 0.63 & 0.142 \\
(110)  & 2.00 &14.0 &  510 & 0.18 & 0.091 \\
(111)  & 2.45 &17.3 &  310 & 0.09 & 0.071 \\
\br
   W   &  &  &  &  &  \\
(100)  & 1.58 &39.8 &  330 & 0.04 & 0.039 \\
(110)  & 2.45 &56.9 &  140 & 0.01 & 0.025 \\
\br
 LiH   &  &  &  &  &  \\
(100)  & 1.90 & 1.8 & 1000 & 3.02 & 0.360 \\
(110)  & 1.30 & 1.0 &  300 &15.50 & 0.675 \\
(111)  & 2.40 & 2.4 &  560 & 1.20 & 0.026 \\
\br
\end{tabular}
\end{indented}
\label{Table1}
\end{table}

Thus, for given $\gamma \ge \gamma_{\rm min}$ the interval
$[a_{\rm min}, a_{\rm max}]$ together with the values of
$\nu$ following from (\ref{stim13aa}) and \eref{AW_8},
represent the ranges of the AW amplitudes and frequencies
 whithin which the emission stimulation occurs with maximal efficiency
and, simultaneously, the physical conditions (\ref{AW_0}),
(\ref{AW_8}) and (\ref{Cond1}) are fulfilled.
For these $a$ and $\nu$ values the expression for $g$
is given by
\begin{equation}
\left[g(L_d)\right]_{x=x_0}  =
10.3\,   r_{\rm cl}\, n\,  {\Theta(p) \over \sqrt{a} }\,
{\alpha^2 \over \gamma \sqrt{R_c}}\,
\label{stim13e}
\end{equation}
Here the coefficient $10.3 = (2\pi)^2\sqrt{\pi/2}\cdot x_0 (1-x_0^2)^2$.

Let us now define the amplitude $a_0 \in [a_{\rm min}, a_{\rm max}]$,
for which the gain factor (\ref{stim13e}) reaches its maximum.
This quantity strongly depends on the factor
$\Theta(p)/\sqrt{a}$, with $\Theta(p)$ from (\ref{stim7a}):
\begin{equation}
{\Theta(p)\over \sqrt{a}} =
\cases{
1/\sqrt{a}
&  for $p^2 \geq 1$ \\
{\pi\, x_0^2 \gamma \over 2 R_c}\cdot \sqrt{a}
&  for $p^2 < 1$ \\}
\label{stim14}
\end{equation}
The latter relation is obtained by replacing $k$ in $p=\gamma k a$
with  $k=x_0\sqrt{\pi/2 R_c a}$ (see (\ref{stim13a}) together with
\eref{AW_8}).

It follows from  (\ref{stim14})  that the maximum point is at
$a_0 = a_{\rm min}$ if $p^2\geq 1$ for all $a$ from the interval
$[a_{\rm min}, a_{\rm max}]$, and $a_0 = a_{\rm max}$ if in the whole
interval of $a$ the inequality $p^2 < 1$ is valid.
A third option appears if the curves representing the dependencies
$\gamma k a = 1$ and $k^2 a = \pi/2\, x_0^2\, R_c^{-1}$ cross
in the point
\begin{equation}
\tilde{a}={2\, R_c \over \pi x_0^2 \gamma^2},
\label{stim17}
\end{equation}
which lies within the interval $[a_{\rm min}, a_{\rm max}]$.
In this case  $a_0 = \gamma^{-2} (2/\pi x_0^2)\, R_c$.

Therefore, the quantity
\begin{equation}
a_0 = \cases{
a_{\rm min}& if $\tilde{a} < a_{\rm min}$ \\
\tilde{a} & if $a_{\rm min} \leq  \tilde{a} \leq a_{\rm max}$
\label{stim15} \\
a_{\rm max}& if $a_{\rm max} < \tilde{a}$ \\}
\end{equation}
defines the AW amplitude at which the gain achieves its {\it maximum
value} which is:
\begin{equation}
 \left[g(L_d)\right]_{\rm max}=
4.3\,  r_{\rm cl}\, n\,
{\alpha^2 \over R_c}\times
\cases{
\sqrt{{\tilde{a} \over a_{\rm min}}}& if $\tilde{a} < a_{\rm min}$ \\
1 & if $a_{\rm min} \leq  \tilde{a} \leq a_{\rm max}$
\label{stim16} \\
\sqrt{{a_{\rm max} \over\tilde{a} }} & if $ a_{\rm max} < \tilde{a}$ \\}
\end{equation}

The first harmonic frequency corresponding to the
maximum gain (\ref{stim16}), is obtained from relation(\ref{omega}),
by expressing $\omega_0$ and $p^2$ through $a_0$ and $\tilde{a}$.
The result for the photon energy, measured in $MeV$ reads
\begin{equation}
\hbar \omega\  {\rm (MeV)} = 3.3\cdot 10^{-11}{ \gamma^2 \over
\sqrt{R_c a_0}}
\, {1 \over 2 + (a_0/\tilde{a})^2}
\label{stim18}
\end{equation}

\begin{figure}
\hspace{3cm}\epsfig{file=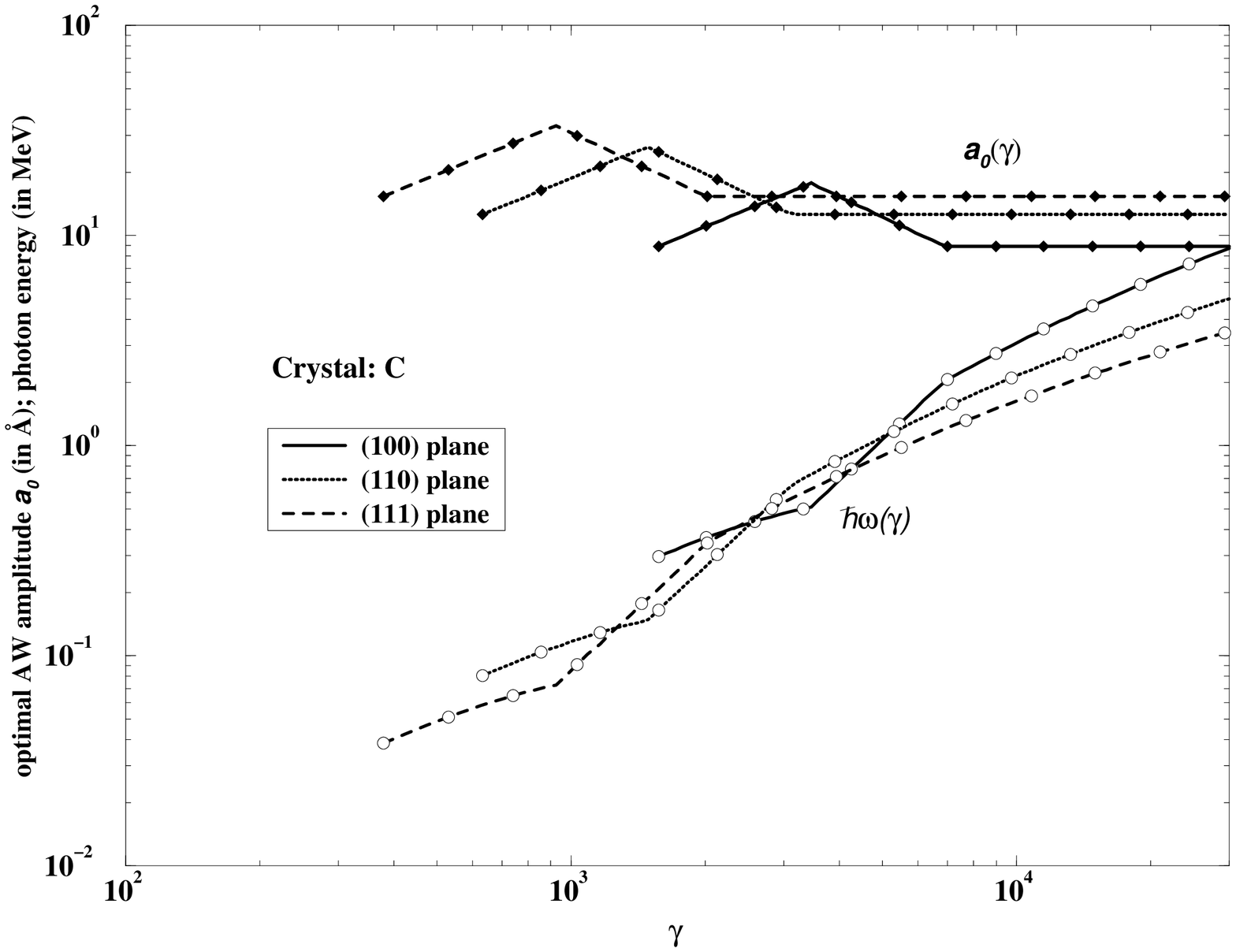,height=11cm,angle=270}
\caption{The magnitudes of the maximum gain
acoustic wave amplitude (in \AA), $a_0$ (see (\ref{stim15})), and the
first harmonic photon energy (in MeV), $\hbar\omega$ (see
(\ref{stim18})), versus the relativistic factor of a positron
channeling in diamond (100), (110) and (111) planar channels (as
indicated).}
\label{Fig5.3}
\end{figure}

Figure \ref{Fig5.3} illustrates the dependence of the optimal
AW amplitude $a_0$
(the diamonds-embedded curves) and of $\hbar \omega$
(the circles-embedded curves)
on $\gamma \ge \gamma_{\rm min}$ in the case of a positron channeling
near the (100), (110) and (111) planes in a diamond taken
at temperature $T=150$ K. The curves corresponds to the
particular values of $a_{\rm min} = 10\, d$ (see \tref{Table1})
and $N_{\rm min} = 10$.  For other crystals the dependencies are
basically the same, differing (not radically)
only in ranges of the parameters.

The curves in figure \ref{Fig5.3} clearly demonstrate, that it
is meaningful to consider the stimulated emission of high energy
photons,  $\hbar\omega \sim 0.01\dots 10$ MeV, during the channeling
of $\varepsilon \sim 0.1\cdots 10$ GeV positrons in a crystal
bent by a high-amplitude ($a\gg d$) transverse acoustic wave
propagating along the channel axis.

The above consideration focused on establishing the role of the
dechanneling effect in the process of
stimulated emission.
The main restriction due to the dechanneling concerns the length of
an undulator, which must be less than the dechanneling length,
$L\leq L_d$.
This, in turn, results in narrowing the ranges of $\gamma$
(eq. (\ref{stim13d})),  $a$ (eq. (\ref{stim13c}))  and $\nu$
for which stimulated emission can effectively occur.

As it will be demonstrated below, small values of the undulator
length, and, additionally, the attenuation of the photon flux in a crystal,
both require high positron beam densities necessary to achieve
a noticeable level of emission amplification. From the theory of
FEL it is known that a model which considers a positron bunch
as a group of independent particles (which has been utilized above)
is adequate provided the following inequality is valid
\cite{Louisell, Fedorov}:
\begin{equation}
\omega_p \, \tau < \gamma^{3/2}
\label{stim19}
\end{equation}
where $\omega_p=(4\pi\, n \e^2 /m)^{1/2}$ is the beam plasma frequency,
and $\tau=L/c$ is the time of flight of the bunch through the crystal.
The physical meaning of (\ref{stim19}) is that
$\tau$ is much smaller then the timescale needed to develop
collective instabilities within the bunch due to its interaction with the
undulator field and that of the radiation \cite{Fedorov}.
Therefore, all effects due to the space-charge dynamics during the
channeling can be neglected if the volume density of the beam
particles satisfies the condition
\begin{equation}
n < {1 \over 4\pi\, r_{\rm cl} } \,
{\gamma^3 \over L^2_d(\gamma, \bar{R}) }
\label{stim20}
\end{equation}
Recalling that $L_d(\gamma, \bar{R})=(1-x_0^2)^2\, \gamma\, \alpha$,
and inserting the values for the reduced dechanneling length $\alpha$
from figures 5.2, one obtains from (\ref{stim20}) an estimation
for $n$ in the range of $\gamma \approx 10^3\dots10^4$:
$n < 10^{22}\dots10^{23}$ cm$^{-3}$.

\begin{figure}
\hspace{3cm}\epsfig{file=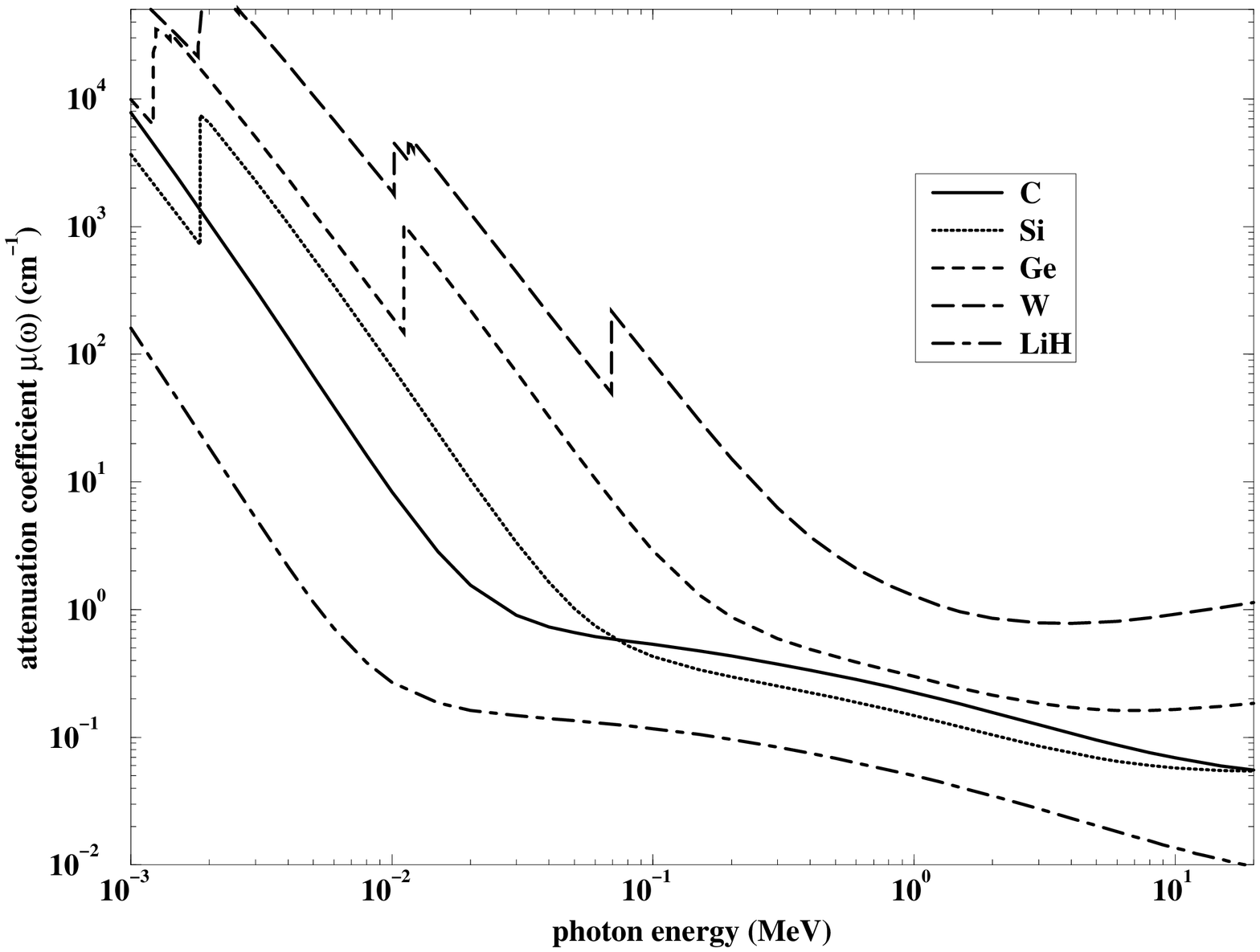,height=11cm,angle=270}
\caption{The mass attenuation coefficient, $\mu$, in dependence on
photon energy for various crystals, as indicated.
The data are taken from \cite{Hubbel}.}
\label{Fig5.4}
\end{figure}

Next we consider {\it photon attenuation}. Its influence can be
accounted for in a rather simple way by suggesting that the increase
per 1 cm in the total number of the emitted photons is given by
\begin{equation}
\d N = (g - \mu)\, N\,\d z
\label{stim21}
\end{equation}
rather than by (\ref{gain}).  Here the quantity $\mu\equiv
\mu(\omega)$ stands for the mass attenuation coefficient.  These mass
attenuation coefficients are tabulated for all elements and for a wide
range of photon frequencies.  For $\hbar\omega > 1$ KeV and up to 100
MeV the $\mu$-values can be found in ref. \cite{Hubbel}. Figure
\ref{Fig5.4} shows the $\mu(\omega)$ dependences for various crystals.
The step-like behaviour of the curves for $Si$, $Ge$ and $W$ is due to
the discontinuity of the atomic photoeffect cross section in the
vicinity of inner-shell thresholds.  For photons of energy less than 1
KeV the main contribution to the photon attenuation arises from the
atomic photoeffect process, and the $\mu(\omega)$ dependence is easily
deduced from ref.
\cite{Henke}.

To provide an increase in the radiation intensity the gain
must satisfy the condition
\begin{equation}
g > \mu(\omega),
\label{stim22}
\end{equation}
which leads to further narrowing of the interval of photon energies.
The large magnitude of $\mu(\omega)$ for $\hbar\omega < 10$ KeV (see
figure \ref{Fig5.4})
makes it almost impossible to obtain stimulated emission in this
$\omega$-region even for crystals with small average atomic number.

To draw a conclusion on what type of a crystal and a plane is
more suitable from the viewpoint of its efficiency to amplify
the spontaneous emission formed in an acoustically based undulator
let us make qualitative estimates of the density of a positron
beam needed to achieve the magnitude of 1 for the total gain.

The {\it total gain}, $G$, which represents by itself the total
increase in the number of photons due to the stimulated emission
on the lengthscale $L\equiv\left[L_d(\gamma, \bar{R})\right]_{x=x_0}$,
equals
\begin{equation}
G = (g(L) - \mu(\omega))\, L
\label{stim23}
\end{equation}
Such a definition of the total gain is valid in the low-gain
limit, $G\ll 1$ (see e/g. \cite{Baier, Fedorov}). Nevertheless, for
estimation purposes, one may use (\ref{stim23}) in the region
$G \leq 1$.

Putting $G=1$ and representing the quantity $g(L)$
in the form $g(L) = \tilde{g}(L)\cdot n$ (the explicit expression for
$\tilde{g}(L)$ is clear from  (\ref{stim13e})), we reach the
following estimate for the volume density of beam particles:
\begin{equation}
n  = {1 +  \mu(\omega)\, L \over  \tilde{g}(L)\, L }
\label{stim24}
\end{equation}

The dependences $n$ versus $\omega$ for various energies of a positron
beam and for various types of planar channels (as indicated) are
presented in figures \ref{Fig5.5}.  For each channel and for
$\varepsilon \ge \varepsilon_{\rm min} = m c^2\gamma_{\rm min}$ (the
definition of $\gamma_{\rm min}$ is given by (\ref{stim13d}) and its
particular values are presented in \tref{Table1}) the quantity
$\tilde{g}(L)$ can be calculated from (\ref{stim13e})), and the
corresponding $\omega$ range was calculated by using the relation
\begin{equation}
\hbar \omega\  {\rm (MeV)} = 3.3\cdot 10^{-11}{ \gamma^2 \over
\sqrt{R_c a}}
\, {1 \over 2 + (a/\tilde{a})^2}
\label{stim25}
\end{equation}
where $\tilde{a}$ is defined in (\ref{stim17}).
In (\ref{stim25}) it is assumed that the AW amplitude $a$ varies
within the interval $a=[a_{\rm min}, a_{\rm max}]$, and both $a$ and the
AW frequency $\nu$ are subject to (\ref{stim13aa}).
The minimum value of the AW amplitude was chosen as $a_{\rm min}= 10\, d$,
the quantity $a_{\rm max}$ was calculated from (\ref{stim13c}) with
the munimum number of the undulator periods $N_{\rm min}= 10$.
The magnitudes of the mass attenuation coefficients, $\mu(\omega)$,
were obtained by interpolating the data from \cite{Hubbel}.

\begin{figure}
\hspace{3cm}\epsfig{file=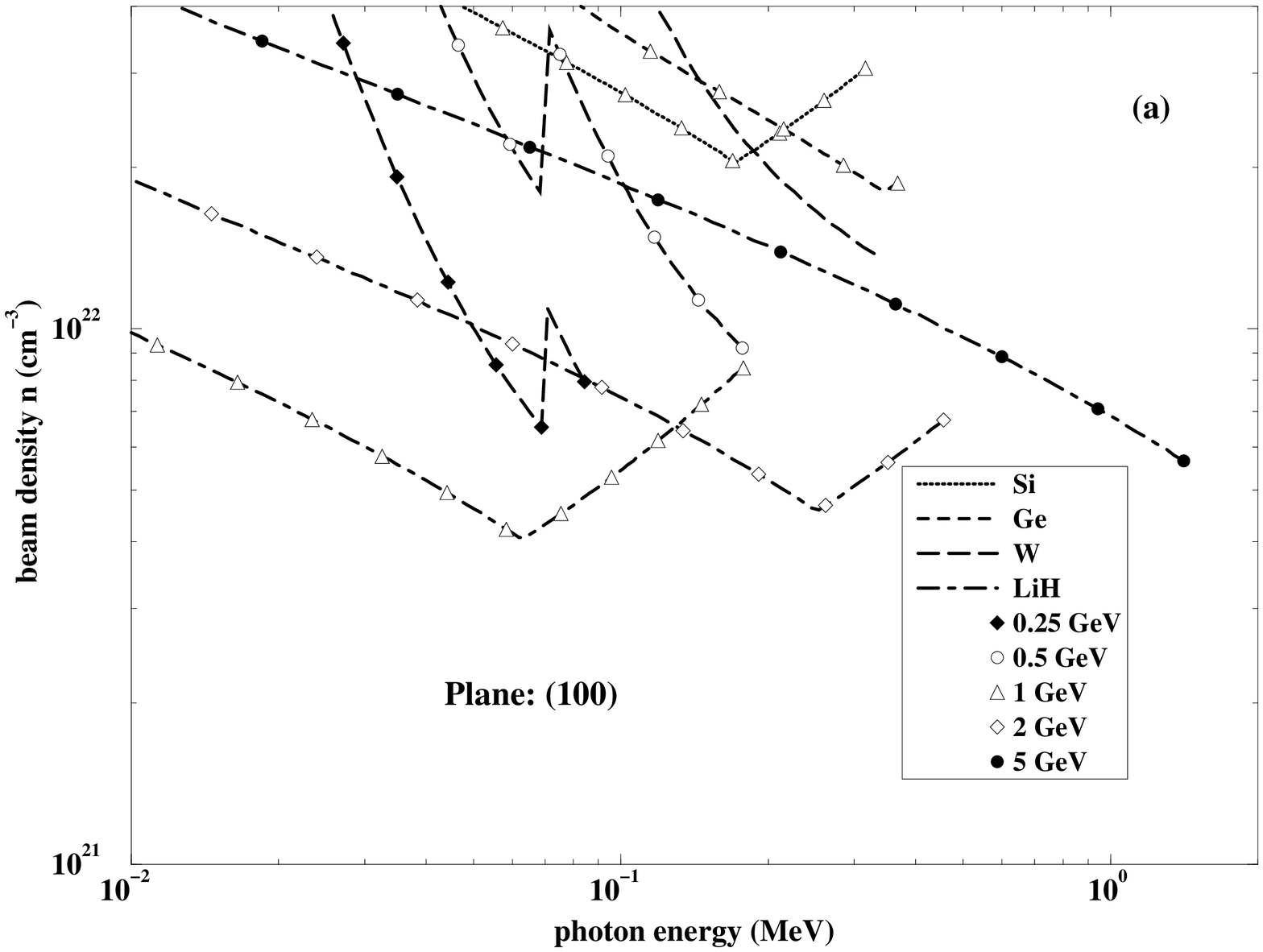,height=11cm,angle=270}
\caption{The volume density needed to achieve the total gain
$G=1$ for a positron beam (for various energies
$\varepsilon$ as indicated)
channeling along
{\bf (a)} (100) plane,
{\bf (b)} (110) plane,
{\bf (c)} (111) plane,
in various crystals as indicated.
The curves correspond to (\ref{stim24}).}
\label{Fig5.5}
\end{figure}

\setcounter{figure}{13}
\begin{figure}
\hspace{3cm}\epsfig{file=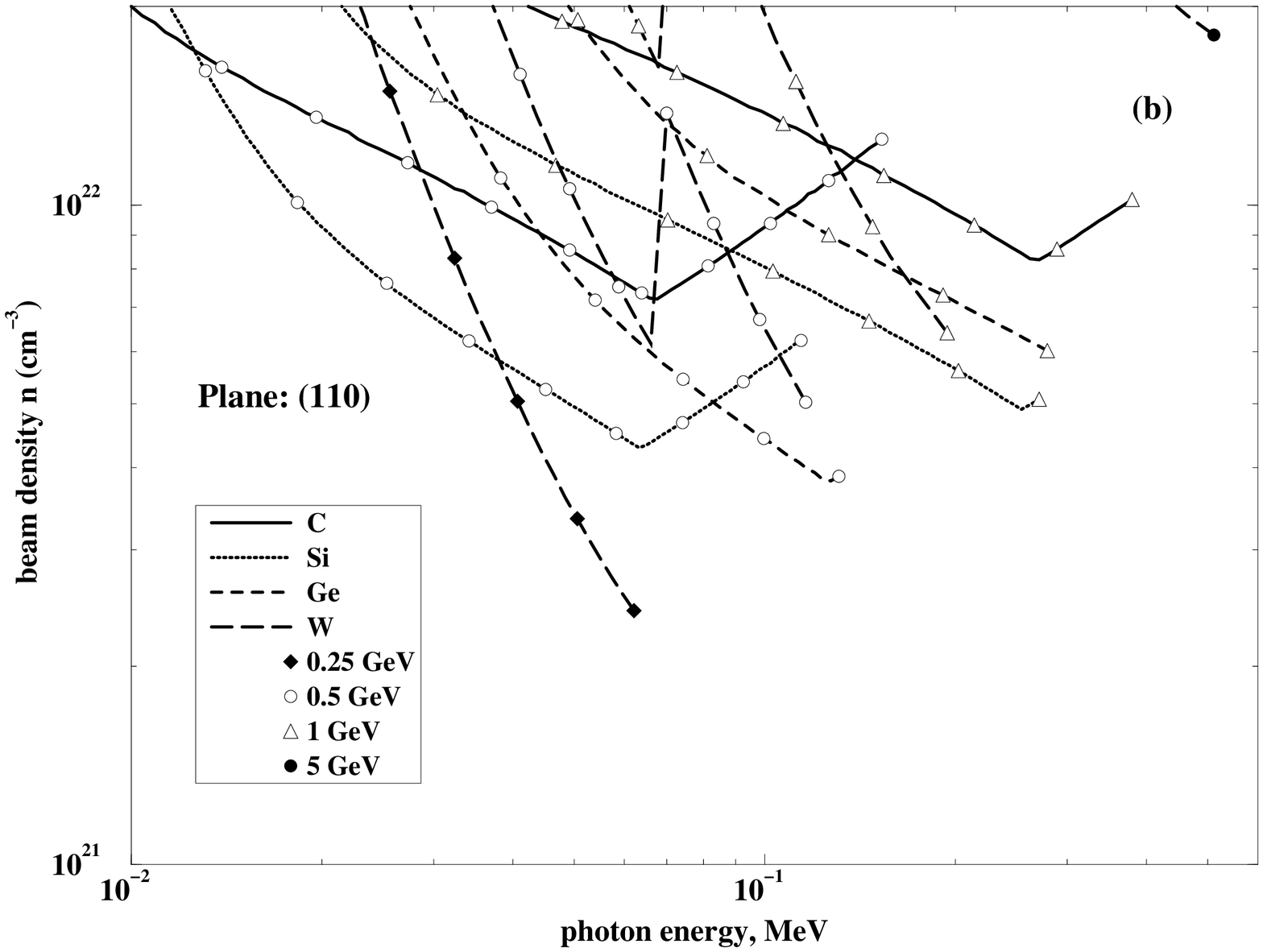,height=11cm,angle=270}
\caption{(\emph{continued})}
\end{figure}

\setcounter{figure}{13}
\begin{figure}
\hspace{3cm}\epsfig{file=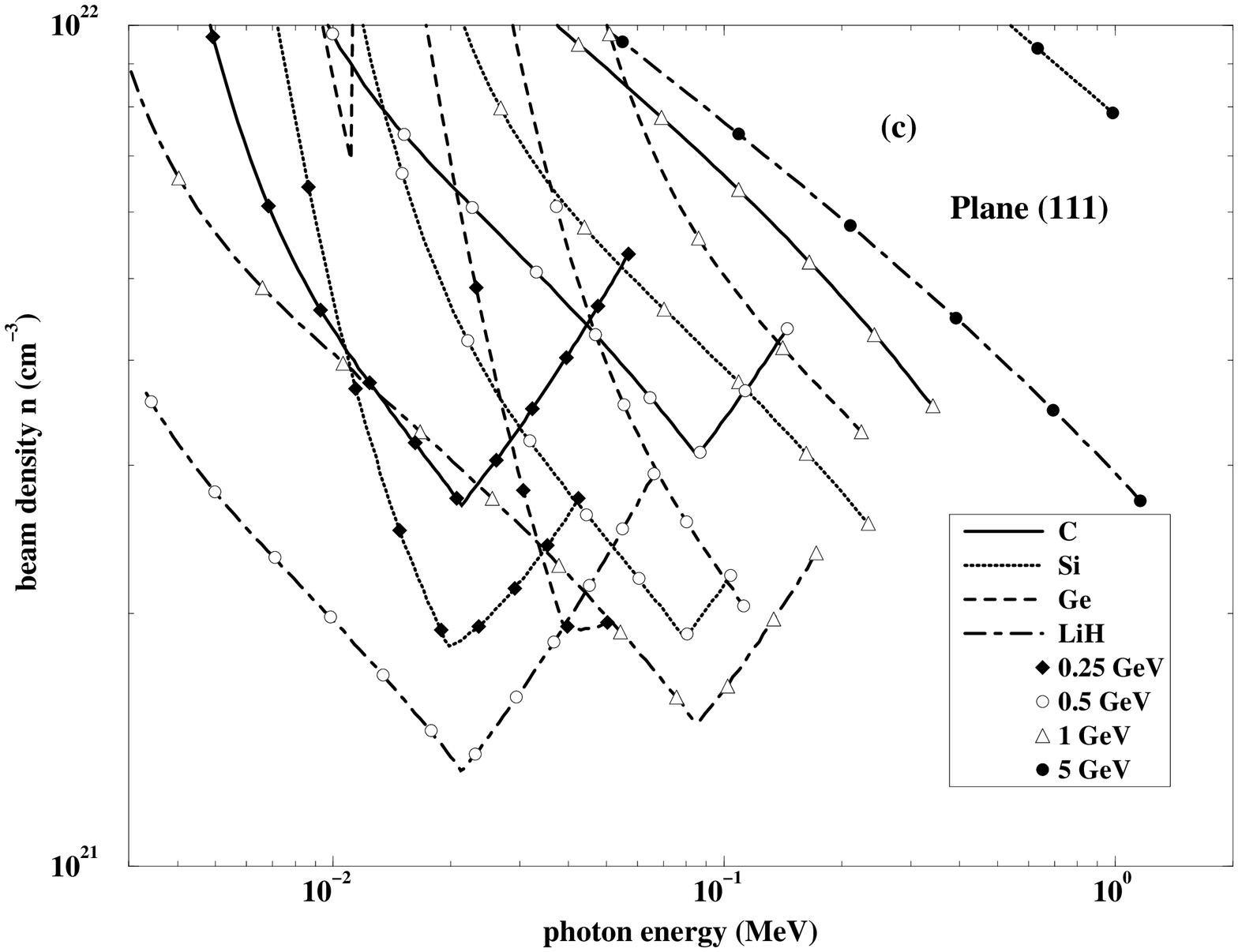,height=11cm,angle=270}
\caption{(\emph{continued})}
\end{figure}

For each curve in figures \ref{Fig5.5} the minimum
value of the beam density is achieved at $a=a_0$ (see (\ref{stim15})) with
the corresponding magnitude of the photon energy given by  (\ref{stim18}).
The only exception is the $n(\omega)$ dependence for a $250$ MeV
positron beam channeling along the (100) plane in $W$
(see \fref{Fig5.5}(a)).
In this case, due to the irregularity in the behaviour
of $\mu(\omega)$ in the vicinity of the $1s$ atomic threshold
($I_{1s} = 69.5$ keV for $W$), -- see figure \ref{Fig5.4},
-- the minimum of the beam density is achieved at
$\hbar\omega \approx 70$ keV.

\Fref{Fig5.6} illustrates the extent to which the photon attenuation
influences the minimum value of the beam density.  For each crystal
(as indicated) the dependence $n(\omega)$ from \eref{stim24} is
represented by the curves without full circles.  The curves embedded
with full circles correspond to the function
$\tilde{n}(\omega)=\left[n(\omega)\right]_{\mu(\omega)\equiv 0}$,
giving thus the minimum values of the volume density which is needed
to achieve the total gain $G=1$ in the case when the photon
attenuation is totally disregarded.  It is seen that for $\hbar\omega
\geq 1$ MeV the role of the attenuation is negligibly small for all
crystals, whereas for less energetic photons it leads to a prominent
increase in $n$.

\begin{figure}
\hspace{3cm}\epsfig{file=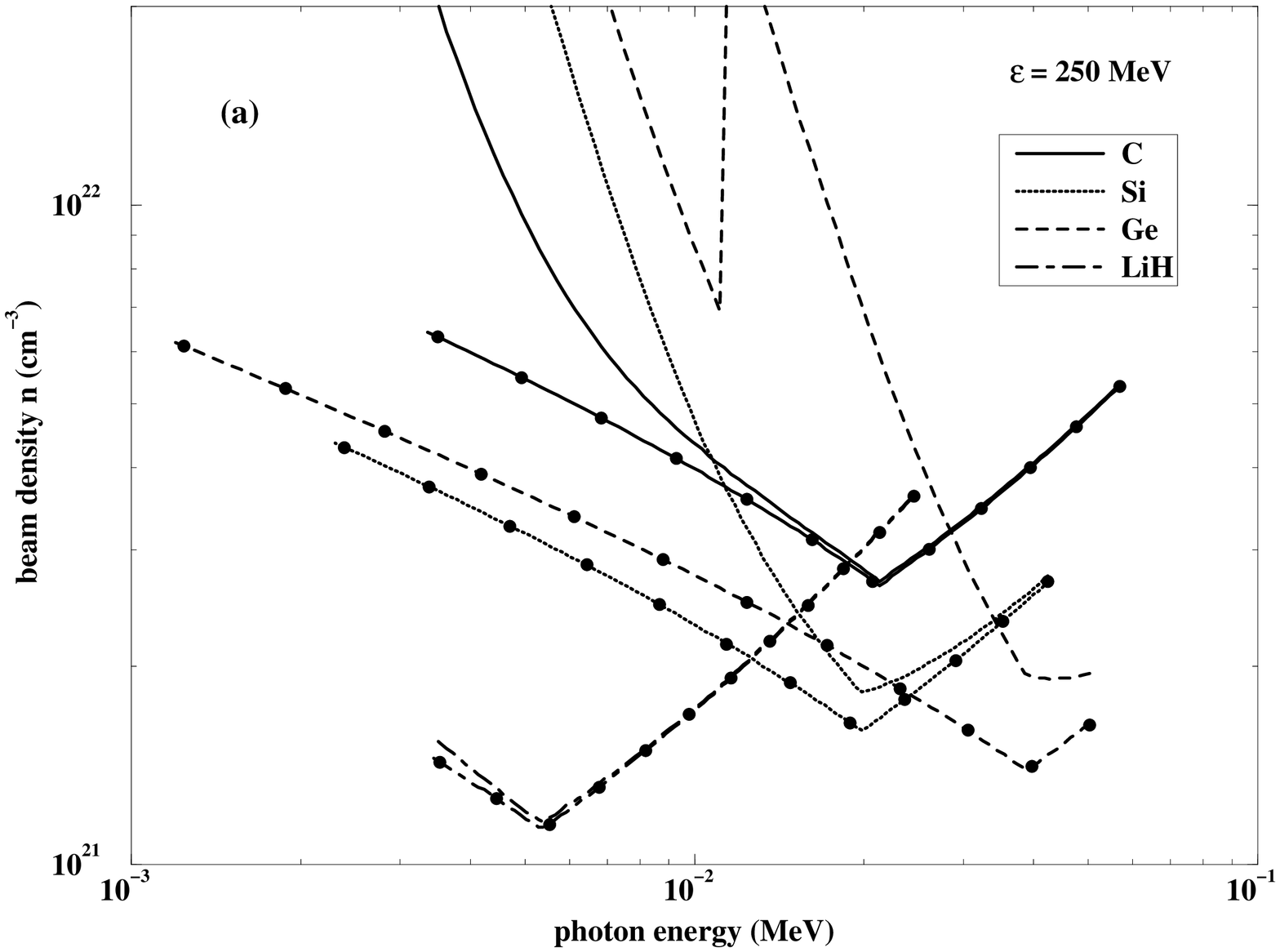,height=11cm,angle=270}
\caption{The volume density needed to achieve the total gain
$G=1$ for a 500 MeV positron beam channeling near the (111) planar
channels in various crystals as indicated.  The curves without full
circles corresponds to (\ref{stim24}).  The curves with full circles
represent the (\ref{stim24}) dependences with $\mu(\omega)\equiv 0$
(i.e. no photon attenuation taken into account).}
\label{Fig5.6}
\end{figure}

\setcounter{figure}{13}
\begin{figure}
\hspace{3cm}\epsfig{file=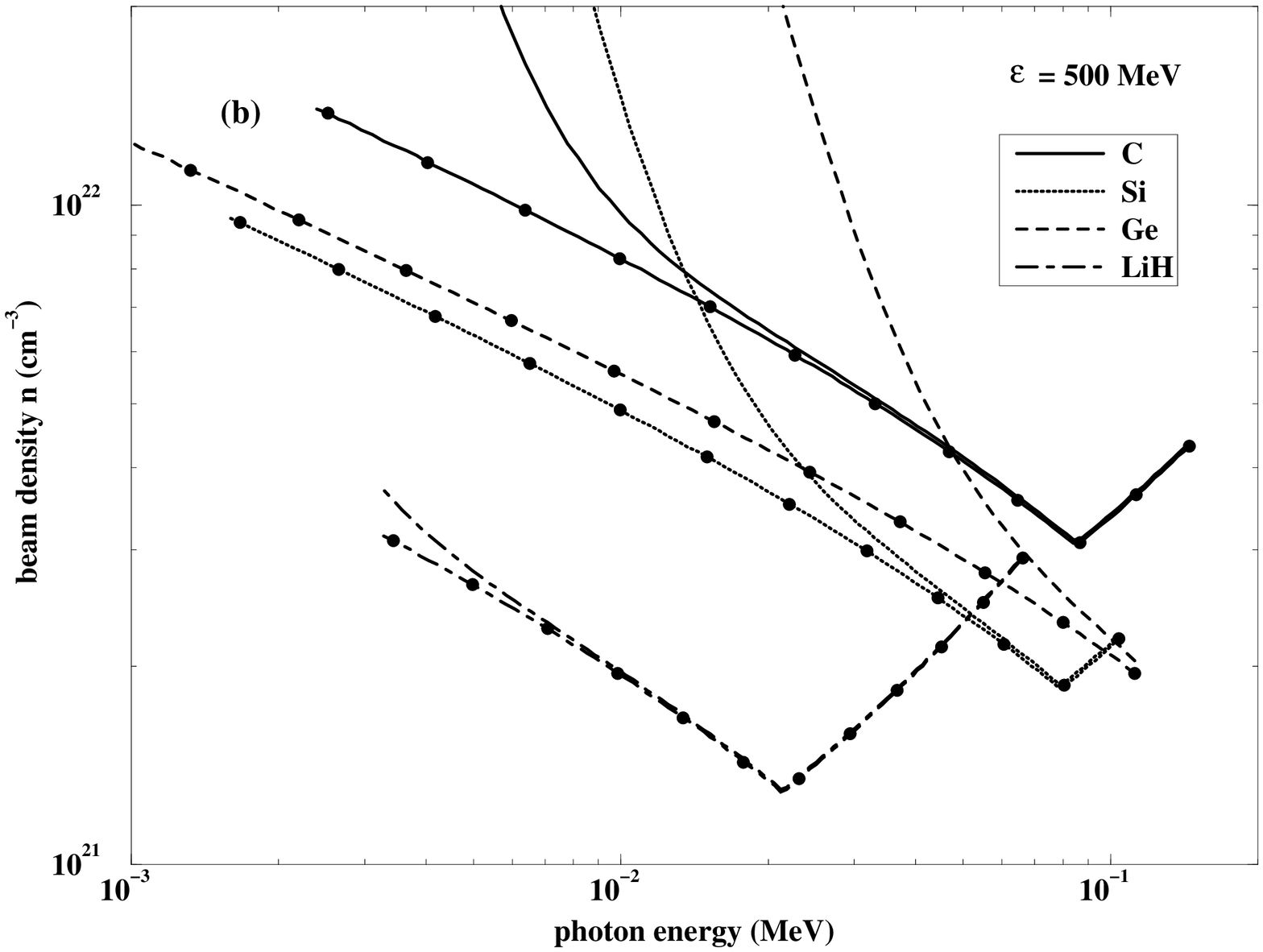,height=11cm,angle=270}
\caption{(\emph{continued})}
\end{figure}

Figures \ref{Fig5.5} and \ref{Fig5.6} show that to achieve a
noticeable level of the amplification of the spontaneous radiation
formed in the acoustically based undulator large magnitudes of a
positron beam volume density, $n \sim 10^{21}\,-\,10^{22}$, are
needed.  This is mainly due to the dechanneling effect which imposes
strong restrictions on the undulator length.

These values of $n$, being high enough, are, nevertheless comparable
with those planned to be achieved within the TESLA project
\cite{TESLA}.

Therefore, we believe that the scheme described above may be
considered as a new feasible source for stimulated emission within the
photon energies range from tens of KeV up to MeV by means of the
positron beam channeling in an acoustically bent crystal.

%%%%%%%%%%%%%%%%%%%%%%%%%%%%%%%%%%%%%%%%%%%%%%%%%%%%%%%%%
\section{Concluding remarks}\label{Conclusions}
%%%%%%%%%%%%%%%%%%%%%%%%%%%%%%%%%%%%%%%%%%%%%%%%%%%%%%%%%

Our investigation shows that a crystal, which is periodically bent
by a transverse acoustic wave, can be used for the construction
of an undulator for a beam of ultra-relativistic particles
channeled in the lattice.
The parameters of this undulator can be tuned by varying the
AW amplitude and frequency, the energy of the projectile, and by using
different types of crystals and its channels.
The suggested undulator can be used for the generation of the
radiation of high energy photons.

Also, it is shown that it is meaningful to discuss the possibility
to create a powerful source of a free-electron laser type stimulated
radiation in the energy range of tens of keV up to the MeV region.

We consider our present research as an initial milestone for
more advanced theoretical treatment of the problem. The goal of this
further investigation is to achieve more accurate quantitative
description of the undulator radiation and the corresponding
laser effect. In out opinion, on this way the following phenomena
must be thoroughly considered.

\begin{itemize}

\item[1.] {\em The AIR mechanism for a wide beam.}
When the wide beam enters the crystal it is spread over many
channels.
Each channel is bent by the AW and, hence, one must (when
calculating spontaneous and stimulated emission) take into
account the photon fluxes formed by these sub-beams.
In this case the coherence effect in the photon emission
can be enhanced due to the interaction of the photons with
the particles moving in many neighbouring channels.
In the present paper we have investigated the undulator and
the laser effects in a single channel.

\item[2.] {\em Kinetics in the AIR problem.}
The more advanced description of the AIR problem must take into
account the kinetic behaviour of the particles in the beam as well
as kinetics of the photon flux. Such a description will treat
properly the dechanneling mechanism, the attenuation phenomenon and
the interaction of photons and particles beams.

\item[3.]
{\em Dechanneling in a periodically bent lattice.}
It is necessary to study dechanneling in a periodically bent channel
in comparison with this effect in a straight channel and/or in a channel
with constant (or/and slowly varied) curvature.
So far nobody studied dechanneling in a periodically bent channel
(the AW bending).
The dechanneling length might increase as compared with the linear
channel case.
The dechanneling length plays the crucial role for the laser effect
which we discuss in our paper (see Section \ref{Stimulated}).
Indeed, increasing of the dechanneling length by factor of 10
results in decreasing of the channeling beam density, necessary
for achievement of large gain $G$, by the  factor of 1000  (see
eqs. (56) and (79)).

Let us stress here that special attention must be paid
the problem of dechanneling of a {\it wide} beam in a
periodically bent crystal. Broadening of the channeled beam
in such a medium, which was not considered before, is important
in our problem.

\item[4.]
{\it The bunching effect.}
The effect of the stimulation of photon emission can be
enhanced by increasing density $n$ of the channeling beam.
From the theory of free electron lasers (e.g. \cite{Revista})
it is known that for high $n$ large gain factors can be
achieved in the regime of collective instabilities in the
beam (the bunching effect).
In our future reaserch we plan to apply the
results of the FEL theory to the description of the
laser-type radiation based on the AIR phenomenon.

\item[5.]
{\it Acoustic waves of various configurations.}
In the present paper we investigated spontaneous and stimulated
AIR in an acoustically bent crystal considering
the most simple example of the acoustic wave, the monochromatic
plane wave.
However, other cases of AW (longitudinal waves, spherical waves,
non-monochromatic waves and various combinations thereof),
interacting with the beam of the  channeling particles,
are worthy to study.
By applying the more complex acoustic waves for the crystal
bending one may construct an undulator with  variable
parameters for the generation of high energy photons in
a wide range.
The use of the crystal-based undulator with variable parameters
may result in noticeable increase of the gain factor for the stimulated
AIR analogously to how it occues in the free electron lasers
based on the tampered magnetic wigglers \cite{KMR}.

\item[6.]
{\it Combination of the ordinary channeling radiation and the AIR.}
The present consideration was focused on the case of the
high amplitude AW (see (1)). This allowed us to disregard the
contribution of the ordinary channeling radiation to the total
spectrum of the emitted photons. More general approach must
include both radiative mechanisms, as well as their interference,
simultaneously.

\end{itemize}

The rigorous treatment of the above outlined problems will be the
subject of further publications.

%%%%%%%%  Acknowledgements
\ack

We express our gratitude to Professor E. Uggerh{\o}j for helpful
discussion.
The authors acknowledge support from the DFG, GSI, BMBF and,
especially, from the Alexander von Humboldt Foundation.

\end{document}